\documentclass[aps,prd,twocolumn,amsmath,amssymb,nofootinbib,superscriptaddress,longbibliography,floatfix]{revtex4-2}

\usepackage{aas_macros}
\usepackage[x11names]{xcolor}
\usepackage{orcidlink}
\usepackage{bm}
\usepackage{graphicx}
\usepackage[normalem]{ulem}
\usepackage{float}
\usepackage{placeins}
\usepackage{needspace}
\usepackage{hyperref}
\usepackage{fontawesome5}
\hypersetup{hidelinks}
\usepackage{booktabs}
\usepackage{nicematrix}
\usepackage{siunitx}
\usepackage{tikz}
\usepackage{siunitx}

\newcommand{\project}[1]{\textsf{#1}}

\DeclareRobustCommand{\codelink}[1]{\,\href{#1}{\mbox{\scalebox{0.65}{\faGithub}}}}

\newcommand{\avi}[1]{\textcolor{orange}{[Avi: #1]}}
\newcommand{\renate}[1]{\textcolor{blue}{[Renate: #1]}}
\newcommand{\pmr}[1]{\textcolor{brown}{[PMR: #1]}}
\newcommand{\todo}[1]{\textcolor{orange}{[\textbf{TODO:} #1]}}
\newcommand{\gm}[1]{\textcolor{red}{[Giorgio: #1]}}

\newif\ifshowold
\showoldtrue

\ifdefined\manuscriptclean
  \showoldfalse
  \renewcommand{\avi}[1]{}
  \renewcommand{\gm}[1]{}
  \renewcommand{\todo}[1]{}
  \renewcommand{\pmr}[1]{}
  \renewcommand{\renate}[1]{}
\fi

\hypersetup{
    colorlinks=true,
    linkcolor=blue!70!black,  
    citecolor=green!50!black, 
    urlcolor=blue!50!white    
}

\newcommand{\aucklandstats}{
  Department of Statistics, The University of Auckland,
  Auckland, New Zealand
}
\newcommand{\glasgowphysics}{
  School of Physics and Astronomy, University of Glasgow,
  Glasgow G12 8QQ, United Kingdom
}
\newcommand{\apcparis}{
  Universit\'e Paris Cit\'e, CNRS,
  Astroparticule et Cosmologie,
  F-75013 Paris, France
}
\newcommand{\autmath}{
  Department of Mathematical Sciences,
  Auckland University of Technology,
  Auckland, New Zealand
}

\IfFileExists{showyourwork.tex}{\usepackage{showyourwork}}{}
\providecommand{\script}[1]{}

\begin{document}
\raggedbottom

\title{Bayesian Time-Varying Spectral Estimation \\for LISA with P-Splines}

\author{Avi Vajpeyi~\orcidlink{0000-0002-4146-1132}}
\affiliation{\aucklandstats}

\author{Giorgio Mentasti~\orcidlink{0000-0003-1115-9220}}
\affiliation{\apcparis}

\author{Ollie Burke~\orcidlink{0000-0003-2393-209X}}
\affiliation{\glasgowphysics}

\author{Renate Meyer~\orcidlink{0000-0003-0268-8569}}
\affiliation{\aucklandstats}

\author{Patricio Maturana-Russel~\orcidlink{0000-0002-5211-9818}}
\affiliation{\aucklandstats}\affiliation{\autmath}

\author{Quentin Baghi~\orcidlink{0000-0002-1555-9283}}
\affiliation{\apcparis}

\date{November 9, 2026}

\begin{abstract}
We model the evolutionary spectrum of a slowly time-varying process by a linear combination of products of time- and frequency-dependent B-splines, placing a Gaussian tensor-product penalty prior on the coefficients. The posterior distribution is obtained by updating the prior using a Whittle-type likelihood approximation of the Wilson--Daubechies--Meyer time--frequency coefficients. Separate regularisation in time and frequency allows non-stationary spectral structure to be inferred. An adaptive coarse-grained likelihood reduces the computational cost while preserving fine resolution around sharp spectral features. We first validate the method using repeated simulations of a locally stationary process with known spectrum, showing lower reconstruction errors and narrower posterior intervals as the sample size increases. We then apply it to a year-long simulated LISA data set with stationary instrumental single link spectra, an evolving time-delay interferometry (TDI) response, and a modulated Galactic foreground. We compare a flexible total power spectral density (PSD) model with models incorporating progressively stronger physical information and examine both continuous and gapped observations. 
Allowing time variation substantially improves reconstruction relative to a stationary model, while incorporating physical information about the TDI response further improves recovery, particularly near its moving spectral features.
The results illustrate the trade-off between spectral flexibility, physical assumptions, and posterior uncertainty in non-stationary LISA noise estimation.
\end{abstract}

\maketitle


\section{Introduction}
\label{sec:intro}

The Laser Interferometer Space Antenna (LISA)~\citep{amaroseoane2017lisa} will observe overlapping gravitational-wave
signals in instrumental noise and an unresolved Galactic foreground.
Separating these contributions calls for a global fit of the sources and
noise~\citep{littenberg2023global,katz2025efficient,strub2024global,deng2025modular}.
The noise spectral model determines the weighting of the data in
likelihood-based inference and must therefore be estimated alongside
the signals.

LISA's antenna motion modulates the anisotropic Galactic foreground
annually~\citep{digman2022cyclostationary,buscicchio2025test,
mathur2026timefrequencyanalysislisagalactic}, while changing arm
lengths modify the time-delay interferometry (TDI) noise response
and shift its spectral dips~\citep{Armstrong1999ApJ,Vallisneri2005PhRvD,tinto2021tdi,bayle2019thesis}.
Thermal evolution can affect optical-bench temperatures, electronics,
and telescope alignment~\citep{Armano2021PhRvL}. Changes in spacecraft
orientation, ageing components, drag-free control and clock
synchronisation, and environmental variations such as solar radiation
pressure and cosmic-ray flux may introduce further non-stationarity
on timescales from days to years~\citep{Armano2017PhRvL,Danzmann2003CQGra,Pagone2024,Dimiccoli2025A}.
A single stationary power spectral density (PSD) can therefore
misrepresent long observations. Noise mismodelling can affect
source-parameter uncertainties even when estimators remain
unbiased~\citep{edy2021mismodeling}. PSD uncertainty can propagate
into source inference~\citep{talbot2020uncertain}, and gaps and noise
mismodelling can jointly affect LISA parameter
uncertainties~\citep{burke2025gaps,Yan:2026:arXiv}.
Realistic noise and foreground modelling also matters for stochastic
gravitational-wave background reconstruction~\citep{Kume:2025:JCAP,karnesis2026noise}.

Flexible spectral estimation has a substantial statistical literature.
Bayesian nonparametric approaches include flexible basis expansions
and mixture representations, such as those based on Bernstein
polynomials and B-splines~\citep{beyond_whittle,edwards_bayesian_2019}.
P-spline methods use a fixed B-spline basis together with a roughness
penalty to control the smoothness of the fitted function~\citep{eilers1996flexible}.
For example,~\citet{maturana2021bayesian} developed a Bayesian
P-spline approach to PSD estimation using the Whittle
likelihood~\citep{whittle1953analysis}.
Penalised Whittle likelihoods are also used in frequentist spectral
estimation~\citep{pawitan1994nonparametric}.
These approaches provide flexible alternatives to fully specified
parametric spectral models, while controlling the effective
complexity of the estimated spectrum through priors or penalties.

In gravitational-wave data analysis, flexible spectral models are
widely used for noise and foreground estimation.
For example, \project{BayesLine} models the broadband noise spectrum
with splines and narrow spectral features with Lorentzian
components~\citep{littenberg_bayesian_2015}.
Spectral fitting is also an important component of LISA foreground
estimation and global analysis~\citep{karnesis2021characterization,rosati2025prototype}.
Recent LISA applications include flexible single-link B-spline noise
models~\citep{Baghi_2023} and TDI-level spline models with
reversible-jump inference of the number and location of
knots~\citep{santini2025sgwb}.
Related approaches include other flexible noise models~\citep{Aimen:2026,hoy2026noisypipelinesiinonparametric},
multivariate spectral estimation for gravitational-wave
detectors~\citep{vajpeyi2026multivar,jiananET,jiananLISA,
yixuanLISAandET, burke2026noisypipelinesiendtoend}, and weakly parametric stochastic-background
inference~\citep{pozzoli2024weakly}.
These developments demonstrate the value of flexible spectral
models for gravitational-wave inference, particularly when the
underlying noise cannot be described adequately by a small
number of physical parameters.

Modelling spectral evolution requires extending these ideas to
non-stationary time series. Bayesian time-varying spectral estimation
builds on the theory of evolutionary spectra~\citep{priestley1965evolutionary} and locally stationary
processes~\citep{dahlhaus1997fitting}.
Existing approaches include segmentation-based
models~\citep{rosen2012adaptspec,rosen2009mixture},
adaptive local Fourier bases~\citep{ombao2001slex,ombao2005slex},
direct time--frequency smoothing~\citep{guo2003smoothing,wolfe2004bayesian},
and locally stationary wavelet
models~\citep{nason2000wavelet}.
Joint time--frequency inference of noise and stochastic backgrounds
has also been explored using Hamiltonian Monte Carlo~\citep{pozzoli2025bahamas,neal2011mcmc}.
Using this framework,~\citet{buscicchio2025firstyear} investigated
the reconstruction of the LISA Galactic foreground, including
the effects of time segmentation and data gaps, and found
strong evidence for non-stationarity.
Related work by~\citet{baghi2026statisticstimefrequencyaveragedspectra} examines the
statistical effects of averaging spectra in time and frequency,
including the validity of independence approximations and
optimal time segmentation for locally stationary processes.

The Wilson--Daubechies--Meyer (WDM) transform provides a
time--frequency representation of the data using real-valued
coefficients localised in both time and frequency~\citep{Necula_2012,cornish2020Wavelets,johnson2026wdmformalism,
vajpeyi2026wdm}.
Other time-frequency representations include moving
periodograms~\citep{HaefnerKirch,tang2026MovingPSD}
and 
multitaper methods~\citep{thomson1982spectrum}.
Multiresolution time--frequency transforms have also been used
in gravitational-wave burst searches~\citep{chatterji2004qtransform}.
Recent WDM work examines approximate diagonalisation of the
noise covariance~\citep{cornishCovar2026wavelet}, relates the
evolutionary spectrum to the full covariance of the WDM
coefficients~\citep{cornish2026modelingnonstationarynoiseapplications},
and applies time--frequency covariance modelling to the
cyclostationary Galactic
foreground~\citep{mathur2026timefrequencyanalysislisagalactic}.

These developments motivate spectral models that can capture
time variation over long observations without requiring a fully
specified physical description of every noise contribution.
For LISA, such models must accommodate the evolving Galactic
foreground and the time-dependent features of the instrumental
TDI response. Computational efficiency is also important given
the duration and sampling rate of the observations.
The combination of a localised time--frequency representation
with a smooth, regularised spectral model offers a practical
approach to these challenges.

\Needspace{3\baselineskip}
In this work, we combine an approximate WDM Whittle likelihood,
which neglects correlations between distinct WDM coefficients,
with a tensor-product P-spline model for the logarithm of the
evolutionary spectral density, with separate regularisation in time
and frequency. Under suitable smoothness assumptions, this spectrum
approximates the normalised WDM coefficient variances.
We use the open-source WDM implementation described by~\citet{vajpeyi2026wdm}, with other implementations available
in~\citep{Moore_WDM_GW_wavelets,matt_code_WDMWaveletTransforms,
cornish_code_WDM_Transform,farr_code_WDMWavelets_jl}.
The model fits either total power or a smooth correction to a
nominal instrumental spectrum, and pools neighbouring WDM
powers to reduce likelihood cost.
We assess reconstruction and posterior uncertainty in repeated
simulations of the locally stationary, time-varying MA(1) process
(LS2, described in Sec.~\ref{sec:sim}), and in continuous and
gapped versions of a simulated LISA record, including comparison
with a parametric model.
We evaluate spectral recovery, not source-parameter recovery
within a global fit.


\section{Bayesian model}
\label{sec:model}

\subsection{WDM likelihood}
\label{sec:obs}

Let $x[k]=x(k\Delta t)$,
$k=0,\ldots,N-1$, denote a real, zero-mean Gaussian
time series with one-sided evolutionary spectral density $S(t,f)$
and duration $T=N\Delta t$. We do not assume global stationarity.
For even $N_t,N_f$ with $N=N_tN_f$, the
Wilson--Daubechies--Meyer tile centres are
\begin{equation}
\begin{aligned}
t_n &= n\Delta T, &
\Delta T &= N_f\Delta t,\\
f_m &= m\Delta F, &
\Delta F &= \frac{1}{2N_f\Delta t}.
\end{aligned}
\codelink{https://github.com/pywavelet/wdm_transform/blob/v0.5.0/src/wdm_transform/datatypes/wdm.py\#L227-L261}
\label{eq:wdm_grid}
\end{equation}
Here $n=0,\ldots,N_t-1$ and $m=1,\ldots,N_f-1$.
We exclude the DC and Nyquist frequency bins from the
analysis.
Following~\citet{vajpeyi2026wdm}, we define the
Fourier and WDM coefficients as
\begin{equation}
\begin{aligned}
\tilde{x}[l]
&=\sum_{k=0}^{N-1}x[k]e^{-2\pi i kl/N},\\
w_{nm}
&=\sum_{l=-N/2}^{N/2-1}
\tilde{x}[l]\tilde{g}_{nm}^{*}[l],
\end{aligned}
\codelink{https://github.com/pywavelet/wdm_transform/blob/v0.5.0/src/wdm_transform/transforms/xp_backend.py\#L65-L113}
\label{eq:wdm_transform}
\end{equation}
where $l=-N/2,\ldots,N/2-1$ indexes the Fourier
frequencies $l/T$.

For completeness, we give the interior Wilson basis,
Meyer window, and phase factor below. We use the
cosine-tapered Meyer window with $A=B=1/3$,
satisfying $2A+B=1$.

\begin{widetext}
\begin{subequations}
\label{eq:wdm_basis_definitions}
\begin{equation}
\tilde{g}_{nm}[l]
=\frac{e^{-2\pi i nl/N_t}}{\sqrt{2}}
\left[
C_{nm}\tilde{\varphi}[l-mN_t/2]
+C_{nm}^{*}\tilde{\varphi}[l+mN_t/2]
\right],
\qquad 1\leq m\leq N_f-1.
\codelink{https://github.com/pywavelet/wdm_transform/blob/v0.5.0/src/wdm_transform/windows.py\#L129-L189}
\label{eq:wdm_basis}
\end{equation}
\begin{equation}
\begin{gathered}
\tilde{\varphi}[l]
=\sqrt{\frac{2}{N_t}}
\begin{cases}
1, & |l_r|<A,\\[2pt]
\displaystyle
\cos\!\left[\frac{\pi}{2}
\frac{|l_r|-A}{B}\right],
& A\leq |l_r|<A+B,\\[2pt]
0, & \text{otherwise},
\end{cases}
\qquad l_r=\frac{2l}{N_t},\\[6pt]
C_{nm}=\begin{cases}
1, & n+m\ \text{even},\\
i, & n+m\ \text{odd}.
\end{cases}
\end{gathered}
\codelink{https://github.com/pywavelet/wdm_transform/blob/v0.5.0/src/wdm_transform/windows.py\#L58-L126}
\label{eq:meyer_window}
\end{equation}
\end{subequations}
\end{widetext}

The interior basis functions satisfy
$\sum_l|\tilde{g}_{nm}[l]|^2=1$.
Since the transform is linear, the real WDM coefficients
$w_{nm}$ are jointly zero-mean Gaussian.

We model the one-sided evolutionary spectral density
$S(t,f)$. If the process varies sufficiently slowly over
the temporal extent of the WDM basis functions, and
the spectrum is approximately constant across their
frequency support, the \emph{midpoint approximation} (see~\citep{JohnsonWDMII2026}) gives
\begin{equation}
\frac{2\Delta t}{N}\mathbb E[w_{nm}^{2}]
\simeq S(t_n,f_m).
\codelink{https://github.com/nz-gravity/lisa_timevarying_logpsplines_study/blob/v0.1.1/src/lisa_psd_analysis/prepare.py\#L137-L144}
\label{eq:wdm_estimand}
\end{equation}
The factor $2\Delta t/N$ converts the coefficient variance
to one-sided PSD units. The \emph{band-averaged approximation},
which accounts for spectral variation across the WDM
frequency window, is described in
Appendix~\ref{app:wdm_projection}~\citep{JohnsonWDMII2026}.

Approximating the marginal WDM coefficient variances and
neglecting correlations between distinct coefficients are
separate assumptions. WDM orthogonality does not generally
imply independent coefficients for coloured or non-stationary
noise~\citep{cornish2020Wavelets,cornishCovar2026wavelet,JohnsonWDMII2026}.
Under the midpoint and diagonal covariance approximations,
\begin{equation}
\operatorname{Cov}(w_{nm},w_{n'm'})
\simeq
\frac{N}{2\Delta t}S(t_n,f_m)
\delta_{(n,m),(n',m')}.
\label{eq:wdm_covariance}
\end{equation}
Here $\delta_{(n,m),(n',m')}$ is the Kronecker delta,
equal to one if $n=n'$ and $m=m'$, and zero otherwise.

For a zero-mean Gaussian coefficient, its squared amplitude
divided by its exact variance follows a $\chi_1^2$
distribution. Under the midpoint approximation, this gives
\begin{equation}
\frac{2\Delta t}{N}
\frac{w_{nm}^{2}}{S(t_n,f_m)}
\overset{\mathrm{approx.}}{\sim}\chi_1^2.
\end{equation}
Neglecting correlations between distinct coefficients
then gives the approximate WDM Whittle likelihood,
\begin{equation}
\begin{split}
&\ln\mathcal L_{\rm WDM}(\mathbf x\mid S)
=\\
&-\frac12\sum_{(n,m)\in\mathcal I}
\left[
\ln S(t_n,f_m)
+\frac{2\Delta t}{N}
\frac{w_{nm}^{2}}{S(t_n,f_m)}
\right]\\
&\quad+\mathrm{const}.
\codelink{https://github.com/nz-gravity/LogPSplinePSD/blob/v0.2.0/src/log_psplines/likelihoods/whittle.py\#L12-L29}
\end{split}
\label{eq:wdm_like}
\end{equation}

Here $\mathcal I$ contains the analysed interior WDM
cells, and the constant includes model-independent terms.

\subsection{Tensor-product P-spline prior}
\label{sec:pspline}
\label{sec:tvpsd}

We model the log evolutionary spectral density with a
tensor-product P-spline~\citep{eilers2003multivariate,wood2017generalized},
\begin{equation}
\ln S(t,f)
=\sum_{k=1}^{K_t}\sum_{l=1}^{K_f}
c_{kl}B_k^{(t)}(t)B_l^{(f)}(f).
\codelink{https://github.com/nz-gravity/LogPSplinePSD/blob/v0.2.0/src/log_psplines/models/spectrum.py\#L62-L73}
\label{eq:pspline_surface}
\end{equation}
Here $B_k^{(t)}$ and $B_l^{(f)}$ are cubic B-spline basis
functions and $c_{kl}$ are unknown coefficients.
We rescale each coordinate to $[0,1]$ and use repeated
boundary knots, giving $K_t-4$ and $K_f-4$ interior
knots, respectively. The basis dimensions and knot
locations are fixed before sampling. Closer knots
permit finer local structure.

The spline defines $S(t,f)$ continuously in time and
frequency. Under the midpoint approximation, its values
at the WDM tile centres enter the likelihood in
Eq.~\eqref{eq:wdm_like}.

Let $N_t^{\rm grid}$ and $N_f^{\rm grid}$ denote the numbers
of time and frequency points on the analysis grid.
With
$[\mathbf B_t]_{nk}=B_k^{(t)}(t_n)$,
$[\mathbf B_f]_{ml}=B_l^{(f)}(f_m)$,
$[\mathbf C]_{kl}=c_{kl}$,
and $[\mathbf S]_{nm}=S(t_n,f_m)$,
the spline model evaluated on this grid becomes
\begin{equation}
\ln\mathbf S=\mathbf B_t\mathbf C\mathbf B_f^\top,
\codelink{https://github.com/nz-gravity/LogPSplinePSD/blob/v0.2.0/src/log_psplines/models/spectrum.py\#L62-L73}
\label{eq:pspline_matrix}
\end{equation}
where the logarithm acts elementwise and
\begin{equation*}
\begin{gathered}
\mathbf B_t\in\mathbb R^{N_t^{\rm grid}\times K_t},
\qquad
\mathbf B_f\in\mathbb R^{N_f^{\rm grid}\times K_f},\\
\mathbf C\in\mathbb R^{K_t\times K_f},
\qquad
\mathbf S\in
\mathbb R^{N_t^{\rm grid}\times N_f^{\rm grid}}.
\end{gathered}
\end{equation*}

Substituting
$S(t_n,f_m)=
\exp\{[\mathbf B_t\mathbf C\mathbf B_f^\top]_{nm}\}$
into Eq.~\eqref{eq:wdm_like} gives the approximate WDM
likelihood in terms of the unknown spline coefficients.
We complete the Bayesian model by assigning a Gaussian
roughness prior to the column-stacked coefficient vector
$\mathbf c=\operatorname{vec}(\mathbf C)$.
The prior penalises curvature separately in time and
frequency, as described below.

To favour smooth surfaces, we penalise curvature in time and
frequency~\citep{wood2017derivative}. For the time basis, we define
the normalised second-derivative penalty as
\begin{equation}
[\mathbf P_t]_{kk'}=
\frac{
\displaystyle\int B_k^{(t)\prime\prime}(t)
B_{k'}^{(t)\prime\prime}(t)\,\mathrm dt
}{
\displaystyle\sum_{j=1}^{K_t}\int
\left[B_j^{(t)\prime\prime}(t)\right]^2\,\mathrm dt
}.
\codelink{https://github.com/nz-gravity/LogPSplinePSD/blob/v0.2.0/src/log_psplines/basis/penalty.py\#L70-L105}
\label{eq:time_penalty}
\end{equation}
The frequency penalty $\mathbf P_f$ is defined analogously.

For $\mathbf c$, the combined
penalty has the Kronecker-sum form~\citep{wood2017generalized}
\begin{equation}
\mathbf{Q}(\sigma_t,\sigma_f)
= \sigma_t^{-2}\,(\mathbf{I}_{K_f}\otimes\mathbf{P}_t)
+ \sigma_f^{-2}\,(\mathbf{P}_f\otimes\mathbf{I}_{K_t}).
\codelink{https://github.com/nz-gravity/LogPSplinePSD/blob/v0.2.0/src/log_psplines/basis/penalty.py\#L44-L67}
\label{eq:penalty}
\end{equation}
Here $\mathbf P_t\in\mathbb R^{K_t\times K_t}$,
$\mathbf P_f\in\mathbb R^{K_f\times K_f}$, $\mathbf I_K$ is the
$K\times K$ identity matrix, and $\otimes$ denotes the Kronecker product.
The first term penalises temporal curvature for each frequency component,
while the second penalises frequency curvature for each temporal component.
The positive roughness scales $\sigma_t$ and $\sigma_f$ control the permitted
curvature in time and frequency: smaller values give stronger smoothing,
while larger values permit more curvature.

To make $\mathbf Q$ invertible, we replace its zero eigenvalues
with the fixed precision $\tau_0=10^{-4}$ and add a ridge
$\epsilon=10^{-6}$ to the remaining eigenvalues
(Appendix~\ref{app:penalty}). The resulting regularised penalty
$\tilde{\mathbf Q}$ is the precision matrix of the coefficient prior
\begin{equation}
\mathbf{c}\mid\sigma_t,\sigma_f \sim
\mathcal{N}\!\left(\mathbf{0},\,\tilde{\mathbf{Q}}^{-1}(\sigma_t,\sigma_f)\right).
\codelink{https://github.com/nz-gravity/LogPSplinePSD/blob/v0.2.0/src/log_psplines/inference/power.py\#L281-L295}
\label{eq:prior}
\end{equation}

We assign independent HalfNormal priors directly to the roughness scales,
\begin{equation}
\sigma_t,\ \sigma_f\sim\operatorname{HalfNormal}(10).
\codelink{https://github.com/nz-gravity/LogPSplinePSD/blob/v0.2.0/src/log_psplines/inference/power.py\#L46-L49}
\label{eq:roughness_prior}
\end{equation}
Here 10 is the scale parameter.
The effect of this prior depends on several factors, including the number and order of the basis functions, the knot locations, and the order of the penalty matrix. Although the basis order and penalty order are fixed across studies, each study explicitly specifies the number of basis functions and the knot layout.

\subsection{Coarse-grained likelihood}
\label{sec:coarse}

We reduce the likelihood cost by grouping neighbouring
WDM cells over which $\ln S$ varies little and
approximating the PSD as constant within each group.
Let $\mathcal T$ and $\mathcal F$ denote partitions
of the time and frequency indices, respectively,
\begin{align}
\mathcal T &= \{\mathcal T_i\}_{i=1}^{I},\\
\mathcal F &= \{\mathcal F_j\}_{j=1}^{J}. \nonumber
\end{align}
Their Cartesian products define rectangular blocks
on the WDM grid,
\begin{equation}
b_{ij}=\mathcal T_i\times\mathcal F_j.
\end{equation}
We evaluate the model at the mean time $\bar t_i$
and frequency $\bar f_j$ of the WDM tile centres
in each group, writing
\begin{equation*}
S_{ij}=S(\bar t_i,\bar f_j).
\end{equation*}

Let $M_{nm}=1$ for cells used in the fit and $0$
otherwise. For each non-empty block, we retain
its mean observed power $\bar P_{ij}$ and the
number of contributing WDM cells $\nu_{ij}$,
\begin{align}
\bar P_{ij}
&=
\frac{2\Delta t}{N\nu_{ij}}
\sum_{(n,m)\in b_{ij}}M_{nm}w_{nm}^2,
\nonumber\\
\nu_{ij}
&=
\sum_{(n,m)\in b_{ij}}M_{nm}.
\codelink{https://github.com/nz-gravity/LogPSplinePSD/blob/v0.2.0/src/log_psplines/preprocessing/power_partition.py\#L146-L171}
\label{eq:masked_pooling}
\end{align}

Under the blockwise-constant PSD approximation,
the coarse-grained WDM likelihood is
\begin{equation}
\ln\mathcal L_{\rm WDM}^{\rm cg}
=
-\frac12
\sum_{\substack{i,j\\\nu_{ij}>0}}
\nu_{ij}
\left[
\ln S_{ij}+\frac{\bar P_{ij}}{S_{ij}}
\right]
+\mathrm{const}.
\codelink{https://github.com/nz-gravity/LogPSplinePSD/blob/v0.2.0/src/log_psplines/likelihoods/whittle.py\#L12-L29}
\label{eq:binned-likelihood}
\end{equation}
Empty blocks, such as those falling entirely within
data gaps, contribute no likelihood term.

To construct the blocks, we first form a smooth
``pilot'' estimate of the local spectrum and use it
to identify regions where $\ln S$ varies slowly.
In frequency, this gives wider bins in smooth regions
and finer bins around sharp spectral features.
Time blocks are approximately fixed in width, with
boundaries adjusted near gaps and missing data.
The resulting partition is fixed before sampling,
and the pilot estimate does not enter the likelihood.

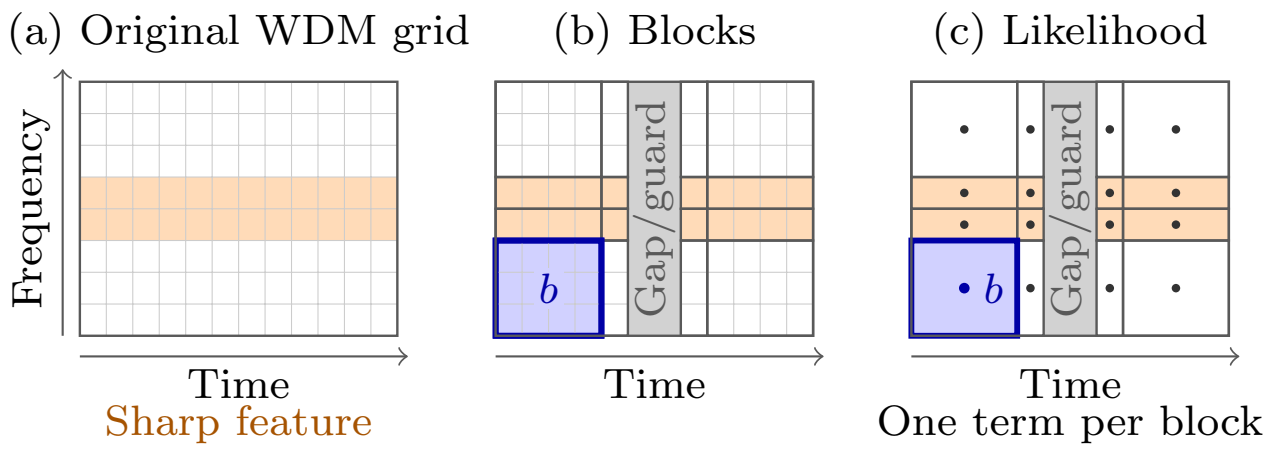
\begin{figure}[h]
\centering
\begin{tikzpicture}[x=0.175cm,y=0.21cm,
  every node/.style={font=\footnotesize,inner sep=1pt},
  native/.style={draw=black!20,line width=0.2pt},
  partition/.style={draw=black!65,line width=0.45pt},
  selected/.style={draw=blue!65!black,fill=blue!18,line width=1.1pt}]
  \foreach \panel/\shift/\heading in {
    1/0/{(a) Original WDM grid},
    2/2.75/{(b) Blocks},
    3/5.5/{(c) Likelihood}} {
    \begin{scope}[xshift=\shift cm]
      \node[anchor=south] at (6,8.65) {\heading};
      \fill[orange!28] (0,3) rectangle (12,5);
      \ifnum\panel<3
        \draw[native,step=1] (0,0) grid (12,8);
      \fi
      \ifnum\panel>1
        \fill[gray!35] (5,0) rectangle (7,8);
        \node[rotate=90,text=black!65] at (6,4) {Gap/guard};
        \foreach \left/\right in {0/4,4/5,7/8,8/12} {
          \foreach \bottom/\top in {0/3,3/4,4/5,5/8} {
            \draw[partition] (\left,\bottom) rectangle (\right,\top);
            \ifnum\panel=3
              \fill[black!80] ({(\left+\right)/2},{(\bottom+\top)/2}) circle (0.8pt);
            \fi
          }
        }
        \draw[selected] (0,0) rectangle (4,3);
        \ifnum\panel=2
          \draw[native,step=1] (0,0) grid (4,3);
          \draw[blue!65!black,line width=1.1pt] (0,0) rectangle (4,3);
          \node[text=blue!65!black,fill=blue!18] at (2,1.5) {$b$};
        \else
          \fill[blue!65!black] (2,1.5) circle (1pt);
          \node[text=blue!65!black,anchor=west] at (2.5,1.5) {$b$};
        \fi
      \fi
      \draw[black!65,line width=0.45pt] (0,0) rectangle (12,8);
      \draw[->,black!65] (0,-0.65) -- (12.4,-0.65);
      \node[anchor=north] at (6,-0.85) {Time};
      \ifnum\panel=1
        \node[anchor=north,text=orange!65!black] at (6,-2.0) {Sharp feature};
      \fi
      \ifnum\panel=3
        \node[anchor=north] at (6,-2.0) {One term per block};
      \fi
      \ifnum\panel=1
        \draw[->,black!65] (-0.65,0) -- (-0.65,8.4);
        \node[rotate=90,anchor=south] at (-0.9,4) {Frequency};
      \fi
    \end{scope}
  }
\end{tikzpicture}
\caption{Likelihood compression starting from
the original WDM time--frequency grid.
(a) The original WDM grid.
(b) Missing data are excluded before pooling.
Frequency bins are wider where the spectrum varies
slowly and narrower around the sharp spectral
feature (orange band), while time blocks remain
approximately regular and stop at data gaps.
(c) Each block contributes one likelihood term.
The highlighted block is summarised by its mean
power $\bar P_{ij}$ and contributing-cell count
$\nu_{ij}$.}
\label{fig:coarse-schematic}
\end{figure}

Figure~\ref{fig:coarse-schematic} illustrates the
construction, while Appendix~\ref{app:coarse}
describes the partitioning procedure, model
conventions, and validation.

\subsection{Posterior inference}
\label{sec:inference}

We perform posterior inference using the No-U-Turn Sampler (NUTS) in
\project{NumPyro}~\citep{hoffman2014_NUTS,phan2019composable}, with
\project{JAX}~\citep{jax2018github} for automatic differentiation.
We sample the spline coefficients and smoothing scales jointly.
These scales are $\sigma_t$ and $\sigma_f$ for the tensor-product model,
and $\sigma_g$ and $\sigma_\eta$ for the orbit-informed model in
Sec.~\ref{sec:lisa_hypotheses}. Each chain is initialised from a penalised
least-squares fit to the WDM powers. The implementation uses standard centred and non-centred
parameterisations of hierarchical Gaussian priors to improve sampling
efficiency~\citep{neal2003slice,papaspiliopoulos2007general,betancourt2015hamiltonian}.
Implementation details are provided in the tagged codebase.

The LS2 fits (Sec.~\ref{sec:sim}) use two chains with $500$ warm-up and $500$ retained draws
per chain. The LISA fits (Sec.~\ref{sec:lisa}) use two sequential chains with $2200$ warm-up
and $4000$ retained draws per chain.

We assess convergence using the rank-normalised potential
scale reduction statistic $\widehat R\leq1.01$~\cite{gelman1992inference}, which
compares agreement across chains, and bulk and tail
effective sample sizes ($\mathrm{ESS}\geq400$), which
measure sampling efficiency in the centre and tails of
the posterior~\citep{vehtari2021rank}. We also require zero divergent transitions, which indicate
numerical integration difficulties, and an energy Bayesian
fraction of missing information (E-BFMI) $>0.3$,
which assesses exploration of the Hamiltonian energy
distribution~\citep{Betancourt:2016:arXiv}.
We additionally monitor maximum tree-depth saturation.
We use these diagnostics to assess sampling reliability.


\subsection{Evaluation criteria}
\label{sec:criteria}

For each posterior draw, we evaluate the model predictions
on the time--frequency grid. At each $(t_n,f_m)$, we
summarise these values by their posterior medians
$\widehat S_{nm}$ and the 5th and 95th percentiles,
$S_{05,nm}$ and $S_{95,nm}$.

We assess reconstruction against a reference spectrum
$S_{nm}^{\rm true}$ constructed from the generating model on the same grid.
This reference approximates the normalised expected WDM power.
Its construction is described with each study in
Secs.~\ref{sec:sim} and~\ref{sec:lisa_data}.

We compare the reconstruction with $S^{\rm true}_{nm}$ using the
root-mean-square log error
\begin{equation}
    {\rm RMSE}_{\log S}
    =
    \sqrt{
        \left\langle
        \left[
        \ln\left(
            \frac{\widehat S_{nm}}{S^{\rm true}_{nm}}
        \right)
        \right]^2
        \right\rangle
    }\ ,
\codelink{https://github.com/nz-gravity/tvpsd_lisa_demo/blob/0603921dce1d7ba0944191295b4df1d09867d710/src/lisa_study/publication_assets.py\#L97-L103}
\end{equation}
where $\langle\cdot\rangle$ denotes an average over the time--frequency grid used for
evaluation. We quantify posterior uncertainty using the logarithmic width of the 
90\% pointwise posterior interval
\begin{equation}
    W_{nm}
    =
    \ln\left(
        \frac{S_{95,nm}}{S_{05,nm}}
    \right),
\codelink{https://github.com/nz-gravity/tvpsd_lisa_demo/blob/0603921dce1d7ba0944191295b4df1d09867d710/src/lisa_study/publication_assets.py\#L97-L103}
\end{equation}
reported as a mean or median over the grid. We also report the fraction of 90\% posterior intervals containing the
reference
\begin{equation}
    {\rm Inc}_{90}
    =
    \frac{
        \#\{S_{05,nm}\leq S^{\rm true}_{nm}\leq S_{95,nm}\}
    }{
        \#\{\text{time-frequency cells}\}
    }\ .
\codelink{https://github.com/nz-gravity/tvpsd_lisa_demo/blob/0603921dce1d7ba0944191295b4df1d09867d710/src/lisa_study/publication_assets.py\#L97-L103}
\end{equation}

For the time-averaged spectrum $\overline S(f_m)$, we average
the grid predictions over the retained evaluation times. For the band-averaged
spectrum $S_{\mathcal B}(t_n)$, we average over the original frequency
cells in a specified band $\mathcal B$. Each average is computed for
each draw in linear PSD units before taking posterior medians and
intervals. These descriptive summaries of an evolving spectrum do not
assume global stationarity and can hide local spectral features.
For a single LISA realisation, truth inclusion is descriptive.
Repeated LS2 simulations allow assessment of interval calibration
relative to the analytic reference.

We also assess how effectively the inferred spectrum whitens
the data, without requiring an analytic reference spectrum.
Using the WDM normalisation of Ref.~\cite{vajpeyi2026wdm},
the standardised coefficients are expected to approximately
follow a standard-normal marginal distribution,
\begin{equation}
    \sqrt{\frac{2\Delta t}{N}}\,
    \frac{w_{nm}}{\sqrt{\widehat S_{nm}}}
    \overset{\mathrm{approx.}}{\sim}
    \mathcal{N}(0,1).
\end{equation}
We quantify agreement using the Anderson--Darling statistic
$A^2$, with the reference mean and variance fixed at zero
and unity. Smaller $A^2$ values indicate closer agreement for a
fixed sample size, with particular sensitivity to deviations
in the tails. For independent samples, the asymptotic
5\% critical value for the fixed $\mathcal{N}(0,1)$
distribution is $A^2 = 2.492$. Values below this threshold
do not reject the standard-normal hypothesis.

Because the spectrum is inferred from the same data and
the coefficients may remain correlated, we interpret
$A^2$ as a descriptive diagnostic of marginal agreement
rather than a formal hypothesis test.


\section{Controlled simulation study}
\label{sec:sim}

Before applying the method to LISA data, we test it in a controlled setting
where the true time-varying spectrum is known. This allows us to measure
reconstruction accuracy and posterior uncertainty directly, and to examine
how both change as the amount of data increases.

Following~\citet{tang2026MovingPSD}, we use the locally stationary
time-varying MA(1) process (LS2)
\begin{equation}
\begin{aligned}
X_{j,N}
&=
\epsilon_j+b_1(t_j)\epsilon_{j-1},
\\
t_j&=\frac{j}{N},
\qquad
j=0,\ldots,N-1,
\end{aligned}
\codelink{https://github.com/nz-gravity/LogPSplinePSD/blob/v0.2.0/src/log_psplines/example_datasets/ls2_data.py\#L143-L161}
\label{eq:ls2}
\end{equation}
where $\epsilon_j$ are independent standard-normal random variables. For this example, time is rescaled to $0\leq t\leq1$, with
\begin{equation}
b_1(t)=1.1\cos(1.5-\cos4\pi t).
\codelink{https://github.com/nz-gravity/LogPSplinePSD/blob/v0.2.0/src/log_psplines/example_datasets/ls2_data.py\#L121-L141}
\end{equation}
The function $b_1(t)$ completes two cycles over the record, causing the
spectral shape to vary continuously with time rather than simply rescaling
a fixed spectrum. 
For comparison with the fitted WDM power, we use the
locally stationary MA(1) spectrum evaluated at the
WDM tile centres as the analytic reference (the midpoint approximation),
\begin{equation}
S^{\rm true}_{nm}
=
2\Delta t
\left[
1+b_1(t_n)^2
+2b_1(t_n)\cos(2\pi f_m\Delta t)
\right],
\label{eq:ls2_true_psd}
\end{equation}
for $0<f_m<1/(2\Delta t)$, where
$\Delta t=0.1\,\mathrm{s}$.
The WDM grid uses $N_f=32$ frequency channels. For
$N=2^{10},\ldots,2^{15}$, the cubic time bases have
$K_t=13,16,20,25,31,40$ functions, respectively. The cubic frequency
basis has $K_f=10$ functions in every case. Interior knots are equally
spaced in each coordinate, with repeated boundary knots.

\begin{figure*}[!t]
  \centering
  \script{plot_ls2_summary.py}
\includegraphics[width=\linewidth]{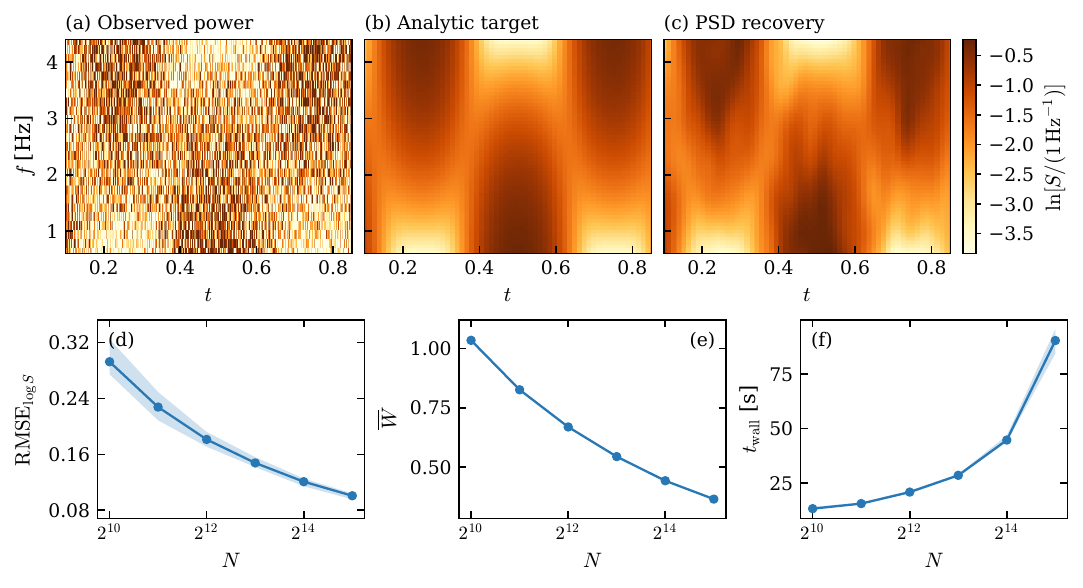}
  \caption{LS2 recovery and performance.
  (a) Observed WDM log power, (b) the scaled analytic target
  $S^{\rm true}_{nm}$, and (c) the recovered log PSD for an $N=2^{15}$
  realisation.
  Panels (d)--(f) show $\mathrm{RMSE}_{\log S}$, mean log-interval width
  $\overline W$, and two-chain NUTS wall time $t_{\rm wall}$.
  Curves and shading show medians and interquartile ranges over 100
  realisations at each record length.
  The metrics are defined in Sec.~\ref{sec:criteria}.}
  \label{fig:ls2-surface}
\end{figure*}

We simulate 100 independent realisations at each
$N\in\{2^{10},2^{11},\ldots,2^{15}\}$. 
Figure~\ref{fig:ls2-surface} shows that reconstruction improves as the
record length increases. From $N=2^{10}$ to $2^{15}$, the median
$\mathrm{RMSE}_{\log S}$ and median $\overline W$ both decrease by about
$65\%$.

Across the 100 independent realisations at each sample size, the median
fraction of evaluated time--frequency cells whose 90\% posterior intervals
contain the analytic reference remains close to 90\%. This supports
approximate interval calibration relative to the reference in this
controlled LS2 study, while reconstruction errors and posterior interval
widths decrease as the sample size increases.

Across all record lengths, Anderson--Darling statistics for $\widehat S_{nm}$ ($A^2=0.65$ to $1.11$) are also comparable to those obtained
using the $S^{\rm true}_{nm}$ ($A^2=0.65$ to $1.07$). 
All reported values remain below the nominal 5\% critical value of
$2.492$, indicating no substantial departures from marginal
standard normality in these summaries.

\section{Application to simulated LISA data}
\label{sec:lisa}

\subsection{Data generation}
\label{sec:lisa_data}

We generate instrumental noise with \project{lisainstrument} and
\project{PyTDI}~\citep{bayle2023lisainstrument,staab2023pytdi},
using ESA trailing orbits with unequal, time-dependent arms.
The simulation includes test-mass acceleration (TM) and optical
metrology system (OMS) noise in the second-generation Michelson
variables $X_2,Y_2,Z_2$. The noise PSDs of the individual
instrument links are stationary, while the corresponding TDI
channel PSDs vary with time due to the evolving arm lengths
and transfer functions (see Appendix~\ref{app:lisa_orb_inference}).

We add a simulated unresolved Galactic foreground to the
instrumental $X_2,Y_2,Z_2$ streams. Rather than simulating
individual Galactic binaries, we generate a correlated
Gaussian realisation from a prescribed strain PSD and a
fixed anisotropic sky distribution, represented by a
\texttt{HEALPix} map~\cite{Gorski2005}.

We use \texttt{backgrounds}~\cite{Baghi_2023} to propagate
this foreground through the orbit-dependent second-generation
TDI response. The resulting XYZ cross-spectral matrix is used
to generate correlated Gaussian Fourier coefficients, which
are transformed into time-domain data. Although the intrinsic
strain PSD and sky distribution are fixed, the observed TDI
foreground evolves with the detector response.
Appendix~\ref{app:lisa_orb_inference} provides further details.

The injected Galactic strain PSD follows the analytic
model of~\citet{karnesis2021characterization},
\begin{equation}
\begin{split}
 S_{\rm gal}(f;\theta)
 &=\frac{A}{2}f^{-7/3}
 \exp[-(f/f_1)^\alpha] \\
 &\quad\times
 [1+\tanh((f_{\rm knee}-f)/f_2)],
\end{split}
\codelink{https://github.com/nz-gravity/lisa_timevarying_logpsplines_study/blob/v0.1.1/src/lisa_psd_analysis/galactic.py\#L50-L65}
\label{eq:karnesis_foreground}
\end{equation}
where $\theta=(A,f_1,f_2,f_{\rm knee},\alpha)$.
We adopt the one-year running-mean prescription of~\citet{karnesis2021characterization}, with a source-resolution
threshold of ${\rm SNR}=7$, to specify the injected PSD
parameters. This prescription does not imply that individual
binaries are simulated or subtracted in our study.
The sky map is normalised to unit total power~\citep{Baghi_2023}.
The same spectral family is used in the parametric model
$\mathcal H_{\rm para}$.
Table~\ref{tab:sim_params} summarises the numerical settings.

\begin{table}[t]
\centering
\caption{LISA simulation and analysis settings.}
\label{tab:sim_params}
\footnotesize
\setlength{\tabcolsep}{4pt}

\begin{NiceTabular}[width=\columnwidth]{@{}X[3,l] X[5,l]@{}}
\CodeBefore
  \rowcolor{gray!6}{2}
  \rowcolor{gray!6}{4}
  \rowcolor{gray!6}{6}
  \rowcolor{gray!6}{8}
  \rowcolor{gray!6}{10}
  \rowcolor{gray!6}{12}
  \rowcolor{gray!6}{14}
  \rowcolor{gray!6}{16}
\Body
\midrule
\multicolumn{2}{@{}l@{}}{\textbf{Data generation}} \\
Analysed duration & $364.09$ days \\
Sampling interval / rate & $2\,\mathrm{s}$ / $0.5\,\mathrm{Hz}$ \\
Orbits & ESA trailing OEM, \texttt{lisaorbits} v2.3 \\
TM / OMS amplitudes
  & $2.4\,\mathrm{fm\,s^{-2}\,Hz^{-1/2}}$ /
    $7.9\,\mathrm{pm\,Hz^{-1/2}}$ \\
TDI / units
  & $X_2,Y_2,Z_2$ / Hz,
    $\nu_0=2.816\times10^{14}\,\mathrm{Hz}$ \\

\midrule
\multicolumn{2}{@{}l@{}}{\textbf{Galactic foreground}} \\
Generation & Correlated Gaussian XYZ, fixed anisotropic sky \\
Strain spectrum & One-year Karnesis prescription~\cite{karnesis2021characterization} \\
$A$ & $1.14\times10^{-44}$ \\
$f_1$ & $1.99526\,\mathrm{mHz}$ \\
$f_2$ & $0.31\,\mathrm{mHz}$ \\
$f_{\rm knee}$ & $3.38844\,\mathrm{mHz}$ \\
$\alpha$ & $1.8$ \\

\midrule
\multicolumn{2}{@{}l@{}}{\textbf{Analysis}} \\
Frequency band & $0.1$--$100\,\mathrm{mHz}$ \\
Original WDM grid & $2046\times3069$ cells \\
WDM time spacing & $4.267$ hours \\
Gap taper & Cosine taper, $1$ hour per side \\
\midrule
\end{NiceTabular}
\end{table}

We transform the Michelson variables to the orthonormal A/E/T basis~\citep{prince2002lisa},
\begin{equation}
\begin{pmatrix}A\\E\\T\end{pmatrix}
=
\underbrace{\begin{pmatrix}
-1/\sqrt2&0&1/\sqrt2\\
1/\sqrt6&-2/\sqrt6&1/\sqrt6\\
1/\sqrt3&1/\sqrt3&1/\sqrt3
\end{pmatrix}}_{\mathbf M}
\begin{pmatrix}X\\Y\\Z\end{pmatrix}\ ,
\codelink{https://github.com/nz-gravity/lisa_timevarying_logpsplines_study/blob/v0.1.1/src/lisa_psd_analysis/lisa_aet.py\#L8-L14}
\label{eq:lisa_aet_matrix}
\end{equation}
where $\mathbf M$ is an orthogonal transformation matrix.
The reference A/E/T spectra are obtained by rotating the full XYZ spectral
covariance and then taking its diagonal. We show A-channel results in the
main text, as E gives similar reconstruction errors and the same model
ordering.

LISA observations will contain scheduled and unscheduled interruptions~\citep{LISA:2024hlh}. Gaps treated via window functions can introduce spectral leakage and
correlations through the observation window~\citep{burke2025gaps,burke2026mindsecondgapimpact}. The time localisation
of the WDM transform allows us to exclude coefficients most affected by
these boundaries.

We compare the continuous simulated LISA data set with a gapped version of
the same realisation.
The illustrative gap schedule combines regular scheduled outages with random
unscheduled interruptions,\footnote{The synthetic scheduled outages occur every 14 days
and last $3.5\,\mathrm{h}$. Unscheduled outages occur at an average rate of
one per week, with random start times and durations drawn log-uniformly
between $0.5$ and $24\,\mathrm{h}$.}
removing $4.00\%$ of the time-domain samples.

\begin{figure}[!tbp]
\centering
\script{plot_gap.py}
\includegraphics[width=\linewidth]{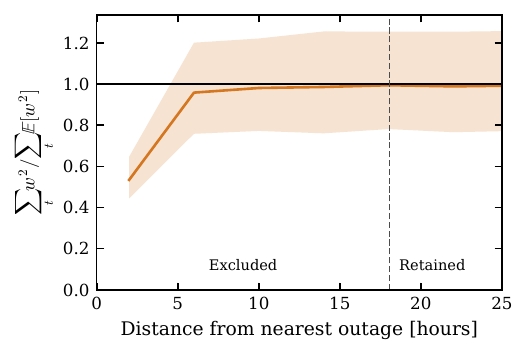}
\caption{A-channel WDM power near outages. For each frequency, observed power is
summed in four-hour distance bins and divided by the summed reference power. The orange line
and shading show the median and 5th--95th percentiles across frequency.
The vertical dashed line marks the fixed boundary cut: the one-hour
taper plus four WDM time spacings ($4\Delta T$) from either outage edge.
Rows beyond this cut are retained.
Shading shows variation across frequencies, not posterior uncertainty.
}
\label{fig:lisa_gap_boundary}
\end{figure}

Gaps are applied after\footnote{Missing data would occur at the level of instrument data and augmented with further drop outs after TDI has been performed. In Ref.~\cite{burke_tdi_gaps}, each gap yields $\sim 90$ seconds of additional missing data. We neglect this augmentation here.} TDI. We zero-fill missing samples and apply a cosine taper to the
neighbouring data over one hour on each side. This brings the data
smoothly to zero at the gap edges, reducing sharp jumps and the
resulting spectral leakage. A WDM coefficient
uses data beyond its tile centre, so leakage can also affect rows
outside the gaps. We therefore exclude rows whose centres lie in an
outage or within the one-hour taper plus four WDM time spacings of
either edge. This four-row guard is about $17.1$ hours on each side.
We also exclude rows within four time spacings of the record boundaries.
The boundary diagnostic in Fig.~\ref{fig:lisa_gap_boundary} checks the
retained coefficients. Appendix~\ref{app:coarse} describes their
coarse-graining.

Because the WDM guards extend beyond the gaps themselves, the gapped analysis
removes $29.18\%$ of the WDM rows, substantially more than the raw downtime.
On the retained rows, the RMSE between gapped
and continuous data, divided by the continuous-coefficient RMSE,
is less than $1\%$ in A and E, with little change in integrated power.
Some residual boundary effects remain (Fig.~\ref{fig:lisa_gap_boundary}).

To assess the effect of tapering, we also compared
with zero-filled gaps without a taper. In the A channel, the RMSE from the continuous data,
normalised by the analytic reference standard deviation
at each WDM cell, is $0.013$ with tapering and $0.090$
without tapering over $0.1$--$100\,\mathrm{mHz}$.
The largest differences occur near the TDI response
nulls, indicating that tapering suppresses leakage
that remains even after the WDM guard is applied.
A systematic investigation of alternative gap-handling
strategies is beyond the scope of this study and is
left for future work.

Both continuous and gapped records use the adaptive time--frequency
coarse-graining described in Sec.~\ref{sec:coarse}. Further details of the
likelihood partitions are given in Appendix~\ref{app:coarse}.

\subsection{Three spectral models}
\label{sec:lisa_hypotheses}
We compare three models for the total channel power from instrumental noise
and the unresolved Galactic foreground. They differ in how much physical
information is supplied, from a flexible spline fit to a fully parametric
instrument-and-foreground model.  Table~\ref{tab:lisa_model_computation}
summarises the three cases. All three models provide spectral predictions for the
WDM likelihood. They differ in whether the spectrum
is modelled directly or incorporates physically
specified instrumental and Galactic contributions.

The spline models $\mathcal H_{\rm agn}$ and $\mathcal H_{\rm orb}$
are applied independently to A, E, and T, with separate parameters
and likelihoods for each channel. The following comparisons focus on A.

The parametric model $\mathcal H_{\rm para}$ instead fits A/E/T
jointly, using shared instrumental and Galactic parameters.
We retain T only above $3\,\mathrm{mHz}$, avoiding the
low-frequency regime where cross-channel correlations can
become significant relative to the T-channel PSD
for unequal arms~\cite{Baghi_2023}. Above this cut, T supplies additional
information on the shared instrumental noise amplitudes.
All models use marginal-power likelihoods and neglect
cross-channel covariance.

\begin{table*}[t]
\centering
\small
\renewcommand{\arraystretch}{1.08}
\setlength{\tabcolsep}{5pt}

\begin{NiceTabularX}{\textwidth}{@{}l X[l] X[l] c X[l] c@{}}
\CodeBefore
  \rowcolor{gray!6}{2}
  \rowcolor{gray!6}{4}
\Body
\toprule
Model
& Physical information
& Fitted quantity
& $N_{\rm par}$
& A/E/T treatment
& Runtime$^{*}${[h]} \\
\midrule

$\mathcal H_{\rm agn}$
& None
& Total-PSD tensor spline
& 1850
& Separate A/E/T fits
& $\sim 1$ \\

\addlinespace[2pt]

$\mathcal H_{\rm orb}$
& \begin{tabular}[t]{@{}l@{}}
    Orbits +\\
    nominal TM/OMS noise
  \end{tabular}
& Smooth correction to instrumental PSD
& 114
& Separate A/E/T fits
& $\sim 2$ \\

\addlinespace[2pt]

$\mathcal H_{\rm para}$
& \begin{tabular}[t]{@{}l@{}}
    Orbits +\\
    nominal TM/OMS noise +\\
    Galactic model
  \end{tabular}
& \begin{tabular}[t]{@{}l@{}}
    2 link-noise amplitudes +\\
    5 Galactic parameters
  \end{tabular}
& 7
& Joint fit
& $\sim 8$ \\

\bottomrule
\end{NiceTabularX}

\caption{
LISA spectral models and inference costs for a
364.09-day observation ($15\,728\,640$ samples per channel).
The spline models fit A, E, and T independently, whereas the
parametric model uses shared parameters across channels.
The main reconstruction comparisons show A.
$^{*}$Approximate runtimes include warm-up and sampling
for two sequential chains. Spline runtimes refer to a
single A-channel fit, whereas the parametric runtime
covers the joint A/E/T fit. These timings are indicative:
computational cost depends on likelihood evaluation and
NUTS sampling efficiency, not only parameter count.
}
\label{tab:lisa_model_computation}
\end{table*}

\par\smallskip
\noindent\textbf{a. $\boldsymbol{\mathcal H_{\rm agn}}$: flexible total spectrum.} The agnostic model fits the total evolutionary PSD using the
tensor-product P-spline of Eq.~\eqref{eq:pspline_surface}. It uses no
orbital information, instrumental reference, or Galactic template.
The midpoint approximation uses $S_c(t_n,f_m)$ directly in the WDM likelihood.
The spline must therefore capture both the foreground modulation and the
moving TDI response features.

\par\smallskip
\noindent\textbf{b. $\boldsymbol{\mathcal H_{\rm orb}}$: orbit-informed spline.}
\label{sec:anova}
The orbit-informed model uses the known spacecraft trajectories and
fixed nominal test-mass (TM) and optical metrology system (OMS)
spectra to construct a time-dependent instrumental reference,
\begin{equation}
R_c(t,f)
=
N_{\mathrm{TM},c}(t,f)
+
N_{\mathrm{OMS},c}(t,f).
\codelink{https://github.com/nz-gravity/lisa_timevarying_logpsplines_study/blob/v0.1.1/src/lisa_psd_analysis/fitting.py\#L39-L62}
\label{eq:lisa_reference_definition}
\end{equation}
Here $N_{\mathrm{TM},c}$ and $N_{\mathrm{OMS},c}$ are the
nominal channel spectra obtained from the TDI response
(Appendix~\ref{app:lisa_orb_inference}).
We hold the band-averaged instrumental reference
$R_{c,nm}$ fixed during inference
(Appendix~\ref{app:wdm_projection})
and infer a smooth log correction,
\begin{equation}
\ln S_{c,nm}
=\ln R_{c,nm}+g_c(f_m)+\eta_c(t_n,f_m).
\codelink{https://github.com/nz-gravity/LogPSplinePSD/blob/v0.2.0/src/log_psplines/models/reconstruction.py\#L342-L363}
\label{eq:h_orb}
\end{equation}
Thus the dimensionless log correction is
\begin{equation}
u_c(t_n,f_m)=g_c(f_m)+\eta_c(t_n,f_m).
\codelink{https://github.com/nz-gravity/LogPSplinePSD/blob/v0.2.0/src/log_psplines/models/anova.py\#L12-L84}
\label{eq:anova_decomposition}
\end{equation}
Following~\citep{lee2011anova,lee2013nested,gu2014ssanova},
$g_c$ is a frequency P-spline and $\eta_c$ a tensor-product P-spline
of the form in Eq.~\eqref{eq:pspline_surface}.
The constraint $\sum_n\eta_c(t_n,f_m)=0$ at each frequency makes
$g_c$ the time-average log correction. Their smoothing amplitudes have
independent priors $\sigma_g\sim\operatorname{HalfNormal}(10)$ and
$\sigma_\eta\sim\operatorname{HalfNormal}(10)$, using the penalty
construction of Appendix~\ref{app:penalty}.
As $\sigma_\eta$ approaches zero, the correction becomes frequency-only,
while the total prediction retains the time dependence of $R_{c,nm}$.
The correction can absorb both instrumental mismatch and Galactic
foreground power, so this model estimates the total spectrum rather
than separating its physical components.

\par\smallskip
\noindent\textbf{c. $\boldsymbol{\mathcal H_{\rm para}}$: parametric instrument and foreground.} The parametric model represents the total spectrum as the sum
of TM noise, OMS noise, and a Galactic foreground.
The instrumental spectral shapes and orbital responses
are fixed. We infer seven parameters: the two link-noise
amplitudes $a_{\rm TM}$ and $a_{\rm OMS}$, and five
Galactic spectral parameters
$\theta=(A,f_1,f_2,f_{\rm knee},\alpha)$ from
Eq.~\eqref{eq:karnesis_foreground}.
The PSD prediction on the WDM grid is
\begin{equation}
\begin{split}
S_{c,nm}
&=
\left(\frac{a_{\rm TM}}{a_{\rm TM}^0}\right)^2
N_{{\rm TM},c,nm}
\\
&\quad+
\left(\frac{a_{\rm OMS}}{a_{\rm OMS}^0}\right)^2
N_{{\rm OMS},c,nm}
\\
&\quad+
S_{{\rm gal},c,nm}(\theta).
\end{split}
\codelink{https://github.com/nz-gravity/lisa_timevarying_logpsplines_study/blob/v0.1.1/src/lisa_psd_analysis/models.py\#L31-L43}
\label{eq:h_para_total}
\end{equation}
Here $a_i^0$ are the nominal link-noise amplitudes used to calculate
$N_{i,c,nm}$. Since PSD scales quadratically with amplitude, the squared
ratios rescale these nominal components to the fitted amplitudes.
The instrumental contributions and Galactic term $S_{{\rm gal},c,nm}(\theta)$
use the band-averaged approximation and are then averaged within
likelihood blocks (Appendices~\ref{app:wdm_projection} and~\ref{app:coarse}).
The physical response models are described in
Appendix~\ref{sec:lisa_response_appendix}.

We assign independent Gaussian priors to the logarithms
of the seven parameters. The Galactic priors are centred
on the injected values from Sec.~\ref{sec:lisa_data},
with standard deviations $(1,0.5,0.5,0.5,0.35)$ for
$(A,f_1,f_2,f_{\rm knee},\alpha)$, respectively.
For the instrumental amplitudes,
$\ln(a_i/a_i^0)\sim\mathcal N(0,0.25^2)$.
The corresponding parameter posteriors are shown in
Fig.~\ref{fig:lisa_corner}.

\par\smallskip

The displayed A-channel fits use $(K_t,K_f)=(14,132)$ for
$\mathcal H_{\rm agn}$ and $(7,16)$ for $\mathcal H_{\rm orb}$.
The agnostic time knots are nonuniform, whereas the smaller
orbit-informed time basis uses equally spaced knots.
Frequency knots follow quantiles of a pilot-based measure of spectral
variation, adapting the quantile approach of~\citet{maturana2021bayesian}. The layouts are fixed before sampling.
Preliminary fits informed these choices. Systematic optimisation of knot
number and placement remains future work.

\subsection{Results across models and observing conditions}
\label{sec:lisa_results}

\begin{figure}[!tbp]
\centering
\script{plot_lisa_spectrum.py}
\includegraphics[width=\linewidth]{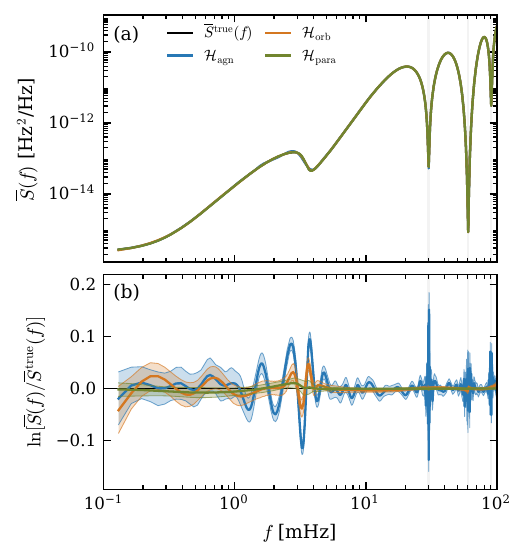}
\caption{Continuous A-channel recovery.
(a) Time-averaged PSD $\overline S(f)$ and
(b) log ratio $\ln(\overline S/\overline S^{\rm true})$.
Curves show posterior medians and shading gives central 90\% intervals
after averaging each draw in linear PSD units. Grey regions mark TDI
response dips.
}
\label{fig:lisa_spectra}
\end{figure}

\begin{figure*}[!t]
\centering
\script{plot_lisa_surface.py}
\includegraphics[width=\linewidth]{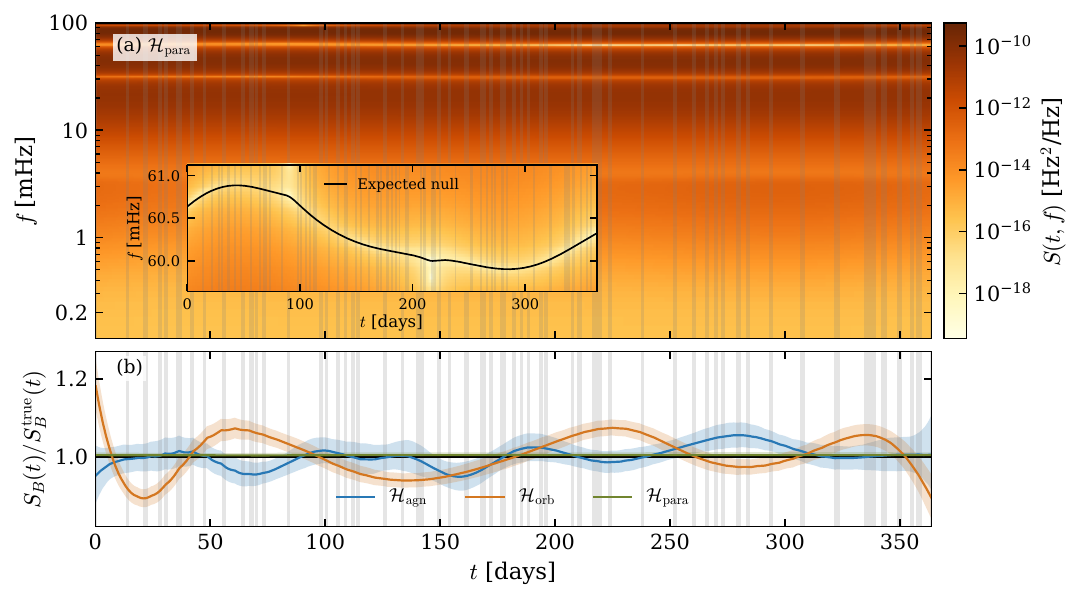}
\caption{Gapped A-channel recovery.
(a) Posterior-median PSD for $\mathcal H_{\rm para}$ on its
model-specific coarse grid, showing the moving TDI response dip.
The inset enlarges the region near $60\,\mathrm{mHz}$, with the expected
continuous-response null shown in black.
(b) Low-frequency ($0.1$--$3\,\mathrm{mHz}$) band-average ratio
$S_{\mathcal B}(t)/S_{\mathcal B}^{\rm true}(t)$ for the three models.
Curves show posterior medians and shading gives central 90\% intervals.
Grey regions mark excluded times.}
\label{fig:lisa_surface}
\end{figure*}

\begin{figure}[!tbp]
\centering
\script{plot_lisa_error.py}
\includegraphics[width=\linewidth]{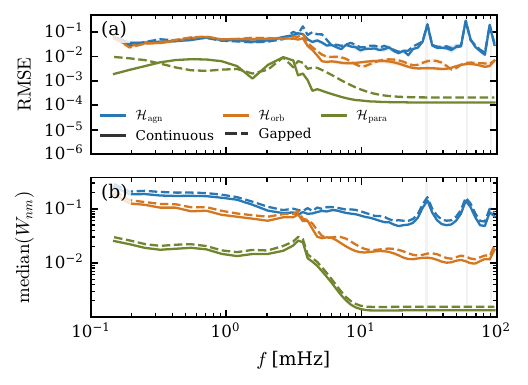}
\caption{A-channel reconstruction error and posterior width.
(a) $\mathrm{RMSE}_{\log S}$ and
(b) $\mathrm{median}(W_{nm})$ in fixed frequency bands.
Solid curves show continuous data and dashed curves show gapped data.
Grey regions mark TDI response dips. Bands in panel (a) show
5th--95th percentiles from paired resampling, not posterior intervals.}
\label{fig:lisa_error_width}
\end{figure}

\begin{figure}[!tbp]
\centering
\script{plot_fractional_error.py}
\includegraphics[width=\linewidth]
{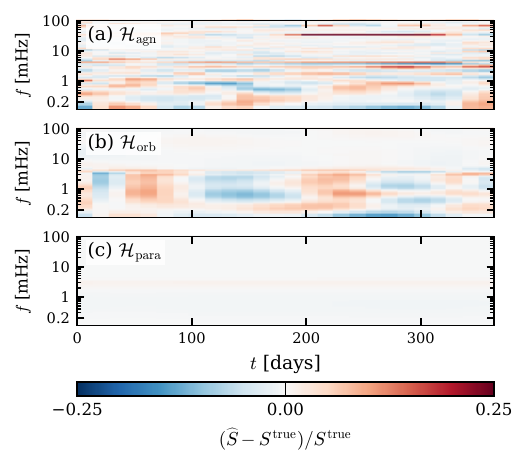}
\caption{Signed fractional error
$(\widehat S-S^{\rm true})/S^{\rm true}$ for continuous A-channel data.
All three models use the same retained time--frequency points and colour
scale, saturated outside $\pm0.25$. For display, errors are summarised
within 26 equal time bins while retaining the original frequency resolution
and TDI response dips.}
\label{fig:lisa_fractional_error}
\end{figure}

We compare the three models on the same retained time--frequency points,
including the TDI response dips. Table~\ref{tab:lisa_results} summarises
the reconstruction error and 90\% truth-inclusion fraction, while
Figs.~\ref{fig:lisa_spectra}--\ref{fig:lisa_fractional_error} show where
the differences arise in frequency and time.

\begin{table*}[t]
\centering
\small
\caption{A-channel reconstruction and whitening diagnostics
for the three models under continuous and gapped observing
conditions. The full band includes the TDI response dips.
The Anderson--Darling statistic $A^2$ assesses marginal
agreement with $\mathcal{N}(0,1)$. The final column uses
the analytic reference spectrum $S^{\rm true}_{nm}$.
For this single LISA realisation, $\mathrm{Inc}_{90}$
is descriptive rather than an ensemble coverage estimate.}
\label{tab:lisa_results}

\begingroup
\setlength{\tabcolsep}{3pt}
\renewcommand{\arraystretch}{1.1}

\begin{tabular*}{\textwidth}{
    @{\extracolsep{\fill}}
    ll
    ccc
    @{\hspace{1.4em}} ccc
    @{\hspace{1.4em}} ccc
    @{\hspace{1.4em}} c
    @{}
}
\toprule
& & \multicolumn{3}{c}{$\mathcal H_{\rm agn}$}
& \multicolumn{3}{c}{$\mathcal H_{\rm orb}$}
& \multicolumn{3}{c}{$\mathcal H_{\rm para}$}
& \multicolumn{1}{c}{$S^{\rm true}_{nm}$} \\
\cmidrule(lr){3-5}
\cmidrule(lr){6-8}
\cmidrule(lr){9-11}
\cmidrule(lr){12-12}
Condition & Band [mHz]
& RMSE & $\mathrm{Inc}_{90}$ [\%] & $A^2$
& RMSE & $\mathrm{Inc}_{90}$ [\%] & $A^2$
& RMSE & $\mathrm{Inc}_{90}$ [\%] & $A^2$
& $A^2$ \\
\midrule
Continuous
& $0.1$--$3$
& 0.0503 & 76.6 & 1.546
& 0.0515 & 50.5 & 1.333
& 0.0063 & 78.9 & 1.232 & 1.238 \\
& $3$--$100$
& 0.0916 & 85.7 & 0.896
& 0.0069 & 83.1 & 0.276
& 0.0004 & 100.0 & 0.326 & 0.296 \\
& $0.1$--$100$
& 0.0907 & 85.4 & 1.025
& 0.0111 & 82.1 & 0.240
& 0.0011 & 99.4 & 0.316 & 0.289 \\
\midrule
Gapped
& $0.1$--$3$
& 0.0463 & 85.8 & 0.531
& 0.0534 & 56.4 & 0.633
& 0.0055 & 99.1 & 0.564 & 0.546 \\
& $3$--$100$
& 0.0933 & 85.7 & 1.134
& 0.0091 & 88.1 & 0.564
& 0.0007 & 100.0 & 0.617 & 0.565 \\
& $0.1$--$100$
& 0.0923 & 85.7 & 1.254
& 0.0127 & 87.2 & 0.665
& 0.0012 & 100.0 & 0.710 & 0.668 \\
\bottomrule
\end{tabular*}
\endgroup
\end{table*}

\par\smallskip
\noindent\textbf{a. $\boldsymbol{\mathcal H_{\rm agn}}$:} The agnostic spline captures the broad time variation in the spectrum but
has difficulty following the sharp, moving TDI response features.
These localised errors dominate the high-frequency reconstruction and are
not always reflected by the posterior interval width. A sensitivity study using $K_f=96$, $132$, and $176$ frequency
basis functions shows decreasing reconstruction error with increasing
basis size, although the improvement beyond our baseline $K_f=132$
is modest. A systematic optimisation of basis size is left for
future work.

\par\smallskip
\noindent\textbf{b. $\boldsymbol{\mathcal H_{\rm orb}}$:} Providing the time-dependent instrumental reference substantially improves
recovery around the response features. For continuous data in the
$3$--$100\,\mathrm{mHz}$ band, the RMSE decreases from $0.0916$ for
$\mathcal H_{\rm agn}$ to $0.0069$ for $\mathcal H_{\rm orb}$.
At $0.1$--$3\,\mathrm{mHz}$, however, the two models
have similar RMSE, while $\mathcal H_{\rm orb}$ has
narrower intervals and a lower truth-inclusion fraction.
The additional physical structure therefore increases
posterior precision without a corresponding improvement
in accuracy on this realisation.

\par
\begin{figure*}[t]
\centering
\script{plot_corner.py}
\includegraphics[width=0.90\textwidth]{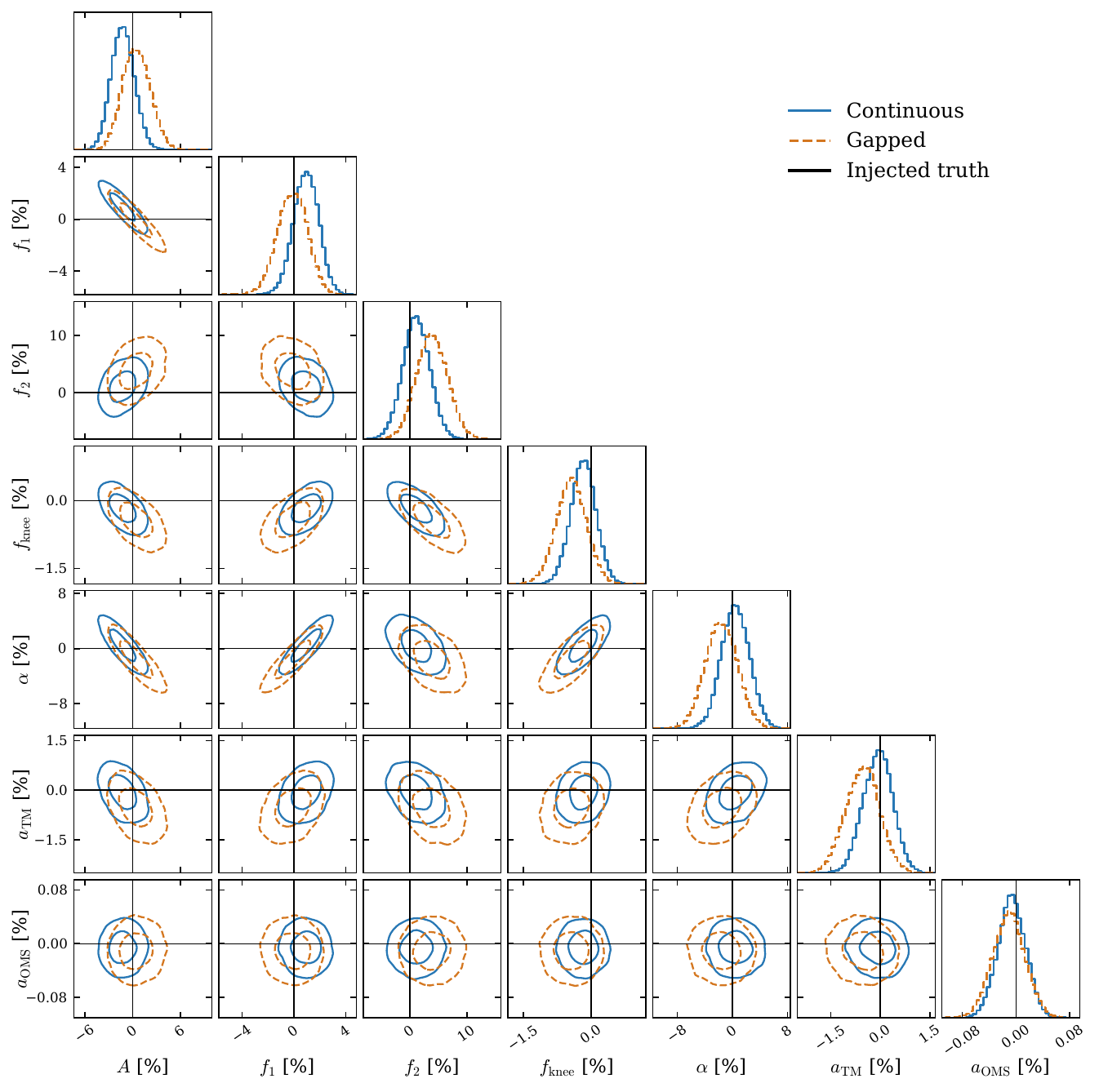}
\caption{Seven-parameter $\mathcal H_{\rm para}$ posterior. Blue solid curves are continuous,
orange dashed curves are gapped, and black lines mark injected values.
The marginals use 8000 posterior draws per condition. Contours enclose
50\% and 90\% of the estimated two-dimensional posterior probability.
Each axis shows the percentage departure from the injected value. The TM
and OMS axes show physical amplitude departures relative to their nominal
values $a_i^0$. These constraints are conditional
on the assumed instrumental and foreground model.
}
\label{fig:lisa_corner}
\end{figure*}

\smallskip
\noindent\textbf{c. $\boldsymbol{\mathcal H_{\rm para}}$:} The parametric model gives the smallest reconstruction errors.
This is expected under the matched simulation, since the foreground family,
Galactic morphology, instrumental spectral shapes, and orbital response
used by the model agree with those used to generate the data.
The fit also uses the T channel. It therefore provides a matched physical
benchmark rather than a test of robustness to model misspecification.
Figure~\ref{fig:lisa_corner} shows the posterior distributions
of the seven parameters under continuous and gapped observations.
The instrumental amplitude departures are shown in physical
amplitude units, rather than squared-amplitude (PSD) units.

\par\smallskip
\noindent\textbf{\textit{Effect of gaps:}}
Recovery on the retained data changes only modestly after
introducing gaps. The full-band RMSE changes from
$0.0907$ to $0.0923$ for $\mathcal H_{\rm agn}$,
from $0.0111$ to $0.0127$ for $\mathcal H_{\rm orb}$,
and from $0.0011$ to $0.0012$ for $\mathcal H_{\rm para}$.
The gapped fits generally have slightly wider posterior
intervals, as shown in Fig.~\ref{fig:lisa_error_width}b, reflecting the reduced
amount of data available for inference.
These comparisons do not assess recovery inside the
excluded gap regions.

Figure~\ref{fig:lisa_fractional_error} shows that the remaining errors are
strongly localised in time and frequency, particularly around the moving
TDI response features. Accurate time- or band-averaged spectra therefore do
not necessarily imply accurate recovery at every
response dip.

The Anderson--Darling statistics in
Table~\ref{tab:lisa_results} provide a complementary
assessment of whitening performance. 
At low frequencies,
all three models yield values comparable to those obtained
using the $S^{\rm true}_{nm}$. At higher frequencies,
the agnostic model exhibits larger deviations, whereas the
orbit-informed and parametric models remain close to the
reference under both continuous and gapped conditions.
All values remain below the nominal 5\% critical value
of $A^2=2.492$.

Appendix~\ref{app:lisa_components} compares the two spline models with
frequency-only fits. For continuous A-channel data, allowing time variation
reduces the low-frequency RMSE of $\mathcal H_{\rm agn}$ from $0.3978$ to
$0.0503$, showing that time dependence is needed to recover the
evolving spectrum.


\section{Discussion and conclusions}
\label{sec:discussion}

We developed a Bayesian P-spline model for a smoothly evolving PSD,
with separate regularisation in time and frequency and adaptive likelihood
compression. In the controlled LS2 study, reconstruction errors and posterior
interval widths decrease as more data are provided for the same evolving
spectral pattern. This is evidence for recovery in that controlled setting,
rather than a general asymptotic result.

The LISA study illustrates the trade-off between spectral flexibility and
physical information. The agnostic spline recovers the broad time variation
but has difficulty following sharp, moving TDI response features. Supplying
a time-dependent instrumental reference substantially improves recovery
around these features while retaining a flexible correction to the total
power. At low frequencies, however, the orbit-informed model gives narrower
intervals without a corresponding reduction in reconstruction error, and
with lower truth inclusion. Additional physical structure can therefore
reduce posterior uncertainty without improving accuracy.

The parametric model gives the smallest errors in this study, but the
comparison is deliberately favourable to it. Its instrumental spectral shapes,
orbital response, Galactic morphology, and foreground spectral family match
those used to generate the data, and the fit also uses the T channel. It
provides a matched physical benchmark rather than a test of robustness to
model misspecification. The spline models also differ in basis size,
regularisation, and likelihood partition, so the comparison does not isolate
the effect of any single modelling choice. Preliminary fits informed their basis choices. Systematic optimisation
of knot number and placement remains future work.

Gaps introduce a separate limitation. The illustrative outages remove
$4.00\%$ of the time-domain samples, but tapering and WDM guards exclude
$29.18\%$ of the WDM rows. The reported errors assess recovery only on the
retained cells. Gap filling~\citep{baghi2019gaps,gapsnextgen2026}, more
localised windows~\citep{pearson2026compact}, and explicit treatment of
window-induced covariance~\citep{burke2026mindsecondgapimpact,burke2025gaps}
could allow more data to be retained. Combining these approaches with the
present spectral model remains to be investigated.

The evolutionary PSD enters the WDM likelihood through
approximations to the marginal coefficient variances.
For the physically specified LISA reference spectra,
we account for variation across the WDM frequency window
(Appendix~\ref{app:wdm_projection}).
The likelihood neglects correlations between distinct
WDM coefficients~\citep{JohnsonWDMII2026}.

The likelihood also neglects cross-channel covariance. Unequal, time-dependent
arms and an anisotropic foreground can produce nonzero A/E/T cross-spectra.
This information can be important for LISA stochastic-background
inference~\citep{Kume:2025:JCAP}, and a product of marginal likelihoods need
not reproduce the posterior uncertainty of a full correlated
likelihood~\citep{ribatet2012bayesian}. The reported intervals are conditional
on both covariance approximations. 

A next step is to account for the cross-channel
correlations neglected in the present analysis.
We plan to extend our method to time-varying spectral-density
matrices, allowing the A/E/T PSD and cross-spectra to be
inferred jointly. The multivariate P-spline framework of~\citet{multivariatepspline2026} provides a starting point,
although efficient likelihood construction and regularisation
remain to be developed. A complementary direction is to model
individual link noises and propagate them through the TDI
response, as in~\citet{Baghi_2023}, to enforce consistency
across channels and help separate instrumental contributions.


\Needspace{16\baselineskip}
\section*{Data and Code availability}

The spectral-estimation library is available as
\href{https://github.com/nz-gravity/LogPSplinePSD/releases/tag/v0.2.0}{\project{LogPSplinePSD}, version \texttt{v0.2.0}}.
The LISA-specific models and analysis scripts are available in the
\href{https://github.com/nz-gravity/lisa_timevarying_logpsplines_study/releases/tag/v0.1.1}{LISA study repository, version \texttt{v0.1.1}}, archived at
\href{https://doi.org/10.5281/zenodo.23093373}{doi:10.5281/zenodo.23093373}.
The GitHub icons (\scalebox{0.65}{\faGithub}) beside equations link to the
corresponding source lines in these tagged releases. The study source
release does not include the large
data sets or posterior artifacts.

The simulated LISA time series, prepared continuous and gapped data sets,
and posterior results for the three models are available on Zenodo at
\href{https://doi.org/10.5281/zenodo.23092878}{doi:10.5281/zenodo.23092878}.
The deposit includes a Jupyter notebook for loading the files and plotting
WDM power, median posterior surfaces and PSD slices.

\begin{acknowledgments}
\enlargethispage{\baselineskip}
\vspace{-4pt}
We thank the members of the LISA Noise Non-Stationarities
Group (part of the ``Deep analysis group'' of the Distributed Data Processing
Centre, DDPC) for helpful discussions.
AV, PMR, and  RM
gratefully acknowledge support from the Marsden Fund Council grant MFP-UOA2531,
funded by the New Zealand Government and managed by the Royal Society Te Apārangi. The work of PMR and RM was further supported by the CNRS International Research Project (IRP) OG-Science FR-
NZ.
GM acknowledges support from the ANR grant AAPG2024 PRC - GalaxyFIT.
O.~Burke acknowledges financial support from the Grant UKRI972 awarded via the UK Space Agency.
This work was performed on the OzSTAR national facility at Swinburne
University of Technology. The OzSTAR program receives funding in part from
the Astronomy National Collaborative Research Infrastructure Strategy (NCRIS)
allocation provided by the Australian Government, and from the Victorian Higher
Education State Investment Fund (VHESIF) provided by the Victorian Government.
\end{acknowledgments}

\appendix
\twocolumngrid

\section{Stationary and time-varying contributions}
\label{app:lisa_components}

We compare each full spline model with a separately fitted restriction:
a frequency-only total PSD for $\mathcal H_{\rm agn}$ and a frequency-only
correction to the time-dependent reference for $\mathcal H_{\rm orb}$. We use the same evaluation cells and
frequency bands as Fig.~\ref{fig:lisa_error_width}. Each restricted fit
uses its corresponding full fit's data masks, frequency pooling, frequency
knots, and sampler settings. The recorded convergence checks pass for all
restricted fits. These comparisons describe reconstruction on this
realisation, rather than evidence for selecting a model.

For continuous A-channel data, adding the time-varying component reduces
low-band RMSE from $0.3730$ to $0.0515$ in $\mathcal H_{\rm orb}$,
and from $0.3978$ to $0.0503$ in $\mathcal H_{\rm agn}$.
The wider full-model intervals reflect the additional freedom.
The agnostic fit also reduces high-band RMSE, from $0.5233$ to $0.0916$, but still struggles near response dips.

\begin{figure*}[t]
\centering
\script{plot_components.py}
\includegraphics[width=\linewidth]{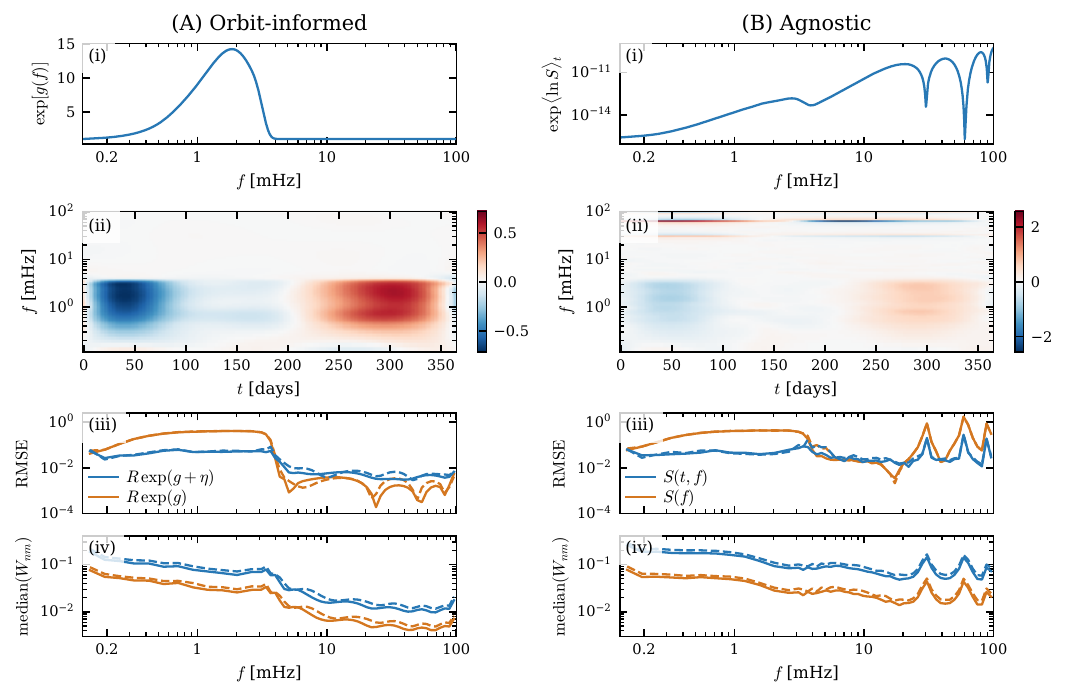}
\caption{A-channel component summaries and matched fit comparisons for
(A) $\mathcal H_{\rm orb}$ and (B) $\mathcal H_{\rm agn}$.
Within each group, (i) shows posterior medians and central $90\%$ intervals
from the joint continuous-data fit: the dimensionless stationary multiplier
$\exp[g(f)]$ in (A), and the geometric time-mean PSD in
$\mathrm{Hz}^2/\mathrm{Hz}$ in (B). The geometric mean is computed for each
draw over the full time grid and differs from the arithmetic average
$\overline S$ in the main results.
(ii) Posterior means of the centred log correction $\eta(t,f)$ in (A)
and the log PSD after subtracting its time mean in (B).
The agnostic summaries are a descriptive decomposition of one fitted tensor
spline, with no separate component regularisation.
(iii) RMSE and (iv) median log interval width $W_{nm}$,
for time-varying and separately fitted frequency-only models on common
evaluation cells. The orbit-informed frequency-only fit has a stationary
correction to a time-dependent reference. The agnostic frequency-only fit
has a stationary total PSD. The legends in (iii) identify the fits compared in (iii)--(iv). Solid curves are continuous and dashed curves
are gapped.}
\label{fig:lisa_components}
\label{fig:lisa_orb_components}
\label{fig:lisa_agn_components}
\end{figure*}


\section{LISA response models and foreground simulation}
\label{app:lisa_orb_inference}
\label{sec:lisa_response_appendix}

We describe the instrumental reference, Galactic response,
and foreground synthesis used in the LISA study. The response
models are evaluated using \texttt{backgrounds}~\cite{Baghi_2023},
with the ESA spacecraft trajectories specified in
Table~\ref{tab:sim_params}.

\subsection{Instrumental reference}

We construct the instrumental reference by propagating
the stationary test-mass (TM) and optical metrology
system (OMS) link-noise spectra through the time-dependent
TDI response.

For each noise component $i\in\{\mathrm{TM},\mathrm{OMS}\}$,
$S_i(f)$ denotes its stationary link-noise PSD.
The TDI response maps this to the time-dependent
A/E/T spectral matrix $\mathbf C_i^{AET}(t,f)$,
whose diagonal gives the PSD $N_{i,c}(t,f)$ of each
channel $c\in\{A,E,T\}$. Summing the TM and OMS
contributions gives the instrumental reference $R_c(t,f)$.
Schematically,
\begin{equation*}
\begin{gathered}
S_i(f)
\xrightarrow{\mathrm{TDI}}
\mathbf C_i^{AET}(t,f)
\\
\xrightarrow{\mathrm{diagonal}}
N_{i,c}(t,f)
\xrightarrow{\sum_i}
R_c(t,f).
\end{gathered}
\end{equation*}

Following~\citet{bayle2023lisainstrument}, the nominal
link-noise spectra are
\begin{align}
S_{\rm TM}(f)
&=
(a_{\rm TM}^{0})^{2}
\left[1+\left(\frac{f_{\rm TM,low}}{f}\right)^2\right]
\nonumber\\
&\quad\times
\left[1+\left(\frac{f}{f_{\rm TM,high}}\right)^4\right]
\left(\frac{1}{2\pi f c_0}\right)^2,
\\
S_{\rm OMS}(f)
&=
(a_{\rm OMS}^{0})^{2}
\left[1+\left(\frac{f_{\rm OMS}}{f}\right)^4\right]
\left(\frac{2\pi f}{c_0}\right)^2,
\end{align}
where $c_0$ is the speed of light. The nominal amplitudes
are given in Table~\ref{tab:sim_params}, with
$f_{\rm TM,low}=4\times10^{-4}\,\mathrm{Hz}$,
$f_{\rm TM,high}=8\times10^{-3}\,\mathrm{Hz}$, and
$f_{\rm OMS}=2\times10^{-3}\,\mathrm{Hz}$.

The spacecraft trajectories determine the directed
light-travel times $L_{pq}(t)$ between spacecraft.
In the locally frozen response calculation, each delay
is represented by the frequency-domain phase factor
\begin{equation}
D_{pq}(t,f)=\exp[-2\pi\mathrm{i}fL_{pq}(t)].
\label{eq:lisa_delay_operator}
\end{equation}
We follow the second-generation TDI conventions of~\citet{tinto2021tdi} and~\citet{bayle2023lisainstrument}.
The transfer matrices $\overline{\mathbf H}_i(t,f)$
map the link noises into $X,Y,Z$, including the fixed
simulator-specific unit conversion and discrete-time filtering.

Let $\mathbf M$ denote the $X,Y,Z\rightarrow A,E,T$
rotation defined in Eq.~\eqref{eq:lisa_aet_matrix}.
For independent link noises with common spectrum $S_i(f)$,
the corresponding A/E/T spectral matrix is
\begin{equation}
\begin{aligned}
\mathbf C_i^{AET}(t,f)
&=
S_i(f)\,\mathbf M
\overline{\mathbf H}_i(t,f)
\\
&\quad\times
\overline{\mathbf H}_i^\dagger(t,f)\mathbf M^\dagger,
\end{aligned}
\codelink{https://github.com/nz-gravity/lisa_timevarying_logpsplines_study/blob/v0.1.1/src/lisa_psd_analysis/lisa_aet.py\#L26-L48}
\label{eq:lisa_aet_covariance}
\end{equation}
where $\dagger$ denotes Hermitian conjugation (complex-conjugate transpose)
and $i\in\{\mathrm{TM},\mathrm{OMS}\}$.
The diagonal elements give the channel PSDs,
\begin{equation}
N_{i,c}(t,f)
=
[\mathbf C_i^{AET}(t,f)]_{cc}.
\end{equation}
The instrumental reference is therefore
\begin{equation}
R_c(t,f)
=
N_{\mathrm{TM},c}(t,f)
+
N_{\mathrm{OMS},c}(t,f).
\codelink{https://github.com/nz-gravity/lisa_timevarying_logpsplines_study/blob/v0.1.1/src/lisa_psd_analysis/fitting.py\#L39-L62}
\label{eq:lisa_reference_components}
\end{equation}
Although the link spectra are stationary, the changing
delays shift the cancellations between TDI terms.
Consequently, the channel spectra and their response
nulls evolve over the orbit.

\subsection{Galactic sky and response}

We construct a fixed anisotropic sky distribution using
the \texttt{Galaxy} model in
\texttt{backgrounds}~\cite{Baghi_2023}.
The map has HEALPix resolution $N_{\rm side}=8$
(768 pixels)~\cite{Gorski2005} and uses 400 radial
integration points. The pixel powers $p_i$ are
normalised so that $\sum_i p_i=1$, and their square
roots $\sqrt{p_i}$ are supplied to the response
calculation. The map specifies the angular distribution,
while $S_{\rm gal}(f;\theta)$ in
Eq.~\eqref{eq:karnesis_foreground} specifies the
intrinsic strain spectrum.

The \texttt{backgrounds} package propagates this sky
distribution through the orbit-dependent second-generation
TDI response. Writing the resulting XYZ response matrix
as $\mathbf R_{\rm gal}^{XYZ}(t,f)$, the injected Galactic
spectral matrix is
\begin{equation}
\mathbf C_{\rm gal}^{XYZ}(t,f)
=
S_{\rm gal}(f;\theta_{\rm inj})
\mathbf R_{\rm gal}^{XYZ}(t,f).
\codelink{https://github.com/nz-gravity/lisa_timevarying_logpsplines_study/blob/v0.1.1/src/lisa_psd_analysis/_generation.py\#L103-L111}
\label{eq:galactic_response_csd}
\end{equation}
The response includes the factor $\nu_0^2$ converting
fractional-frequency spectra to the TDI units of
Table~\ref{tab:sim_params}. Rotating the full response
matrix into A/E/T and taking its diagonal gives the
channel response,
\begin{equation}
R_{{\rm gal},c}(t,f)
=
\left[
\mathbf M
\mathbf R_{\rm gal}^{XYZ}(t,f)
\mathbf M^\dagger
\right]_{cc}.
\codelink{https://github.com/nz-gravity/lisa_timevarying_logpsplines_study/blob/v0.1.1/src/lisa_psd_analysis/lisa_aet.py\#L26-L48}
\label{eq:galactic_channel_response}
\end{equation}
The sky distribution is fixed throughout the observation.
Its observed modulation arises from the evolving
detector response.

\subsection{Gaussian foreground synthesis}

We interpolate the Galactic response matrices linearly
in time and log frequency, then multiply by
$S_{\rm gal}(f;\theta_{\rm inj})$.
For each synthesis segment, the resulting spectral matrix
is evaluated at the segment's reference epoch. Its Cholesky
factor is used to generate correlated complex Gaussian
XYZ Fourier coefficients. Inverse Fourier transforms
give real time-domain segments.

Each segment contains $2^{19}$ samples at the
$2\,\mathrm{s}$ cadence, corresponding to approximately
$12.14$ days. The segments are multiplied by a sine window
and combined using 50\% overlap. The full XYZ spectral
matrices are retained when generating the foreground,
including cross-channel correlations.
The foreground is added to the independently generated
instrumental XYZ time series before rotation to A/E/T.

The total pointwise channel spectrum used to construct
the analytic reference is
\begin{equation}
S_c^{\rm true}(t,f)
=
R_c(t,f)
+
R_{{\rm gal},c}(t,f)
S_{\rm gal}(f;\theta_{\rm inj}).
\codelink{https://github.com/nz-gravity/lisa_timevarying_logpsplines_study/blob/v0.1.1/src/lisa_psd_analysis/_generation.py\#L225-L255}
\label{eq:lisa_true_spectrum}
\end{equation}
Appendix~\ref{app:wdm_projection} describes how the normalised expected
squared WDM coefficients approximate band-averaged versions of these
physical PSDs under slow temporal evolution.


\section{Band-averaged and midpoint PSD approximations}
\label{app:wdm_projection}

The LISA response models in Appendix~\ref{sec:lisa_response_appendix}
give pointwise evolutionary PSDs, not exact WDM coefficient variances.
For an interior frequency channel, define
\begin{equation}
K_m(f)=\frac{q[(f-f_m)/\Delta F]}{\Delta F},\qquad
\int_0^{f_{\rm Ny}}K_m(f)\,\mathrm df=1,
\codelink{https://github.com/nz-gravity/lisa_timevarying_logpsplines_study/blob/v0.1.1/src/lisa_psd_analysis/wdm_projection.py\#L39-L65}
\end{equation}
where $f_{\rm Ny}=1/(2\Delta t)$ and the squared cosine-Meyer window
corresponding to Eq.~\eqref{eq:meyer_window} is
\begin{equation}
q(u)=
\begin{cases}
1,& |u|<1/3,\\
\cos^2[\tfrac{3\pi}{2}(|u|-1/3)],&1/3\leq|u|<2/3,\\
0,&\text{otherwise}.
\end{cases}
\codelink{https://github.com/nz-gravity/lisa_timevarying_logpsplines_study/blob/v0.1.1/src/lisa_psd_analysis/wdm_projection.py\#L24-L36}
\end{equation}
Here $\int q(u)\,\mathrm du=1$, so $K_m$ has unit integral.
DC and Nyquist channels, which require separate endpoint conventions,
are excluded throughout our analysis.

Following~\citet{JohnsonWDMII2026}, if the process
admits an oscillatory spectral representation whose
complex amplitude varies slowly over the temporal extent
of the WDM window, the normalised expected coefficient
power is approximately the band-averaged evolutionary PSD,
\begin{equation}
\frac{2\Delta t}{N}\mathbb E[w_{nm}^{2}]
\simeq
\int_0^{f_{\rm Ny}}
K_m(f)S(t_n,f)\,\mathrm df.
\codelink{https://github.com/nz-gravity/lisa_timevarying_logpsplines_study/blob/v0.1.1/src/lisa_psd_analysis/wdm_projection.py\#L39-L117}
\label{eq:wdm_band_average}
\end{equation}
This band-averaging relation assumes slow evolution over the temporal
support of the WDM window; it is not an identity for arbitrary
non-stationary processes~\citep{priestley1965evolutionary,JohnsonWDMII2026}.
If $S(t_n,f)$ also varies little across the frequency support of $K_m(f)$,
the unit-integral normalisation allows the integral to be approximated
by $S(t_n,f_m)$, giving the \emph{midpoint approximation} in
Eq.~\eqref{eq:wdm_estimand}~\citep[Eq.~(39)]{JohnsonWDMII2026}.
For the same sampled interior window, Johnson
et al.'s coefficients obey
$w^{\rm J}_{mn}=\sqrt{\Delta t/N}\,w_{nm}$;
their Eq.~(36) therefore gives half the
band-averaged one-sided PSD in our convention.

Near sharp LISA TDI response nulls, the centre-frequency PSD can be
unrepresentative of the power sampled by the finite WDM window.
We therefore define the reference predictions by band averaging,
\begin{align}
R_{c,nm}&=\int_0^{f_{\rm Ny}}K_m(f)R_c(t_n,f)\,\mathrm df,\\
S_{c,nm}^{\rm true}&=\int_0^{f_{\rm Ny}}K_m(f)S_c^{\rm true}(t_n,f)\,\mathrm df.
\codelink{https://github.com/nz-gravity/lisa_timevarying_logpsplines_study/blob/v0.1.1/src/lisa_psd_analysis/_preparation.py\#L150-L179}
\label{eq:wdm_quadrature}
\end{align}
These frequency-window averages generally differ from the pointwise
physical PSDs. We evaluate the integrals using
16-node Gauss--Legendre quadrature, with non-negative squared-window
weights normalised to sum to one.
The fixed $R_{c,nm}$ enters $\mathcal H_{\rm orb}$, whereas
$\mathcal H_{\rm agn}$ fits a smooth midpoint surface directly.

For $\mathcal H_{\rm para}$, the nominal instrumental components are
band averaged in the same way, and the Galactic contribution is
\begin{equation}
S_{{\rm gal},c,nm}(\theta)
=\int_0^{f_{\rm Ny}}K_m(f)R_{{\rm gal},c}(t_n,f)
S_{\rm gal}(f;\theta)\,\mathrm df.
\codelink{https://github.com/nz-gravity/lisa_timevarying_logpsplines_study/blob/v0.1.1/src/lisa_psd_analysis/parametric_response.py\#L16-L51}
\label{eq:galactic_wdm_projection}
\end{equation}
The parametric model combines this frequency projection with
likelihood-block averaging (Appendix~\ref{app:coarse}).
Frequency projection improves the approximation to marginal WDM powers.
It does not recover off-diagonal WDM covariance and is distinct from
coarse-graining the observed WDM powers.

\section{Likelihood coarse-graining}
\label{sec:coarse_appendix}
\label{app:coarse}

Section~\ref{sec:coarse} defines the coarse-grained likelihood. Here we
summarise the partition used in the reported analyses and assess its impact
in the controlled LS2 study. The tagged analysis code provides the full
implementation of the partitioning procedure.

\paragraph*{Partition construction.}

The partition is constructed before sampling from a smooth pilot estimate
of the WDM power. The pilot determines only the block boundaries. All
likelihood calculations use the original, unsmoothed WDM powers.

For the LISA analysis, we divide the data into 32 broad time intervals and
compute the median log power at each frequency. A 31-frequency-cell running
median is then used to obtain a smoother pilot spectrum, following the idea
of a robust initial spectral estimate in~\citet{gupta2024bayesian}.
\codelink{https://github.com/nz-gravity/lisa_timevarying_logpsplines_study/blob/v0.1.1/src/lisa_psd_analysis/_preparation.py\#L283-L336}
Starting from low frequency, neighbouring frequency cells are grouped while
the pilot log power varies by no more than $0.25$ within the bin, with a
maximum width of 24 original frequency cells. This gives wider bins in smooth
regions and finer bins around sharp spectral features.
\codelink{https://github.com/nz-gravity/LogPSplinePSD/blob/v0.2.0/src/log_psplines/preprocessing/power_partition.py\#L104-L121}

The frequency bins are combined with groups of four neighbouring WDM time
rows. For each rectangular block $b_{ij}=\mathcal T_i\times\mathcal F_j$,
the mean power $\bar P_{ij}$ and contributing-cell count $\nu_{ij}$
are computed from the original WDM powers as defined in
Eq.~\eqref{eq:masked_pooling}. The spline is evaluated at the mean time and
frequency of the original cells defining the block. The implementation
stores summed power $\nu_{ij}\bar P_{ij}$ and counts, giving the same
likelihood as Eq.~\eqref{eq:binned-likelihood}.

\begin{figure}[tb]
\centering
\script{plot_lisa_coarse_graining.py}
\includegraphics[width=\linewidth]{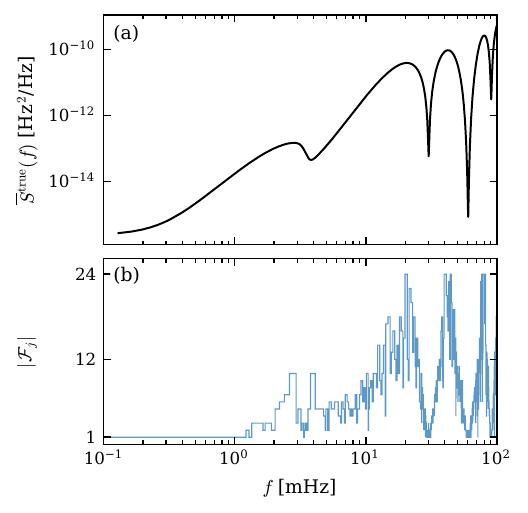}
\caption{Frequency coarse-graining for the continuous A-channel
$\mathcal H_{\rm agn}$ fit.
(a) Time-averaged true PSD from Fig.~\ref{fig:lisa_spectra}, with the same
axis limits. (b) Number of original frequency cells per coarse frequency bin,
$|\mathcal F_j|$, ranging from 1 to the imposed maximum of 24 cells.
The likelihood also groups four time
rows, so fully retained blocks have $\nu_{ij}=4|\mathcal F_j|$ contributing
WDM cells. The exact frequency-cell counts are not smoothed.
One original frequency cell spans
$32.55\,\mu\mathrm{Hz}$. The partition is fixed from the pilot estimate
before sampling.}
\label{fig:lisa_coarse_frequency}
\end{figure}

For the continuous A-channel data, the 3069 original frequency cells are
reduced to 676 frequency bins. Together with the four-row time grouping,
this corresponds to approximately an 18-fold reduction in the number of
likelihood terms away from boundaries and missing-data regions.
Figure~\ref{fig:lisa_coarse_frequency} shows the frequency dependence of
this partition alongside the time-averaged PSD. The partition responds to
frequency variation in the pilot log power across all 32 time profiles,
rather than to the level of the time-averaged PSD. The plotted quantity
$|\mathcal F_j|$ is the number of original frequency cells in each coarse
frequency bin. For a fully retained four-row time block, the corresponding
likelihood count is $\nu_{ij}=4|\mathcal F_j|$. Boundaries and exclusions
can reduce the contributing-cell count.

\paragraph*{Model-specific coarse-graining.}

For the LS2 and $\mathcal H_{\rm agn}$ fits, the PSD itself is approximated
as constant within each block. For $\mathcal H_{\rm orb}$, each original power
is first divided by the fixed instrumental reference before pooling. The
blockwise-constant approximation therefore applies to the smooth correction
rather than to the sharp instrumental response
(Sec.~\ref{sec:lisa_hypotheses}). Thus Eq.~\eqref{eq:binned-likelihood}
uses mean reference-normalised powers and the predicted correction for
this model. For $\mathcal H_{\rm para}$, the
deterministic model prediction is averaged over the original cells belonging
to each block. Separate partitions are constructed for the continuous and gapped data.
The joint A/E/T parametric fit uses a shared frequency partition.

\paragraph*{Impact of coarse-graining.}

We assess the particular coarse-graining rule used in this work with paired
LS2 simulations. Time is pooled in groups of four WDM rows, while the
adaptive frequency bins span 1--5 original cells, with median width one.
Relative to the original likelihood, this partition gives a median
$3.40\times$ reduction in sampling time and changes the RMSE by
$+0.0067$ nats, with a $95\%$ bootstrap interval of
$[+0.0040,+0.0128]$ nats. The corresponding truth-inclusion fractions are
$0.942$ and $0.937$.

These checks characterise the coarse-graining used in this analysis rather
than coarse-graining in general. Here coarse-graining is used primarily to
reduce the computational cost of the time-varying fits, and we do not attempt
to optimise the partitioning rule. The trade-off between block size,
computational cost, and reconstruction accuracy, together with alternative
pilot estimates and adaptive partitioning strategies, is left for future
work.

\section{Penalty construction and regularisation}
\label{app:penalty}

The regularised penalty $\tilde{\mathbf Q}$ in
Eq.~\eqref{eq:prior} follows~\citet{lang2004bayesian,maturana2021bayesian,boer2023tensor}. The joint time--frequency penalty is
\begin{equation}
\mathbf Q
=
\sigma_t^{-2}(\mathbf I_{K_f}\otimes\mathbf P_t)
+
\sigma_f^{-2}(\mathbf P_f\otimes\mathbf I_{K_t}),
\codelink{https://github.com/nz-gravity/LogPSplinePSD/blob/v0.2.0/src/log_psplines/basis/penalty.py\#L44-L67}
\end{equation}
as defined in Eq.~\eqref{eq:penalty}. Since $\mathbf P_t$ and $\mathbf P_f$ are symmetric, they admit the following eigendecompositions
\begin{equation}
\mathbf P_t
=
\mathbf U_t\boldsymbol\Lambda_t\mathbf U_t^\top,
\qquad
\mathbf P_f
=
\mathbf U_f\boldsymbol\Lambda_f\mathbf U_f^\top,
\codelink{https://github.com/nz-gravity/LogPSplinePSD/blob/v0.2.0/src/log_psplines/basis/penalty.py\#L17-L40}
\end{equation}
where $\mathbf U_t$ and $\mathbf U_f$ are orthogonal matrices whose columns are the corresponding eigenvectors, and $\boldsymbol\Lambda_t$ and $\boldsymbol\Lambda_f$ are diagonal matrices containing the corresponding eigenvalues.

The Kronecker-sum penalty is diagonal in the basis
$\mathbf U=\mathbf U_f\otimes\mathbf U_t$,
\begin{equation}
\mathbf Q
=
\mathbf U\mathbf D\mathbf U^\top,
\qquad
d_{ab}
=
\sigma_t^{-2}\lambda_a^t
+
\sigma_f^{-2}\lambda_b^f,
\codelink{https://github.com/nz-gravity/LogPSplinePSD/blob/v0.2.0/src/log_psplines/basis/penalty.py\#L17-L40}
\label{eq:penalty_eigendecomposition}
\end{equation}
where $\lambda_a^t$ and $\lambda_b^f$ are the eigenvalues of the time and
frequency penalties.

Some $d_{ab}$ are zero because the second-derivative penalties leave
low-order modes unpenalised, making $\mathbf Q$ singular. We assign these
modes a fixed weak precision $\tau_0$ and add a small ridge $\epsilon$ to
the remaining modes,
\begin{equation}
\tilde{\mathbf D}
=
\mathbf D
+
\epsilon(\mathbf I-\mathbf J_0)
+
\tau_0\mathbf J_0,
\qquad
\tilde{\mathbf Q}
=
\mathbf U\tilde{\mathbf D}\mathbf U^\top,
\codelink{https://github.com/nz-gravity/LogPSplinePSD/blob/v0.2.0/src/log_psplines/basis/penalty.py\#L44-L67}
\label{eq:proper_penalty_eigendecomposition}
\end{equation}
where $\mathbf J_0$ selects the zero-eigenvalue modes. We use $\tau_0=10^{-4}$ and $\epsilon=10^{-6}$. The resulting
$\tilde{\mathbf Q}$ is invertible and defines the proper Gaussian prior in
Eq.~\eqref{eq:prior}.

\newpage


\begin{thebibliography}{100}%
\makeatletter
\providecommand \@ifxundefined [1]{%
 \@ifx{#1\undefined}
}%
\providecommand \@ifnum [1]{%
 \ifnum #1\expandafter \@firstoftwo
 \else \expandafter \@secondoftwo
 \fi
}%
\providecommand \@ifx [1]{%
 \ifx #1\expandafter \@firstoftwo
 \else \expandafter \@secondoftwo
 \fi
}%
\providecommand \natexlab [1]{#1}%
\providecommand \enquote  [1]{``#1''}%
\providecommand \bibnamefont  [1]{#1}%
\providecommand \bibfnamefont [1]{#1}%
\providecommand \citenamefont [1]{#1}%
\providecommand \href@noop [0]{\@secondoftwo}%
\providecommand \href [0]{\begingroup \@sanitize@url \@href}%
\providecommand \@href[1]{\@@startlink{#1}\@@href}%
\providecommand \@@href[1]{\endgroup#1\@@endlink}%
\providecommand \@sanitize@url [0]{\catcode `\\12\catcode `\$12\catcode
  `\&12\catcode `\#12\catcode `\^12\catcode `\_12\catcode `\%12\relax}%
\providecommand \@@startlink[1]{}%
\providecommand \@@endlink[0]{}%
\providecommand \url  [0]{\begingroup\@sanitize@url \@url }%
\providecommand \@url [1]{\endgroup\@href {#1}{\urlprefix }}%
\providecommand \urlprefix  [0]{URL }%
\providecommand \Eprint [0]{\href }%
\providecommand \doibase [0]{https://doi.org/}%
\providecommand \selectlanguage [0]{\@gobble}%
\providecommand \bibinfo  [0]{\@secondoftwo}%
\providecommand \bibfield  [0]{\@secondoftwo}%
\providecommand \translation [1]{[#1]}%
\providecommand \BibitemOpen [0]{}%
\providecommand \bibitemStop [0]{}%
\providecommand \bibitemNoStop [0]{.\EOS\space}%
\providecommand \EOS [0]{\spacefactor3000\relax}%
\providecommand \BibitemShut  [1]{\csname bibitem#1\endcsname}%
\let\auto@bib@innerbib\@empty
\bibitem [{\citenamefont {Amaro-Seoane}\ \emph {et~al.}(2017)\citenamefont
  {Amaro-Seoane} \emph {et~al.}}]{amaroseoane2017lisa}%
  \BibitemOpen
  \bibfield  {author} {\bibinfo {author} {\bibfnamefont {P.}~\bibnamefont
  {Amaro-Seoane}} \emph {et~al.},\ }\href@noop {} {\bibinfo {title} {Laser
  interferometer space antenna}} (\bibinfo {year} {2017}),\ \bibinfo {note}
  {{LISA} mission proposal},\ \Eprint {https://arxiv.org/abs/1702.00786}
  {arXiv:1702.00786 [astro-ph.IM]} \BibitemShut {NoStop}%
\bibitem [{\citenamefont {Littenberg}\ and\ \citenamefont
  {Cornish}(2023)}]{littenberg2023global}%
  \BibitemOpen
  \bibfield  {author} {\bibinfo {author} {\bibfnamefont {T.~B.}\ \bibnamefont
  {Littenberg}}\ and\ \bibinfo {author} {\bibfnamefont {N.~J.}\ \bibnamefont
  {Cornish}},\ }\href {https://doi.org/10.1103/PhysRevD.107.063004} {\bibfield
  {journal} {\bibinfo  {journal} {Physical Review D}\ }\textbf {\bibinfo
  {volume} {107}},\ \bibinfo {pages} {063004} (\bibinfo {year} {2023})},\
  \Eprint {https://arxiv.org/abs/2301.03673} {arXiv:2301.03673 [gr-qc]}
  \BibitemShut {NoStop}%
\bibitem [{\citenamefont {Katz}\ \emph {et~al.}(2025)\citenamefont {Katz},
  \citenamefont {Karnesis}, \citenamefont {Korsakova}, \citenamefont {Gair},\
  and\ \citenamefont {Stergioulas}}]{katz2025efficient}%
  \BibitemOpen
  \bibfield  {author} {\bibinfo {author} {\bibfnamefont {M.~L.}\ \bibnamefont
  {Katz}}, \bibinfo {author} {\bibfnamefont {N.}~\bibnamefont {Karnesis}},
  \bibinfo {author} {\bibfnamefont {N.}~\bibnamefont {Korsakova}}, \bibinfo
  {author} {\bibfnamefont {J.~R.}\ \bibnamefont {Gair}},\ and\ \bibinfo
  {author} {\bibfnamefont {N.}~\bibnamefont {Stergioulas}},\ }\href
  {https://doi.org/10.1103/PhysRevD.111.024060} {\bibfield  {journal} {\bibinfo
   {journal} {Physical Review D}\ }\textbf {\bibinfo {volume} {111}},\ \bibinfo
  {pages} {024060} (\bibinfo {year} {2025})}\BibitemShut {NoStop}%
\bibitem [{\citenamefont {Strub}\ \emph {et~al.}(2024)\citenamefont {Strub},
  \citenamefont {Ferraioli}, \citenamefont {Schmelzbach}, \citenamefont
  {St{\"a}hler},\ and\ \citenamefont {Giardini}}]{strub2024global}%
  \BibitemOpen
  \bibfield  {author} {\bibinfo {author} {\bibfnamefont {S.~H.}\ \bibnamefont
  {Strub}}, \bibinfo {author} {\bibfnamefont {L.}~\bibnamefont {Ferraioli}},
  \bibinfo {author} {\bibfnamefont {C.}~\bibnamefont {Schmelzbach}}, \bibinfo
  {author} {\bibfnamefont {S.~C.}\ \bibnamefont {St{\"a}hler}},\ and\ \bibinfo
  {author} {\bibfnamefont {D.}~\bibnamefont {Giardini}},\ }\href
  {https://arxiv.org/abs/2403.15318} {\bibinfo {title} {Global analysis of
  {LISA} data with galactic binaries and massive black hole binaries}}
  (\bibinfo {year} {2024}),\ \Eprint {https://arxiv.org/abs/2403.15318}
  {arXiv:2403.15318 [gr-qc]} \BibitemShut {NoStop}%
\bibitem [{\citenamefont {Deng}\ \emph {et~al.}(2025)\citenamefont {Deng},
  \citenamefont {Babak}, \citenamefont {Le~Jeune}, \citenamefont {Marsat},
  \citenamefont {Plagnol},\ and\ \citenamefont {Sartirana}}]{deng2025modular}%
  \BibitemOpen
  \bibfield  {author} {\bibinfo {author} {\bibfnamefont {S.}~\bibnamefont
  {Deng}}, \bibinfo {author} {\bibfnamefont {S.}~\bibnamefont {Babak}},
  \bibinfo {author} {\bibfnamefont {M.}~\bibnamefont {Le~Jeune}}, \bibinfo
  {author} {\bibfnamefont {S.}~\bibnamefont {Marsat}}, \bibinfo {author}
  {\bibfnamefont {{\'E}.}\ \bibnamefont {Plagnol}},\ and\ \bibinfo {author}
  {\bibfnamefont {A.}~\bibnamefont {Sartirana}},\ }\href
  {https://arxiv.org/abs/2501.10277} {\bibinfo {title} {Modular global-fit
  pipeline for {LISA} data analysis}} (\bibinfo {year} {2025}),\ \Eprint
  {https://arxiv.org/abs/2501.10277} {arXiv:2501.10277 [gr-qc]} \BibitemShut
  {NoStop}%
\bibitem [{\citenamefont {Digman}\ and\ \citenamefont
  {Cornish}(2022)}]{digman2022cyclostationary}%
  \BibitemOpen
  \bibfield  {author} {\bibinfo {author} {\bibfnamefont {M.~C.}\ \bibnamefont
  {Digman}}\ and\ \bibinfo {author} {\bibfnamefont {N.~J.}\ \bibnamefont
  {Cornish}},\ }\href {https://doi.org/10.3847/1538-4357/ac9139} {\bibfield
  {journal} {\bibinfo  {journal} {The Astrophysical Journal}\ }\textbf
  {\bibinfo {volume} {940}},\ \bibinfo {pages} {10} (\bibinfo {year}
  {2022})}\BibitemShut {NoStop}%
\bibitem [{\citenamefont {Buscicchio}\ \emph
  {et~al.}(2025{\natexlab{a}})\citenamefont {Buscicchio}, \citenamefont
  {Klein}, \citenamefont {Korol}, \citenamefont {Di~Renzo}, \citenamefont
  {Moore}, \citenamefont {Gerosa},\ and\ \citenamefont
  {Carzaniga}}]{buscicchio2025test}%
  \BibitemOpen
  \bibfield  {author} {\bibinfo {author} {\bibfnamefont {R.}~\bibnamefont
  {Buscicchio}}, \bibinfo {author} {\bibfnamefont {A.}~\bibnamefont {Klein}},
  \bibinfo {author} {\bibfnamefont {V.}~\bibnamefont {Korol}}, \bibinfo
  {author} {\bibfnamefont {F.}~\bibnamefont {Di~Renzo}}, \bibinfo {author}
  {\bibfnamefont {C.~J.}\ \bibnamefont {Moore}}, \bibinfo {author}
  {\bibfnamefont {D.}~\bibnamefont {Gerosa}},\ and\ \bibinfo {author}
  {\bibfnamefont {A.}~\bibnamefont {Carzaniga}},\ }\href
  {https://doi.org/10.1140/s10052-025-14616-w} {\bibfield  {journal} {\bibinfo
  {journal} {The European Physical Journal C}\ }\textbf {\bibinfo {volume}
  {85}},\ \bibinfo {pages} {887} (\bibinfo {year}
  {2025}{\natexlab{a}})}\BibitemShut {NoStop}%
\bibitem [{\citenamefont {Mathur}\ and\ \citenamefont
  {Cornish}(2026)}]{mathur2026timefrequencyanalysislisagalactic}%
  \BibitemOpen
  \bibfield  {author} {\bibinfo {author} {\bibfnamefont {S.}~\bibnamefont
  {Mathur}}\ and\ \bibinfo {author} {\bibfnamefont {N.~J.}\ \bibnamefont
  {Cornish}},\ }\href {https://arxiv.org/abs/2610.07977} {\bibinfo {title}
  {Time-frequency analysis for lisa: Galactic foreground modulation}} (\bibinfo
  {year} {2026}),\ \Eprint {https://arxiv.org/abs/2610.07977} {arXiv:2610.07977
  [gr-qc]} \BibitemShut {NoStop}%
\bibitem [{\citenamefont {{Armstrong}}\ \emph {et~al.}(1999)\citenamefont
  {{Armstrong}}, \citenamefont {{Estabrook}},\ and\ \citenamefont
  {{Tinto}}}]{Armstrong1999ApJ}%
  \BibitemOpen
  \bibfield  {author} {\bibinfo {author} {\bibfnamefont {J.~W.}\ \bibnamefont
  {{Armstrong}}}, \bibinfo {author} {\bibfnamefont {F.~B.}\ \bibnamefont
  {{Estabrook}}},\ and\ \bibinfo {author} {\bibfnamefont {M.}~\bibnamefont
  {{Tinto}}},\ }\href {https://doi.org/10.1086/308110} {\bibfield  {journal}
  {\bibinfo  {journal} {\apj}\ }\textbf {\bibinfo {volume} {527}},\ \bibinfo
  {pages} {814} (\bibinfo {year} {1999})}\BibitemShut {NoStop}%
\bibitem [{\citenamefont {{Vallisneri}}(2005)}]{Vallisneri2005PhRvD}%
  \BibitemOpen
  \bibfield  {author} {\bibinfo {author} {\bibfnamefont {M.}~\bibnamefont
  {{Vallisneri}}},\ }\href {https://doi.org/10.1103/PhysRevD.72.042003}
  {\bibfield  {journal} {\bibinfo  {journal} {\prd}\ }\textbf {\bibinfo
  {volume} {72}},\ \bibinfo {eid} {042003} (\bibinfo {year} {2005})},\ \Eprint
  {https://arxiv.org/abs/gr-qc/0504145} {arXiv:gr-qc/0504145 [gr-qc]}
  \BibitemShut {NoStop}%
\bibitem [{\citenamefont {{Tinto}}\ and\ \citenamefont
  {{Dhurandhar}}(2021)}]{tinto2021tdi}%
  \BibitemOpen
  \bibfield  {author} {\bibinfo {author} {\bibfnamefont {M.}~\bibnamefont
  {{Tinto}}}\ and\ \bibinfo {author} {\bibfnamefont {S.~V.}\ \bibnamefont
  {{Dhurandhar}}},\ }\href {https://doi.org/10.1007/s41114-020-00029-6}
  {\bibfield  {journal} {\bibinfo  {journal} {Living Reviews in Relativity}\
  }\textbf {\bibinfo {volume} {24}},\ \bibinfo {eid} {1} (\bibinfo {year}
  {2021})},\ \Eprint {https://arxiv.org/abs/gr-qc/0409034} {arXiv:gr-qc/0409034
  [gr-qc]} \BibitemShut {NoStop}%
\bibitem [{\citenamefont {Bayle}(2019)}]{bayle2019thesis}%
  \BibitemOpen
  \bibfield  {author} {\bibinfo {author} {\bibfnamefont {J.-B.}\ \bibnamefont
  {Bayle}},\ }\emph {\bibinfo {title} {Simulation and Data Analysis for LISA:
  Instrumental Modeling, Time-Delay Interferometry, Noise-Reduction Performance
  Study, and Discrimination of Transient Gravitational Signals}},\ \href
  {https://doi.org/10.70675/ab35754bz520bz49e8zad08zad6c974b4e5b} {\bibinfo
  {type} {{PhD} thesis}},\ \bibinfo  {school} {Universit{\'e} de Paris}
  (\bibinfo {year} {2019})\BibitemShut {NoStop}%
\bibitem [{\citenamefont {{Armano}}\ \emph {et~al.}(2021)\citenamefont
  {{Armano}}, \citenamefont {{Audley}}, \citenamefont {{Baird}}, \citenamefont
  {{Binetruy}}, \citenamefont {{Born}}, \citenamefont {{Bortoluzzi}},
  \citenamefont {{Brandt}}, \citenamefont {{Castelli}}, \citenamefont
  {{Cavalleri}}, \citenamefont {{Cesarini}}, \citenamefont {{Cruise}},
  \citenamefont {{Danzmann}}, \citenamefont {{de Deus Silva}}, \citenamefont
  {{Diepholz}}, \citenamefont {{Dixon}}, \citenamefont {{Dolesi}},
  \citenamefont {{Ferraioli}}, \citenamefont {{Ferroni}}, \citenamefont
  {{Fitzsimons}}, \citenamefont {{Flatscher}}, \citenamefont {{Freschi}},
  \citenamefont {{Garc{\'\i}a}}, \citenamefont {{Gerndt}}, \citenamefont
  {{Gesa}}, \citenamefont {{Giardini}}, \citenamefont {{Gibert}}, \citenamefont
  {{Giusteri}}, \citenamefont {{Grimani}}, \citenamefont {{Grzymisch}},
  \citenamefont {{Guzman}}, \citenamefont {{Harrison}}, \citenamefont
  {{Hartig}}, \citenamefont {{Heinzel}}, \citenamefont {{Hewitson}},
  \citenamefont {{Hollington}}, \citenamefont {{Hoyland}}, \citenamefont
  {{Hueller}}, \citenamefont {{Inchausp{\'e}}}, \citenamefont {{Jennrich}},
  \citenamefont {{Jetzer}}, \citenamefont {{Johann}}, \citenamefont
  {{Johlander}}, \citenamefont {{Karnesis}}, \citenamefont {{Kaune}},
  \citenamefont {{Killow}}, \citenamefont {{Korsakova}}, \citenamefont
  {{Lobo}}, \citenamefont {{Liu}}, \citenamefont {{L{\'o}pez-Zaragoza}},
  \citenamefont {{Maarschalkerweerd}}, \citenamefont {{Mance}}, \citenamefont
  {{Mart{\'\i}n}}, \citenamefont {{Martin-Polo}}, \citenamefont
  {{Martin-Porqueras}}, \citenamefont {{Martino}}, \citenamefont {{McNamara}},
  \citenamefont {{Mendes}}, \citenamefont {{Mendes}}, \citenamefont
  {{Meshksar}}, \citenamefont {{Monsky}}, \citenamefont {{Nofrarias}},
  \citenamefont {{Paczkowski}}, \citenamefont {{Perreur-Lloyd}}, \citenamefont
  {{Petiteau}}, \citenamefont {{Pivato}}, \citenamefont {{Plagnol}},
  \citenamefont {{Ramos-Castro}}, \citenamefont {{Reiche}}, \citenamefont
  {{Rivas}}, \citenamefont {{Robertson}}, \citenamefont {{Russano}},
  \citenamefont {{Sanjuan}}, \citenamefont {{Slutsky}}, \citenamefont
  {{Sopuerta}}, \citenamefont {{Steier}}, \citenamefont {{Sumner}},
  \citenamefont {{Texier}}, \citenamefont {{Thorpe}}, \citenamefont
  {{Vetrugno}}, \citenamefont {{Vitale}}, \citenamefont {{Wand}}, \citenamefont
  {{Wanner}}, \citenamefont {{Ward}}, \citenamefont {{Wass}}, \citenamefont
  {{Weber}}, \citenamefont {{Wissel}}, \citenamefont {{Wittchen}},\ and\
  \citenamefont {{Zweifel}}}]{Armano2021PhRvL}%
  \BibitemOpen
  \bibfield  {author} {\bibinfo {author} {\bibfnamefont {M.}~\bibnamefont
  {{Armano}}}, \bibinfo {author} {\bibfnamefont {H.}~\bibnamefont {{Audley}}},
  \bibinfo {author} {\bibfnamefont {J.}~\bibnamefont {{Baird}}}, \bibinfo
  {author} {\bibfnamefont {P.}~\bibnamefont {{Binetruy}}}, \bibinfo {author}
  {\bibfnamefont {M.}~\bibnamefont {{Born}}}, \bibinfo {author} {\bibfnamefont
  {D.}~\bibnamefont {{Bortoluzzi}}}, \bibinfo {author} {\bibfnamefont
  {N.}~\bibnamefont {{Brandt}}}, \bibinfo {author} {\bibfnamefont
  {E.}~\bibnamefont {{Castelli}}}, \bibinfo {author} {\bibfnamefont
  {A.}~\bibnamefont {{Cavalleri}}}, \bibinfo {author} {\bibfnamefont
  {A.}~\bibnamefont {{Cesarini}}}, \bibinfo {author} {\bibfnamefont {A.~M.}\
  \bibnamefont {{Cruise}}}, \bibinfo {author} {\bibfnamefont {K.}~\bibnamefont
  {{Danzmann}}}, \bibinfo {author} {\bibfnamefont {M.}~\bibnamefont {{de Deus
  Silva}}}, \bibinfo {author} {\bibfnamefont {I.}~\bibnamefont {{Diepholz}}},
  \bibinfo {author} {\bibfnamefont {G.}~\bibnamefont {{Dixon}}}, \bibinfo
  {author} {\bibfnamefont {R.}~\bibnamefont {{Dolesi}}}, \bibinfo {author}
  {\bibfnamefont {L.}~\bibnamefont {{Ferraioli}}}, \bibinfo {author}
  {\bibfnamefont {V.}~\bibnamefont {{Ferroni}}}, \bibinfo {author}
  {\bibfnamefont {E.~D.}\ \bibnamefont {{Fitzsimons}}}, \bibinfo {author}
  {\bibfnamefont {R.}~\bibnamefont {{Flatscher}}}, \bibinfo {author}
  {\bibfnamefont {M.}~\bibnamefont {{Freschi}}}, \bibinfo {author}
  {\bibfnamefont {A.}~\bibnamefont {{Garc{\'\i}a}}}, \bibinfo {author}
  {\bibfnamefont {R.}~\bibnamefont {{Gerndt}}}, \bibinfo {author}
  {\bibfnamefont {L.}~\bibnamefont {{Gesa}}}, \bibinfo {author} {\bibfnamefont
  {D.}~\bibnamefont {{Giardini}}}, \bibinfo {author} {\bibfnamefont
  {F.}~\bibnamefont {{Gibert}}}, \bibinfo {author} {\bibfnamefont
  {R.}~\bibnamefont {{Giusteri}}}, \bibinfo {author} {\bibfnamefont
  {C.}~\bibnamefont {{Grimani}}}, \bibinfo {author} {\bibfnamefont
  {J.}~\bibnamefont {{Grzymisch}}}, \bibinfo {author} {\bibfnamefont
  {F.}~\bibnamefont {{Guzman}}}, \bibinfo {author} {\bibfnamefont
  {I.}~\bibnamefont {{Harrison}}}, \bibinfo {author} {\bibfnamefont {M.-S.}\
  \bibnamefont {{Hartig}}}, \bibinfo {author} {\bibfnamefont {G.}~\bibnamefont
  {{Heinzel}}}, \bibinfo {author} {\bibfnamefont {M.}~\bibnamefont
  {{Hewitson}}}, \bibinfo {author} {\bibfnamefont {D.}~\bibnamefont
  {{Hollington}}}, \bibinfo {author} {\bibfnamefont {D.}~\bibnamefont
  {{Hoyland}}}, \bibinfo {author} {\bibfnamefont {M.}~\bibnamefont
  {{Hueller}}}, \bibinfo {author} {\bibfnamefont {H.}~\bibnamefont
  {{Inchausp{\'e}}}}, \bibinfo {author} {\bibfnamefont {O.}~\bibnamefont
  {{Jennrich}}}, \bibinfo {author} {\bibfnamefont {P.}~\bibnamefont
  {{Jetzer}}}, \bibinfo {author} {\bibfnamefont {U.}~\bibnamefont {{Johann}}},
  \bibinfo {author} {\bibfnamefont {B.}~\bibnamefont {{Johlander}}}, \bibinfo
  {author} {\bibfnamefont {N.}~\bibnamefont {{Karnesis}}}, \bibinfo {author}
  {\bibfnamefont {B.}~\bibnamefont {{Kaune}}}, \bibinfo {author} {\bibfnamefont
  {C.~J.}\ \bibnamefont {{Killow}}}, \bibinfo {author} {\bibfnamefont
  {N.}~\bibnamefont {{Korsakova}}}, \bibinfo {author} {\bibfnamefont {J.~A.}\
  \bibnamefont {{Lobo}}}, \bibinfo {author} {\bibfnamefont {L.}~\bibnamefont
  {{Liu}}}, \bibinfo {author} {\bibfnamefont {J.~P.}\ \bibnamefont
  {{L{\'o}pez-Zaragoza}}}, \bibinfo {author} {\bibfnamefont {R.}~\bibnamefont
  {{Maarschalkerweerd}}}, \bibinfo {author} {\bibfnamefont {D.}~\bibnamefont
  {{Mance}}}, \bibinfo {author} {\bibfnamefont {V.}~\bibnamefont
  {{Mart{\'\i}n}}}, \bibinfo {author} {\bibfnamefont {L.}~\bibnamefont
  {{Martin-Polo}}}, \bibinfo {author} {\bibfnamefont {F.}~\bibnamefont
  {{Martin-Porqueras}}}, \bibinfo {author} {\bibfnamefont {J.}~\bibnamefont
  {{Martino}}}, \bibinfo {author} {\bibfnamefont {P.~W.}\ \bibnamefont
  {{McNamara}}}, \bibinfo {author} {\bibfnamefont {J.}~\bibnamefont
  {{Mendes}}}, \bibinfo {author} {\bibfnamefont {L.}~\bibnamefont {{Mendes}}},
  \bibinfo {author} {\bibfnamefont {N.}~\bibnamefont {{Meshksar}}}, \bibinfo
  {author} {\bibfnamefont {A.}~\bibnamefont {{Monsky}}}, \bibinfo {author}
  {\bibfnamefont {M.}~\bibnamefont {{Nofrarias}}}, \bibinfo {author}
  {\bibfnamefont {S.}~\bibnamefont {{Paczkowski}}}, \bibinfo {author}
  {\bibfnamefont {M.}~\bibnamefont {{Perreur-Lloyd}}}, \bibinfo {author}
  {\bibfnamefont {A.}~\bibnamefont {{Petiteau}}}, \bibinfo {author}
  {\bibfnamefont {P.}~\bibnamefont {{Pivato}}}, \bibinfo {author}
  {\bibfnamefont {E.}~\bibnamefont {{Plagnol}}}, \bibinfo {author}
  {\bibfnamefont {J.}~\bibnamefont {{Ramos-Castro}}}, \bibinfo {author}
  {\bibfnamefont {J.}~\bibnamefont {{Reiche}}}, \bibinfo {author}
  {\bibfnamefont {F.}~\bibnamefont {{Rivas}}}, \bibinfo {author} {\bibfnamefont
  {D.~I.}\ \bibnamefont {{Robertson}}}, \bibinfo {author} {\bibfnamefont
  {G.}~\bibnamefont {{Russano}}}, \bibinfo {author} {\bibfnamefont
  {J.}~\bibnamefont {{Sanjuan}}}, \bibinfo {author} {\bibfnamefont
  {J.}~\bibnamefont {{Slutsky}}}, \bibinfo {author} {\bibfnamefont {C.~F.}\
  \bibnamefont {{Sopuerta}}}, \bibinfo {author} {\bibfnamefont
  {F.}~\bibnamefont {{Steier}}}, \bibinfo {author} {\bibfnamefont
  {T.}~\bibnamefont {{Sumner}}}, \bibinfo {author} {\bibfnamefont
  {D.}~\bibnamefont {{Texier}}}, \bibinfo {author} {\bibfnamefont {J.~I.}\
  \bibnamefont {{Thorpe}}}, \bibinfo {author} {\bibfnamefont {D.}~\bibnamefont
  {{Vetrugno}}}, \bibinfo {author} {\bibfnamefont {S.}~\bibnamefont
  {{Vitale}}}, \bibinfo {author} {\bibfnamefont {V.}~\bibnamefont {{Wand}}},
  \bibinfo {author} {\bibfnamefont {G.}~\bibnamefont {{Wanner}}}, \bibinfo
  {author} {\bibfnamefont {H.}~\bibnamefont {{Ward}}}, \bibinfo {author}
  {\bibfnamefont {P.~J.}\ \bibnamefont {{Wass}}}, \bibinfo {author}
  {\bibfnamefont {W.~J.}\ \bibnamefont {{Weber}}}, \bibinfo {author}
  {\bibfnamefont {L.}~\bibnamefont {{Wissel}}}, \bibinfo {author}
  {\bibfnamefont {A.}~\bibnamefont {{Wittchen}}},\ and\ \bibinfo {author}
  {\bibfnamefont {P.}~\bibnamefont {{Zweifel}}},\ }\href
  {https://doi.org/10.1103/PhysRevLett.126.131103} {\bibfield  {journal}
  {\bibinfo  {journal} {\prl}\ }\textbf {\bibinfo {volume} {126}},\ \bibinfo
  {eid} {131103} (\bibinfo {year} {2021})}\BibitemShut {NoStop}%
\bibitem [{\citenamefont {{Armano}}\ \emph {et~al.}(2017)\citenamefont
  {{Armano}}, \citenamefont {{Audley}}, \citenamefont {{Auger}}, \citenamefont
  {{Baird}}, \citenamefont {{Binetruy}}, \citenamefont {{Born}}, \citenamefont
  {{Bortoluzzi}}, \citenamefont {{Brandt}}, \citenamefont {{Bursi}},
  \citenamefont {{Caleno}}, \citenamefont {{Cavalleri}}, \citenamefont
  {{Cesarini}}, \citenamefont {{Cruise}}, \citenamefont {{Danzmann}},
  \citenamefont {{de Deus Silva}}, \citenamefont {{Diepholz}}, \citenamefont
  {{Dolesi}}, \citenamefont {{Dunbar}}, \citenamefont {{Ferraioli}},
  \citenamefont {{Ferroni}}, \citenamefont {{Fitzsimons}}, \citenamefont
  {{Flatscher}}, \citenamefont {{Freschi}}, \citenamefont {{Gallegos}},
  \citenamefont {{Garc{\'\i}a Marirrodriga}}, \citenamefont {{Gerndt}},
  \citenamefont {{Gesa}}, \citenamefont {{Gibert}}, \citenamefont {{Giardini}},
  \citenamefont {{Giusteri}}, \citenamefont {{Grimani}}, \citenamefont
  {{Grzymisch}}, \citenamefont {{Harrison}}, \citenamefont {{Heinzel}},
  \citenamefont {{Hewitson}}, \citenamefont {{Hollington}}, \citenamefont
  {{Hueller}}, \citenamefont {{Huesler}}, \citenamefont {{Inchausp{\'e}}},
  \citenamefont {{Jennrich}}, \citenamefont {{Jetzer}}, \citenamefont
  {{Johlander}}, \citenamefont {{Karnesis}}, \citenamefont {{Kaune}},
  \citenamefont {{Killow}}, \citenamefont {{Korsakova}}, \citenamefont
  {{Lloro}}, \citenamefont {{Liu}}, \citenamefont {{L{\'o}pez-Zaragoza}},
  \citenamefont {{Maarschalkerweerd}}, \citenamefont {{Madden}}, \citenamefont
  {{Mance}}, \citenamefont {{Mart{\'\i}n}}, \citenamefont {{Martin-Polo}},
  \citenamefont {{Martino}}, \citenamefont {{Martin-Porqueras}}, \citenamefont
  {{Mateos}}, \citenamefont {{McNamara}}, \citenamefont {{Mendes}},
  \citenamefont {{Mendes}}, \citenamefont {{Moroni}}, \citenamefont
  {{Nofrarias}}, \citenamefont {{Paczkowski}}, \citenamefont {{Perreur-Lloyd}},
  \citenamefont {{Petiteau}}, \citenamefont {{Pivato}}, \citenamefont
  {{Plagnol}}, \citenamefont {{Prat}}, \citenamefont {{Ragnit}}, \citenamefont
  {{Ramos-Castro}}, \citenamefont {{Reiche}}, \citenamefont {{Romera Perez}},
  \citenamefont {{Robertson}}, \citenamefont {{Rozemeijer}}, \citenamefont
  {{Rivas}}, \citenamefont {{Russano}}, \citenamefont {{Sarra}}, \citenamefont
  {{Schleicher}}, \citenamefont {{Slutsky}}, \citenamefont {{Sopuerta}},
  \citenamefont {{Sumner}}, \citenamefont {{Texier}}, \citenamefont {{Thorpe}},
  \citenamefont {{Trenkel}}, \citenamefont {{Vetrugno}}, \citenamefont
  {{Vitale}}, \citenamefont {{Wanner}}, \citenamefont {{Ward}}, \citenamefont
  {{Wass}}, \citenamefont {{Wealthy}}, \citenamefont {{Weber}}, \citenamefont
  {{Wittchen}}, \citenamefont {{Zanoni}}, \citenamefont {{Ziegler}},
  \citenamefont {{Zweifel}},\ and\ \citenamefont {{LISA Pathfinder
  Collaboration}}}]{Armano2017PhRvL}%
  \BibitemOpen
  \bibfield  {author} {\bibinfo {author} {\bibfnamefont {M.}~\bibnamefont
  {{Armano}}}, \bibinfo {author} {\bibfnamefont {H.}~\bibnamefont {{Audley}}},
  \bibinfo {author} {\bibfnamefont {G.}~\bibnamefont {{Auger}}}, \bibinfo
  {author} {\bibfnamefont {J.~T.}\ \bibnamefont {{Baird}}}, \bibinfo {author}
  {\bibfnamefont {P.}~\bibnamefont {{Binetruy}}}, \bibinfo {author}
  {\bibfnamefont {M.}~\bibnamefont {{Born}}}, \bibinfo {author} {\bibfnamefont
  {D.}~\bibnamefont {{Bortoluzzi}}}, \bibinfo {author} {\bibfnamefont
  {N.}~\bibnamefont {{Brandt}}}, \bibinfo {author} {\bibfnamefont
  {A.}~\bibnamefont {{Bursi}}}, \bibinfo {author} {\bibfnamefont
  {M.}~\bibnamefont {{Caleno}}}, \bibinfo {author} {\bibfnamefont
  {A.}~\bibnamefont {{Cavalleri}}}, \bibinfo {author} {\bibfnamefont
  {A.}~\bibnamefont {{Cesarini}}}, \bibinfo {author} {\bibfnamefont
  {M.}~\bibnamefont {{Cruise}}}, \bibinfo {author} {\bibfnamefont
  {K.}~\bibnamefont {{Danzmann}}}, \bibinfo {author} {\bibfnamefont
  {M.}~\bibnamefont {{de Deus Silva}}}, \bibinfo {author} {\bibfnamefont
  {I.}~\bibnamefont {{Diepholz}}}, \bibinfo {author} {\bibfnamefont
  {R.}~\bibnamefont {{Dolesi}}}, \bibinfo {author} {\bibfnamefont
  {N.}~\bibnamefont {{Dunbar}}}, \bibinfo {author} {\bibfnamefont
  {L.}~\bibnamefont {{Ferraioli}}}, \bibinfo {author} {\bibfnamefont
  {V.}~\bibnamefont {{Ferroni}}}, \bibinfo {author} {\bibfnamefont {E.~D.}\
  \bibnamefont {{Fitzsimons}}}, \bibinfo {author} {\bibfnamefont
  {R.}~\bibnamefont {{Flatscher}}}, \bibinfo {author} {\bibfnamefont
  {M.}~\bibnamefont {{Freschi}}}, \bibinfo {author} {\bibfnamefont
  {J.}~\bibnamefont {{Gallegos}}}, \bibinfo {author} {\bibfnamefont
  {C.}~\bibnamefont {{Garc{\'\i}a Marirrodriga}}}, \bibinfo {author}
  {\bibfnamefont {R.}~\bibnamefont {{Gerndt}}}, \bibinfo {author}
  {\bibfnamefont {L.}~\bibnamefont {{Gesa}}}, \bibinfo {author} {\bibfnamefont
  {F.}~\bibnamefont {{Gibert}}}, \bibinfo {author} {\bibfnamefont
  {D.}~\bibnamefont {{Giardini}}}, \bibinfo {author} {\bibfnamefont
  {R.}~\bibnamefont {{Giusteri}}}, \bibinfo {author} {\bibfnamefont
  {C.}~\bibnamefont {{Grimani}}}, \bibinfo {author} {\bibfnamefont
  {J.}~\bibnamefont {{Grzymisch}}}, \bibinfo {author} {\bibfnamefont
  {I.}~\bibnamefont {{Harrison}}}, \bibinfo {author} {\bibfnamefont
  {G.}~\bibnamefont {{Heinzel}}}, \bibinfo {author} {\bibfnamefont
  {M.}~\bibnamefont {{Hewitson}}}, \bibinfo {author} {\bibfnamefont
  {D.}~\bibnamefont {{Hollington}}}, \bibinfo {author} {\bibfnamefont
  {M.}~\bibnamefont {{Hueller}}}, \bibinfo {author} {\bibfnamefont
  {J.}~\bibnamefont {{Huesler}}}, \bibinfo {author} {\bibfnamefont
  {H.}~\bibnamefont {{Inchausp{\'e}}}}, \bibinfo {author} {\bibfnamefont
  {O.}~\bibnamefont {{Jennrich}}}, \bibinfo {author} {\bibfnamefont
  {P.}~\bibnamefont {{Jetzer}}}, \bibinfo {author} {\bibfnamefont
  {B.}~\bibnamefont {{Johlander}}}, \bibinfo {author} {\bibfnamefont
  {N.}~\bibnamefont {{Karnesis}}}, \bibinfo {author} {\bibfnamefont
  {B.}~\bibnamefont {{Kaune}}}, \bibinfo {author} {\bibfnamefont {C.~J.}\
  \bibnamefont {{Killow}}}, \bibinfo {author} {\bibfnamefont {N.}~\bibnamefont
  {{Korsakova}}}, \bibinfo {author} {\bibfnamefont {I.}~\bibnamefont
  {{Lloro}}}, \bibinfo {author} {\bibfnamefont {L.}~\bibnamefont {{Liu}}},
  \bibinfo {author} {\bibfnamefont {J.~P.}\ \bibnamefont
  {{L{\'o}pez-Zaragoza}}}, \bibinfo {author} {\bibfnamefont {R.}~\bibnamefont
  {{Maarschalkerweerd}}}, \bibinfo {author} {\bibfnamefont {S.}~\bibnamefont
  {{Madden}}}, \bibinfo {author} {\bibfnamefont {D.}~\bibnamefont {{Mance}}},
  \bibinfo {author} {\bibfnamefont {V.}~\bibnamefont {{Mart{\'\i}n}}}, \bibinfo
  {author} {\bibfnamefont {L.}~\bibnamefont {{Martin-Polo}}}, \bibinfo {author}
  {\bibfnamefont {J.}~\bibnamefont {{Martino}}}, \bibinfo {author}
  {\bibfnamefont {F.}~\bibnamefont {{Martin-Porqueras}}}, \bibinfo {author}
  {\bibfnamefont {I.}~\bibnamefont {{Mateos}}}, \bibinfo {author}
  {\bibfnamefont {P.~W.}\ \bibnamefont {{McNamara}}}, \bibinfo {author}
  {\bibfnamefont {J.}~\bibnamefont {{Mendes}}}, \bibinfo {author}
  {\bibfnamefont {L.}~\bibnamefont {{Mendes}}}, \bibinfo {author}
  {\bibfnamefont {A.}~\bibnamefont {{Moroni}}}, \bibinfo {author}
  {\bibfnamefont {M.}~\bibnamefont {{Nofrarias}}}, \bibinfo {author}
  {\bibfnamefont {S.}~\bibnamefont {{Paczkowski}}}, \bibinfo {author}
  {\bibfnamefont {M.}~\bibnamefont {{Perreur-Lloyd}}}, \bibinfo {author}
  {\bibfnamefont {A.}~\bibnamefont {{Petiteau}}}, \bibinfo {author}
  {\bibfnamefont {P.}~\bibnamefont {{Pivato}}}, \bibinfo {author}
  {\bibfnamefont {E.}~\bibnamefont {{Plagnol}}}, \bibinfo {author}
  {\bibfnamefont {P.}~\bibnamefont {{Prat}}}, \bibinfo {author} {\bibfnamefont
  {U.}~\bibnamefont {{Ragnit}}}, \bibinfo {author} {\bibfnamefont
  {J.}~\bibnamefont {{Ramos-Castro}}}, \bibinfo {author} {\bibfnamefont
  {J.}~\bibnamefont {{Reiche}}}, \bibinfo {author} {\bibfnamefont {J.~A.}\
  \bibnamefont {{Romera Perez}}}, \bibinfo {author} {\bibfnamefont {D.~I.}\
  \bibnamefont {{Robertson}}}, \bibinfo {author} {\bibfnamefont
  {H.}~\bibnamefont {{Rozemeijer}}}, \bibinfo {author} {\bibfnamefont
  {F.}~\bibnamefont {{Rivas}}}, \bibinfo {author} {\bibfnamefont
  {G.}~\bibnamefont {{Russano}}}, \bibinfo {author} {\bibfnamefont
  {P.}~\bibnamefont {{Sarra}}}, \bibinfo {author} {\bibfnamefont
  {A.}~\bibnamefont {{Schleicher}}}, \bibinfo {author} {\bibfnamefont
  {J.}~\bibnamefont {{Slutsky}}}, \bibinfo {author} {\bibfnamefont
  {C.}~\bibnamefont {{Sopuerta}}}, \bibinfo {author} {\bibfnamefont {T.~J.}\
  \bibnamefont {{Sumner}}}, \bibinfo {author} {\bibfnamefont {D.}~\bibnamefont
  {{Texier}}}, \bibinfo {author} {\bibfnamefont {J.~I.}\ \bibnamefont
  {{Thorpe}}}, \bibinfo {author} {\bibfnamefont {C.}~\bibnamefont {{Trenkel}}},
  \bibinfo {author} {\bibfnamefont {D.}~\bibnamefont {{Vetrugno}}}, \bibinfo
  {author} {\bibfnamefont {S.}~\bibnamefont {{Vitale}}}, \bibinfo {author}
  {\bibfnamefont {G.}~\bibnamefont {{Wanner}}}, \bibinfo {author}
  {\bibfnamefont {H.}~\bibnamefont {{Ward}}}, \bibinfo {author} {\bibfnamefont
  {P.~J.}\ \bibnamefont {{Wass}}}, \bibinfo {author} {\bibfnamefont
  {D.}~\bibnamefont {{Wealthy}}}, \bibinfo {author} {\bibfnamefont {W.~J.}\
  \bibnamefont {{Weber}}}, \bibinfo {author} {\bibfnamefont {A.}~\bibnamefont
  {{Wittchen}}}, \bibinfo {author} {\bibfnamefont {C.}~\bibnamefont
  {{Zanoni}}}, \bibinfo {author} {\bibfnamefont {T.}~\bibnamefont {{Ziegler}}},
  \bibinfo {author} {\bibfnamefont {P.}~\bibnamefont {{Zweifel}}},\ and\
  \bibinfo {author} {\bibnamefont {{LISA Pathfinder Collaboration}}},\ }\href
  {https://doi.org/10.1103/PhysRevLett.118.171101} {\bibfield  {journal}
  {\bibinfo  {journal} {\prl}\ }\textbf {\bibinfo {volume} {118}},\ \bibinfo
  {eid} {171101} (\bibinfo {year} {2017})},\ \Eprint
  {https://arxiv.org/abs/1702.04633} {arXiv:1702.04633 [astro-ph.IM]}
  \BibitemShut {NoStop}%
\bibitem [{\citenamefont {{Danzmann}}\ and\ \citenamefont
  {{R{\"u}diger}}(2003)}]{Danzmann2003CQGra}%
  \BibitemOpen
  \bibfield  {author} {\bibinfo {author} {\bibfnamefont {K.}~\bibnamefont
  {{Danzmann}}}\ and\ \bibinfo {author} {\bibfnamefont {A.}~\bibnamefont
  {{R{\"u}diger}}},\ }\href {https://doi.org/10.1088/0264-9381/20/10/301}
  {\bibfield  {journal} {\bibinfo  {journal} {Classical and Quantum Gravity}\
  }\textbf {\bibinfo {volume} {20}},\ \bibinfo {pages} {S1} (\bibinfo {year}
  {2003})}\BibitemShut {NoStop}%
\bibitem [{\citenamefont {{Pagone}}\ and\ \citenamefont
  {{Novara}}(2024)}]{Pagone2024}%
  \BibitemOpen
  \bibfield  {author} {\bibinfo {author} {\bibfnamefont {M.}~\bibnamefont
  {{Pagone}}}\ and\ \bibinfo {author} {\bibfnamefont {C.}~\bibnamefont
  {{Novara}}},\ }\href {https://doi.org/10.48550/arXiv.2405.02339} {\bibfield
  {journal} {\bibinfo  {journal} {arXiv e-prints}\ ,\ \bibinfo {eid}
  {arXiv:2405.02339}} (\bibinfo {year} {2024})},\ \Eprint
  {https://arxiv.org/abs/2405.02339} {arXiv:2405.02339 [astro-ph.IM]}
  \BibitemShut {NoStop}%
\bibitem [{\citenamefont {{Dimiccoli}}\ \emph {et~al.}(2025)\citenamefont
  {{Dimiccoli}}, \citenamefont {{Dolesi}}, \citenamefont {{Fabi}},
  \citenamefont {{Ferroni}}, \citenamefont {{Grimani}}, \citenamefont
  {{Muratore}}, \citenamefont {{Sarra}}, \citenamefont {{Villani}},\ and\
  \citenamefont {{Weber}}}]{Dimiccoli2025A}%
  \BibitemOpen
  \bibfield  {author} {\bibinfo {author} {\bibfnamefont {F.}~\bibnamefont
  {{Dimiccoli}}}, \bibinfo {author} {\bibfnamefont {R.}~\bibnamefont
  {{Dolesi}}}, \bibinfo {author} {\bibfnamefont {M.}~\bibnamefont {{Fabi}}},
  \bibinfo {author} {\bibfnamefont {V.}~\bibnamefont {{Ferroni}}}, \bibinfo
  {author} {\bibfnamefont {C.}~\bibnamefont {{Grimani}}}, \bibinfo {author}
  {\bibfnamefont {M.}~\bibnamefont {{Muratore}}}, \bibinfo {author}
  {\bibfnamefont {P.}~\bibnamefont {{Sarra}}}, \bibinfo {author} {\bibfnamefont
  {M.}~\bibnamefont {{Villani}}},\ and\ \bibinfo {author} {\bibfnamefont
  {W.~J.}\ \bibnamefont {{Weber}}},\ }\href
  {https://doi.org/10.1051/0004-6361/202453076} {\bibfield  {journal} {\bibinfo
   {journal} {\aap}\ }\textbf {\bibinfo {volume} {700}},\ \bibinfo {eid} {A102}
  (\bibinfo {year} {2025})},\ \Eprint {https://arxiv.org/abs/2411.18030}
  {arXiv:2411.18030 [astro-ph.IM]} \BibitemShut {NoStop}%
\bibitem [{\citenamefont {Edy}\ \emph {et~al.}(2021)\citenamefont {Edy},
  \citenamefont {Lundgren},\ and\ \citenamefont
  {Nuttall}}]{edy2021mismodeling}%
  \BibitemOpen
  \bibfield  {author} {\bibinfo {author} {\bibfnamefont {O.}~\bibnamefont
  {Edy}}, \bibinfo {author} {\bibfnamefont {A.}~\bibnamefont {Lundgren}},\ and\
  \bibinfo {author} {\bibfnamefont {L.~K.}\ \bibnamefont {Nuttall}},\ }\href
  {https://doi.org/10.1103/PhysRevD.103.124061} {\bibfield  {journal} {\bibinfo
   {journal} {Physical Review D}\ }\textbf {\bibinfo {volume} {103}},\ \bibinfo
  {pages} {124061} (\bibinfo {year} {2021})},\ \Eprint
  {https://arxiv.org/abs/2101.07743} {arXiv:2101.07743 [astro-ph.IM]}
  \BibitemShut {NoStop}%
\bibitem [{\citenamefont {Talbot}\ and\ \citenamefont
  {Thrane}(2020)}]{talbot2020uncertain}%
  \BibitemOpen
  \bibfield  {author} {\bibinfo {author} {\bibfnamefont {C.}~\bibnamefont
  {Talbot}}\ and\ \bibinfo {author} {\bibfnamefont {E.}~\bibnamefont
  {Thrane}},\ }\href {https://doi.org/10.1103/PhysRevResearch.2.043298}
  {\bibfield  {journal} {\bibinfo  {journal} {Physical Review Research}\
  }\textbf {\bibinfo {volume} {2}},\ \bibinfo {pages} {043298} (\bibinfo {year}
  {2020})},\ \Eprint {https://arxiv.org/abs/2006.05292} {arXiv:2006.05292
  [astro-ph.IM]} \BibitemShut {NoStop}%
\bibitem [{\citenamefont {Burke}\ \emph {et~al.}(2025)\citenamefont {Burke},
  \citenamefont {Marsat}, \citenamefont {Gair},\ and\ \citenamefont
  {Katz}}]{burke2025gaps}%
  \BibitemOpen
  \bibfield  {author} {\bibinfo {author} {\bibfnamefont {O.}~\bibnamefont
  {Burke}}, \bibinfo {author} {\bibfnamefont {S.}~\bibnamefont {Marsat}},
  \bibinfo {author} {\bibfnamefont {J.~R.}\ \bibnamefont {Gair}},\ and\
  \bibinfo {author} {\bibfnamefont {M.~L.}\ \bibnamefont {Katz}},\ }\href
  {https://doi.org/10.1103/5jr8-k2ss} {\bibfield  {journal} {\bibinfo
  {journal} {Physical Review D}\ }\textbf {\bibinfo {volume} {111}},\ \bibinfo
  {pages} {124053} (\bibinfo {year} {2025})},\ \Eprint
  {https://arxiv.org/abs/2502.17426} {arXiv:2502.17426 [gr-qc]} \BibitemShut
  {NoStop}%
\bibitem [{\citenamefont {{Yan}}\ and\ \citenamefont
  {{Guo}}(2026)}]{Yan:2026:arXiv}%
  \BibitemOpen
  \bibfield  {author} {\bibinfo {author} {\bibfnamefont {J.-M.}\ \bibnamefont
  {{Yan}}}\ and\ \bibinfo {author} {\bibfnamefont {Z.-K.}\ \bibnamefont
  {{Guo}}},\ }\href {https://doi.org/10.48550/arXiv.2609.37248} {\bibfield
  {journal} {\bibinfo  {journal} {arXiv e-prints}\ ,\ \bibinfo {eid}
  {arXiv:2609.37248}} (\bibinfo {year} {2026})},\ \Eprint
  {https://arxiv.org/abs/2609.37248} {arXiv:2609.37248 [gr-qc]} \BibitemShut
  {NoStop}%
\bibitem [{\citenamefont {{Kume}}\ \emph {et~al.}(2025)\citenamefont {{Kume}},
  \citenamefont {{Peloso}}, \citenamefont {{Pieroni}},\ and\ \citenamefont
  {{Ricciardone}}}]{Kume:2025:JCAP}%
  \BibitemOpen
  \bibfield  {author} {\bibinfo {author} {\bibfnamefont {J.}~\bibnamefont
  {{Kume}}}, \bibinfo {author} {\bibfnamefont {M.}~\bibnamefont {{Peloso}}},
  \bibinfo {author} {\bibfnamefont {M.}~\bibnamefont {{Pieroni}}},\ and\
  \bibinfo {author} {\bibfnamefont {A.}~\bibnamefont {{Ricciardone}}},\ }\href
  {https://doi.org/10.1088/1475-7516/2025/06/030} {\bibfield  {journal}
  {\bibinfo  {journal} {\jcap}\ }\textbf {\bibinfo {volume} {2025}},\ \bibinfo
  {eid} {030} (\bibinfo {year} {2025})},\ \Eprint
  {https://arxiv.org/abs/2410.10342} {arXiv:2410.10342 [gr-qc]} \BibitemShut
  {NoStop}%
\bibitem [{\citenamefont {Karnesis}\ \emph {et~al.}(2026)\citenamefont
  {Karnesis}, \citenamefont {Baghi}, \citenamefont {Bayle},\ and\ \citenamefont
  {Galanis}}]{karnesis2026noise}%
  \BibitemOpen
  \bibfield  {author} {\bibinfo {author} {\bibfnamefont {N.}~\bibnamefont
  {Karnesis}}, \bibinfo {author} {\bibfnamefont {Q.}~\bibnamefont {Baghi}},
  \bibinfo {author} {\bibfnamefont {J.-B.}\ \bibnamefont {Bayle}},\ and\
  \bibinfo {author} {\bibfnamefont {N.}~\bibnamefont {Galanis}},\ }\href@noop
  {} {\bibfield  {journal} {\bibinfo  {journal} {arXiv}\ } (\bibinfo {year}
  {2026})},\ \bibinfo {note} {arXiv:2601.19741}\BibitemShut {NoStop}%
\bibitem [{\citenamefont {Kirch}\ \emph {et~al.}(2019)\citenamefont {Kirch},
  \citenamefont {Edwards}, \citenamefont {Meier},\ and\ \citenamefont
  {Meyer}}]{beyond_whittle}%
  \BibitemOpen
  \bibfield  {author} {\bibinfo {author} {\bibfnamefont {C.}~\bibnamefont
  {Kirch}}, \bibinfo {author} {\bibfnamefont {M.~C.}\ \bibnamefont {Edwards}},
  \bibinfo {author} {\bibfnamefont {A.}~\bibnamefont {Meier}},\ and\ \bibinfo
  {author} {\bibfnamefont {R.}~\bibnamefont {Meyer}},\ }\href
  {https://doi.org/10.1214/18-BA1126} {\bibfield  {journal} {\bibinfo
  {journal} {Bayesian Analysis}\ }\textbf {\bibinfo {volume} {14}},\ \bibinfo
  {pages} {1037 } (\bibinfo {year} {2019})}\BibitemShut {NoStop}%
\bibitem [{\citenamefont {Edwards}\ \emph {et~al.}(2019)\citenamefont
  {Edwards}, \citenamefont {Meyer},\ and\ \citenamefont
  {Christensen}}]{edwards_bayesian_2019}%
  \BibitemOpen
  \bibfield  {author} {\bibinfo {author} {\bibfnamefont {M.~C.}\ \bibnamefont
  {Edwards}}, \bibinfo {author} {\bibfnamefont {R.}~\bibnamefont {Meyer}},\
  and\ \bibinfo {author} {\bibfnamefont {N.}~\bibnamefont {Christensen}},\
  }\href {https://doi.org/10.1007/s11222-017-9796-9} {\bibfield  {journal}
  {\bibinfo  {journal} {Statistics and Computing}\ }\textbf {\bibinfo {volume}
  {29}},\ \bibinfo {pages} {67} (\bibinfo {year} {2019})}\BibitemShut {NoStop}%
\bibitem [{\citenamefont {Eilers}\ and\ \citenamefont
  {Marx}(1996)}]{eilers1996flexible}%
  \BibitemOpen
  \bibfield  {author} {\bibinfo {author} {\bibfnamefont {P.~H.~C.}\
  \bibnamefont {Eilers}}\ and\ \bibinfo {author} {\bibfnamefont {B.~D.}\
  \bibnamefont {Marx}},\ }\href {https://doi.org/10.1214/ss/1038425655}
  {\bibfield  {journal} {\bibinfo  {journal} {Statistical Science}\ }\textbf
  {\bibinfo {volume} {11}},\ \bibinfo {pages} {89} (\bibinfo {year}
  {1996})}\BibitemShut {NoStop}%
\bibitem [{\citenamefont {Maturana-Russel}\ and\ \citenamefont
  {Meyer}(2021)}]{maturana2021bayesian}%
  \BibitemOpen
  \bibfield  {author} {\bibinfo {author} {\bibfnamefont {P.}~\bibnamefont
  {Maturana-Russel}}\ and\ \bibinfo {author} {\bibfnamefont {R.}~\bibnamefont
  {Meyer}},\ }\href@noop {} {\bibfield  {journal} {\bibinfo  {journal}
  {Computational Statistics}\ }\textbf {\bibinfo {volume} {36}},\ \bibinfo
  {pages} {2055} (\bibinfo {year} {2021})}\BibitemShut {NoStop}%
\bibitem [{\citenamefont {Whittle}(1953)}]{whittle1953analysis}%
  \BibitemOpen
  \bibfield  {author} {\bibinfo {author} {\bibfnamefont {P.}~\bibnamefont
  {Whittle}},\ }\href@noop {} {\bibfield  {journal} {\bibinfo  {journal}
  {Journal of the Royal Statistical Society: Series B}\ }\textbf {\bibinfo
  {volume} {15}},\ \bibinfo {pages} {125} (\bibinfo {year} {1953})}\BibitemShut
  {NoStop}%
\bibitem [{\citenamefont {Pawitan}\ and\ \citenamefont
  {O'Sullivan}(1994)}]{pawitan1994nonparametric}%
  \BibitemOpen
  \bibfield  {author} {\bibinfo {author} {\bibfnamefont {Y.}~\bibnamefont
  {Pawitan}}\ and\ \bibinfo {author} {\bibfnamefont {F.}~\bibnamefont
  {O'Sullivan}},\ }\href@noop {} {\bibfield  {journal} {\bibinfo  {journal}
  {Journal of the American Statistical Association}\ }\textbf {\bibinfo
  {volume} {89}},\ \bibinfo {pages} {600} (\bibinfo {year} {1994})}\BibitemShut
  {NoStop}%
\bibitem [{\citenamefont {Littenberg}\ and\ \citenamefont
  {Cornish}(2015)}]{littenberg_bayesian_2015}%
  \BibitemOpen
  \bibfield  {author} {\bibinfo {author} {\bibfnamefont {T.~B.}\ \bibnamefont
  {Littenberg}}\ and\ \bibinfo {author} {\bibfnamefont {N.~J.}\ \bibnamefont
  {Cornish}},\ }\href {https://doi.org/10.1103/PhysRevD.91.084034} {\bibfield
  {journal} {\bibinfo  {journal} {Physical Review D}\ }\textbf {\bibinfo
  {volume} {91}},\ \bibinfo {pages} {084034} (\bibinfo {year}
  {2015})}\BibitemShut {NoStop}%
\bibitem [{\citenamefont {Karnesis}\ \emph {et~al.}(2021)\citenamefont
  {Karnesis}, \citenamefont {Babak}, \citenamefont {Pieroni}, \citenamefont
  {Cornish},\ and\ \citenamefont {Littenberg}}]{karnesis2021characterization}%
  \BibitemOpen
  \bibfield  {author} {\bibinfo {author} {\bibfnamefont {N.}~\bibnamefont
  {Karnesis}}, \bibinfo {author} {\bibfnamefont {S.}~\bibnamefont {Babak}},
  \bibinfo {author} {\bibfnamefont {M.}~\bibnamefont {Pieroni}}, \bibinfo
  {author} {\bibfnamefont {N.}~\bibnamefont {Cornish}},\ and\ \bibinfo {author}
  {\bibfnamefont {T.}~\bibnamefont {Littenberg}},\ }\href
  {https://doi.org/10.1103/PhysRevD.104.043019} {\bibfield  {journal} {\bibinfo
   {journal} {Physical Review D}\ }\textbf {\bibinfo {volume} {104}},\ \bibinfo
  {pages} {043019} (\bibinfo {year} {2021})}\BibitemShut {NoStop}%
\bibitem [{\citenamefont {Rosati}\ and\ \citenamefont
  {Littenberg}(2025)}]{rosati2025prototype}%
  \BibitemOpen
  \bibfield  {author} {\bibinfo {author} {\bibfnamefont {R.}~\bibnamefont
  {Rosati}}\ and\ \bibinfo {author} {\bibfnamefont {T.~B.}\ \bibnamefont
  {Littenberg}},\ }\href {https://doi.org/10.1103/g48w-lblc} {\bibfield
  {journal} {\bibinfo  {journal} {Physical Review D}\ }\textbf {\bibinfo
  {volume} {112}},\ \bibinfo {pages} {084060} (\bibinfo {year}
  {2025})}\BibitemShut {NoStop}%
\bibitem [{\citenamefont {{Baghi}}\ \emph {et~al.}(2023)\citenamefont
  {{Baghi}}, \citenamefont {{Karnesis}}, \citenamefont {{Bayle}}, \citenamefont
  {{Besan{\c{c}}on}},\ and\ \citenamefont {{Inchausp{\'e}}}}]{Baghi_2023}%
  \BibitemOpen
  \bibfield  {author} {\bibinfo {author} {\bibfnamefont {Q.}~\bibnamefont
  {{Baghi}}}, \bibinfo {author} {\bibfnamefont {N.}~\bibnamefont {{Karnesis}}},
  \bibinfo {author} {\bibfnamefont {J.-B.}\ \bibnamefont {{Bayle}}}, \bibinfo
  {author} {\bibfnamefont {M.}~\bibnamefont {{Besan{\c{c}}on}}},\ and\ \bibinfo
  {author} {\bibfnamefont {H.}~\bibnamefont {{Inchausp{\'e}}}},\ }\href
  {https://doi.org/10.1088/1475-7516/2023/04/066} {\bibfield  {journal}
  {\bibinfo  {journal} {\jcap}\ }\textbf {\bibinfo {volume} {2023}},\ \bibinfo
  {eid} {066} (\bibinfo {year} {2023})},\ \Eprint
  {https://arxiv.org/abs/2302.12573} {arXiv:2302.12573 [gr-qc]} \BibitemShut
  {NoStop}%
\bibitem [{\citenamefont {{Santini}}\ \emph {et~al.}(2025)\citenamefont
  {{Santini}}, \citenamefont {{Muratore}}, \citenamefont {{Gair}},\ and\
  \citenamefont {{Hartwig}}}]{santini2025sgwb}%
  \BibitemOpen
  \bibfield  {author} {\bibinfo {author} {\bibfnamefont {A.}~\bibnamefont
  {{Santini}}}, \bibinfo {author} {\bibfnamefont {M.}~\bibnamefont
  {{Muratore}}}, \bibinfo {author} {\bibfnamefont {J.}~\bibnamefont {{Gair}}},\
  and\ \bibinfo {author} {\bibfnamefont {O.}~\bibnamefont {{Hartwig}}},\ }\href
  {https://doi.org/10.1103/csx9-9trp} {\bibfield  {journal} {\bibinfo
  {journal} {\prd}\ }\textbf {\bibinfo {volume} {112}},\ \bibinfo {eid}
  {084050} (\bibinfo {year} {2025})},\ \Eprint
  {https://arxiv.org/abs/2507.06300} {arXiv:2507.06300 [gr-qc]} \BibitemShut
  {NoStop}%
\bibitem [{\citenamefont {Aimen}\ \emph {et~al.}(2026)\citenamefont {Aimen},
  \citenamefont {Maturana-Russel}, \citenamefont {Vajpeyi}, \citenamefont
  {Christensen},\ and\ \citenamefont {Meyer}}]{Aimen:2026}%
  \BibitemOpen
  \bibfield  {author} {\bibinfo {author} {\bibfnamefont {N.}~\bibnamefont
  {Aimen}}, \bibinfo {author} {\bibfnamefont {P.}~\bibnamefont
  {Maturana-Russel}}, \bibinfo {author} {\bibfnamefont {A.}~\bibnamefont
  {Vajpeyi}}, \bibinfo {author} {\bibfnamefont {N.}~\bibnamefont
  {Christensen}},\ and\ \bibinfo {author} {\bibfnamefont {R.}~\bibnamefont
  {Meyer}},\ }\href {https://doi.org/10.1103/dcb6-1jsl} {\bibfield  {journal}
  {\bibinfo  {journal} {Phys. Rev. D}\ }\textbf {\bibinfo {volume} {113}},\
  \bibinfo {pages} {024022} (\bibinfo {year} {2026})}\BibitemShut {NoStop}%
\bibitem [{\citenamefont {Hoy}\ \emph {et~al.}(2026)\citenamefont {Hoy},
  \citenamefont {Bayley}, \citenamefont {Burke},\ and\ \citenamefont
  {Nuttall}}]{hoy2026noisypipelinesiinonparametric}%
  \BibitemOpen
  \bibfield  {author} {\bibinfo {author} {\bibfnamefont {C.}~\bibnamefont
  {Hoy}}, \bibinfo {author} {\bibfnamefont {J.}~\bibnamefont {Bayley}},
  \bibinfo {author} {\bibfnamefont {O.}~\bibnamefont {Burke}},\ and\ \bibinfo
  {author} {\bibfnamefont {L.~K.}\ \bibnamefont {Nuttall}},\ }\href
  {https://arxiv.org/abs/2611.03911} {\bibinfo {title} {Noisy pipelines ii:
  Non-parametric noise inference for lisa}} (\bibinfo {year} {2026}),\ \Eprint
  {https://arxiv.org/abs/2611.03911} {arXiv:2611.03911 [gr-qc]} \BibitemShut
  {NoStop}%
\bibitem [{\citenamefont {Vajpeyi}\ \emph
  {et~al.}(2026{\natexlab{a}})\citenamefont {Vajpeyi}, \citenamefont {Meyer},
  \citenamefont {Maturana-Russel},\ and\ \citenamefont
  {Liu}}]{vajpeyi2026multivar}%
  \BibitemOpen
  \bibfield  {author} {\bibinfo {author} {\bibfnamefont {A.}~\bibnamefont
  {Vajpeyi}}, \bibinfo {author} {\bibfnamefont {R.}~\bibnamefont {Meyer}},
  \bibinfo {author} {\bibfnamefont {P.}~\bibnamefont {Maturana-Russel}},\ and\
  \bibinfo {author} {\bibfnamefont {J.}~\bibnamefont {Liu}},\ }\href
  {https://arxiv.org/abs/2607.04833} {\bibinfo {title} {Multivariate bayesian
  p-spline estimation of spectral density matrices, with application to lisa
  tdi noise}} (\bibinfo {year} {2026}{\natexlab{a}}),\ \Eprint
  {https://arxiv.org/abs/2607.04833} {arXiv:2607.04833 [stat.ME]} \BibitemShut
  {NoStop}%
\bibitem [{\citenamefont {Liu}\ \emph {et~al.}(2024)\citenamefont {Liu},
  \citenamefont {Vajpeyi}, \citenamefont {Meyer}, \citenamefont {Janssens},
  \citenamefont {Lee}, \citenamefont {Maturana-Russel}, \citenamefont
  {Christensen},\ and\ \citenamefont {Liu}}]{jiananET}%
  \BibitemOpen
  \bibfield  {author} {\bibinfo {author} {\bibfnamefont {J.}~\bibnamefont
  {Liu}}, \bibinfo {author} {\bibfnamefont {A.}~\bibnamefont {Vajpeyi}},
  \bibinfo {author} {\bibfnamefont {R.}~\bibnamefont {Meyer}}, \bibinfo
  {author} {\bibfnamefont {K.}~\bibnamefont {Janssens}}, \bibinfo {author}
  {\bibfnamefont {J.~E.}\ \bibnamefont {Lee}}, \bibinfo {author} {\bibfnamefont
  {P.}~\bibnamefont {Maturana-Russel}}, \bibinfo {author} {\bibfnamefont
  {N.}~\bibnamefont {Christensen}},\ and\ \bibinfo {author} {\bibfnamefont
  {Y.}~\bibnamefont {Liu}},\ }\href {https://arxiv.org/abs/2409.13224}
  {\bibfield  {journal} {\bibinfo  {journal} {arXiv preprint}\ }\textbf
  {\bibinfo {volume} {2409.13224}} (\bibinfo {year} {2024})}\BibitemShut
  {NoStop}%
\bibitem [{\citenamefont {{Liu}}\ \emph {et~al.}(2026)\citenamefont {{Liu}},
  \citenamefont {{Vajpeyi}}, \citenamefont {{Meyer}}, \citenamefont {{Lee}},\
  and\ \citenamefont {{Maturana-Russel}}}]{jiananLISA}%
  \BibitemOpen
  \bibfield  {author} {\bibinfo {author} {\bibfnamefont {J.}~\bibnamefont
  {{Liu}}}, \bibinfo {author} {\bibfnamefont {A.}~\bibnamefont {{Vajpeyi}}},
  \bibinfo {author} {\bibfnamefont {R.}~\bibnamefont {{Meyer}}}, \bibinfo
  {author} {\bibfnamefont {J.~E.}\ \bibnamefont {{Lee}}},\ and\ \bibinfo
  {author} {\bibfnamefont {P.}~\bibnamefont {{Maturana-Russel}}},\ }\href
  {https://doi.org/10.48550/arXiv.2608.22245} {\bibfield  {journal} {\bibinfo
  {journal} {arXiv e-prints}\ ,\ \bibinfo {eid} {arXiv:2608.22245}} (\bibinfo
  {year} {2026})},\ \Eprint {https://arxiv.org/abs/2608.22245}
  {arXiv:2608.22245 [astro-ph.IM]} \BibitemShut {NoStop}%
\bibitem [{\citenamefont {Liu}\ \emph {et~al.}(2026)\citenamefont {Liu},
  \citenamefont {Meyer}, \citenamefont {Christensen}, \citenamefont {Lee},
  \citenamefont {Liu}, \citenamefont {Maturana-Russel},\ and\ \citenamefont
  {Vajpeyi}}]{yixuanLISAandET}%
  \BibitemOpen
  \bibfield  {author} {\bibinfo {author} {\bibfnamefont {Y.}~\bibnamefont
  {Liu}}, \bibinfo {author} {\bibfnamefont {R.}~\bibnamefont {Meyer}}, \bibinfo
  {author} {\bibfnamefont {N.}~\bibnamefont {Christensen}}, \bibinfo {author}
  {\bibfnamefont {J.~E.}\ \bibnamefont {Lee}}, \bibinfo {author} {\bibfnamefont
  {J.}~\bibnamefont {Liu}}, \bibinfo {author} {\bibfnamefont {P.}~\bibnamefont
  {Maturana-Russel}},\ and\ \bibinfo {author} {\bibfnamefont {A.}~\bibnamefont
  {Vajpeyi}},\ }\href {https://arxiv.org/abs/2607.26619} {\bibfield  {journal}
  {\bibinfo  {journal} {Phys. Rev. D}\ } (\bibinfo {year} {2026})},\ \bibinfo
  {note} {submitted},\ \Eprint {https://arxiv.org/abs/2607.26619}
  {arXiv:2607.26619 [gr-qc]} \BibitemShut {NoStop}%
\bibitem [{\citenamefont {Burke}\ \emph
  {et~al.}(2026{\natexlab{a}})\citenamefont {Burke}, \citenamefont {Bayley},
  \citenamefont {Hoy},\ and\ \citenamefont
  {Nuttall}}]{burke2026noisypipelinesiendtoend}%
  \BibitemOpen
  \bibfield  {author} {\bibinfo {author} {\bibfnamefont {O.}~\bibnamefont
  {Burke}}, \bibinfo {author} {\bibfnamefont {J.}~\bibnamefont {Bayley}},
  \bibinfo {author} {\bibfnamefont {C.}~\bibnamefont {Hoy}},\ and\ \bibinfo
  {author} {\bibfnamefont {L.}~\bibnamefont {Nuttall}},\ }\href
  {https://arxiv.org/abs/2610.28726} {\bibinfo {title} {Noisy pipelines i: An
  end-to-end pipeline to perform multivariate noise inference for lisa}}
  (\bibinfo {year} {2026}{\natexlab{a}}),\ \Eprint
  {https://arxiv.org/abs/2610.28726} {arXiv:2610.28726 [gr-qc]} \BibitemShut
  {NoStop}%
\bibitem [{\citenamefont {Pozzoli}\ \emph {et~al.}(2024)\citenamefont
  {Pozzoli}, \citenamefont {Buscicchio}, \citenamefont {Moore}, \citenamefont
  {Haardt},\ and\ \citenamefont {Sesana}}]{pozzoli2024weakly}%
  \BibitemOpen
  \bibfield  {author} {\bibinfo {author} {\bibfnamefont {F.}~\bibnamefont
  {Pozzoli}}, \bibinfo {author} {\bibfnamefont {R.}~\bibnamefont {Buscicchio}},
  \bibinfo {author} {\bibfnamefont {C.~J.}\ \bibnamefont {Moore}}, \bibinfo
  {author} {\bibfnamefont {F.}~\bibnamefont {Haardt}},\ and\ \bibinfo {author}
  {\bibfnamefont {A.}~\bibnamefont {Sesana}},\ }\href
  {https://doi.org/10.1103/PhysRevD.109.083029} {\bibfield  {journal} {\bibinfo
   {journal} {Physical Review D}\ }\textbf {\bibinfo {volume} {109}},\ \bibinfo
  {pages} {083029} (\bibinfo {year} {2024})}\BibitemShut {NoStop}%
\bibitem [{\citenamefont {Priestley}(1965)}]{priestley1965evolutionary}%
  \BibitemOpen
  \bibfield  {author} {\bibinfo {author} {\bibfnamefont {M.~B.}\ \bibnamefont
  {Priestley}},\ }\href {https://doi.org/10.1111/j.2517-6161.1965.tb01488.x}
  {\bibfield  {journal} {\bibinfo  {journal} {Journal of the Royal Statistical
  Society: Series B}\ }\textbf {\bibinfo {volume} {27}},\ \bibinfo {pages}
  {204} (\bibinfo {year} {1965})}\BibitemShut {NoStop}%
\bibitem [{\citenamefont {Dahlhaus}(1997)}]{dahlhaus1997fitting}%
  \BibitemOpen
  \bibfield  {author} {\bibinfo {author} {\bibfnamefont {R.}~\bibnamefont
  {Dahlhaus}},\ }\href {https://doi.org/10.1214/aos/1034276620} {\bibfield
  {journal} {\bibinfo  {journal} {The Annals of Statistics}\ }\textbf {\bibinfo
  {volume} {25}},\ \bibinfo {pages} {1} (\bibinfo {year} {1997})}\BibitemShut
  {NoStop}%
\bibitem [{\citenamefont {Rosen}\ \emph {et~al.}(2012)\citenamefont {Rosen},
  \citenamefont {Wood},\ and\ \citenamefont {Stoffer}}]{rosen2012adaptspec}%
  \BibitemOpen
  \bibfield  {author} {\bibinfo {author} {\bibfnamefont {O.}~\bibnamefont
  {Rosen}}, \bibinfo {author} {\bibfnamefont {S.}~\bibnamefont {Wood}},\ and\
  \bibinfo {author} {\bibfnamefont {D.~S.}\ \bibnamefont {Stoffer}},\ }\href
  {https://doi.org/10.1080/01621459.2012.716340} {\bibfield  {journal}
  {\bibinfo  {journal} {Journal of the American Statistical Association}\
  }\textbf {\bibinfo {volume} {107}},\ \bibinfo {pages} {1575} (\bibinfo {year}
  {2012})}\BibitemShut {NoStop}%
\bibitem [{\citenamefont {Rosen}\ \emph {et~al.}(2009)\citenamefont {Rosen},
  \citenamefont {Stoffer},\ and\ \citenamefont {Wood}}]{rosen2009mixture}%
  \BibitemOpen
  \bibfield  {author} {\bibinfo {author} {\bibfnamefont {O.}~\bibnamefont
  {Rosen}}, \bibinfo {author} {\bibfnamefont {D.~S.}\ \bibnamefont {Stoffer}},\
  and\ \bibinfo {author} {\bibfnamefont {S.}~\bibnamefont {Wood}},\ }\href@noop
  {} {\bibfield  {journal} {\bibinfo  {journal} {Journal of the American
  Statistical Association}\ }\textbf {\bibinfo {volume} {104}},\ \bibinfo
  {pages} {249} (\bibinfo {year} {2009})}\BibitemShut {NoStop}%
\bibitem [{\citenamefont {Ombao}\ \emph {et~al.}(2001)\citenamefont {Ombao},
  \citenamefont {Raz}, \citenamefont {von Sachs},\ and\ \citenamefont
  {Malow}}]{ombao2001slex}%
  \BibitemOpen
  \bibfield  {author} {\bibinfo {author} {\bibfnamefont {H.~C.}\ \bibnamefont
  {Ombao}}, \bibinfo {author} {\bibfnamefont {J.~A.}\ \bibnamefont {Raz}},
  \bibinfo {author} {\bibfnamefont {R.}~\bibnamefont {von Sachs}},\ and\
  \bibinfo {author} {\bibfnamefont {B.~A.}\ \bibnamefont {Malow}},\ }\href
  {https://doi.org/10.1198/016214501753168244} {\bibfield  {journal} {\bibinfo
  {journal} {Journal of the American Statistical Association}\ }\textbf
  {\bibinfo {volume} {96}},\ \bibinfo {pages} {543} (\bibinfo {year}
  {2001})}\BibitemShut {NoStop}%
\bibitem [{\citenamefont {Ombao}\ \emph {et~al.}(2005)\citenamefont {Ombao},
  \citenamefont {von Sachs},\ and\ \citenamefont {Guo}}]{ombao2005slex}%
  \BibitemOpen
  \bibfield  {author} {\bibinfo {author} {\bibfnamefont {H.}~\bibnamefont
  {Ombao}}, \bibinfo {author} {\bibfnamefont {R.}~\bibnamefont {von Sachs}},\
  and\ \bibinfo {author} {\bibfnamefont {W.}~\bibnamefont {Guo}},\ }\href@noop
  {} {\bibfield  {journal} {\bibinfo  {journal} {Journal of the American
  Statistical Association}\ }\textbf {\bibinfo {volume} {100}},\ \bibinfo
  {pages} {519} (\bibinfo {year} {2005})}\BibitemShut {NoStop}%
\bibitem [{\citenamefont {Guo}\ \emph {et~al.}(2003)\citenamefont {Guo},
  \citenamefont {Dai}, \citenamefont {Ombao},\ and\ \citenamefont {von
  Sachs}}]{guo2003smoothing}%
  \BibitemOpen
  \bibfield  {author} {\bibinfo {author} {\bibfnamefont {W.}~\bibnamefont
  {Guo}}, \bibinfo {author} {\bibfnamefont {M.}~\bibnamefont {Dai}}, \bibinfo
  {author} {\bibfnamefont {H.~C.}\ \bibnamefont {Ombao}},\ and\ \bibinfo
  {author} {\bibfnamefont {R.}~\bibnamefont {von Sachs}},\ }\href@noop {}
  {\bibfield  {journal} {\bibinfo  {journal} {Journal of the American
  Statistical Association}\ }\textbf {\bibinfo {volume} {98}},\ \bibinfo
  {pages} {643} (\bibinfo {year} {2003})}\BibitemShut {NoStop}%
\bibitem [{\citenamefont {Wolfe}\ \emph {et~al.}(2004)\citenamefont {Wolfe},
  \citenamefont {Godsill},\ and\ \citenamefont {Ng}}]{wolfe2004bayesian}%
  \BibitemOpen
  \bibfield  {author} {\bibinfo {author} {\bibfnamefont {P.~J.}\ \bibnamefont
  {Wolfe}}, \bibinfo {author} {\bibfnamefont {S.~J.}\ \bibnamefont {Godsill}},\
  and\ \bibinfo {author} {\bibfnamefont {W.-J.}\ \bibnamefont {Ng}},\
  }\href@noop {} {\bibfield  {journal} {\bibinfo  {journal} {Journal of the
  Royal Statistical Society: Series B}\ }\textbf {\bibinfo {volume} {66}},\
  \bibinfo {pages} {575} (\bibinfo {year} {2004})}\BibitemShut {NoStop}%
\bibitem [{\citenamefont {Nason}\ \emph {et~al.}(2000)\citenamefont {Nason},
  \citenamefont {von Sachs},\ and\ \citenamefont
  {Kroisandt}}]{nason2000wavelet}%
  \BibitemOpen
  \bibfield  {author} {\bibinfo {author} {\bibfnamefont {G.~P.}\ \bibnamefont
  {Nason}}, \bibinfo {author} {\bibfnamefont {R.}~\bibnamefont {von Sachs}},\
  and\ \bibinfo {author} {\bibfnamefont {G.}~\bibnamefont {Kroisandt}},\ }\href
  {https://doi.org/10.1111/1467-9868.00231} {\bibfield  {journal} {\bibinfo
  {journal} {Journal of the Royal Statistical Society: Series B}\ }\textbf
  {\bibinfo {volume} {62}},\ \bibinfo {pages} {271} (\bibinfo {year}
  {2000})}\BibitemShut {NoStop}%
\bibitem [{\citenamefont {Pozzoli}\ \emph {et~al.}(2025)\citenamefont
  {Pozzoli}, \citenamefont {Buscicchio}, \citenamefont {Klein},\ and\
  \citenamefont {Chirico}}]{pozzoli2025bahamas}%
  \BibitemOpen
  \bibfield  {author} {\bibinfo {author} {\bibfnamefont {F.}~\bibnamefont
  {Pozzoli}}, \bibinfo {author} {\bibfnamefont {R.}~\bibnamefont {Buscicchio}},
  \bibinfo {author} {\bibfnamefont {A.}~\bibnamefont {Klein}},\ and\ \bibinfo
  {author} {\bibfnamefont {D.}~\bibnamefont {Chirico}},\ }\href
  {https://arxiv.org/abs/2506.22542} {\bibinfo {title} {{Bahamas}: {BAyesian}
  inference with {HAmiltonian} {Montecarlo} for {Astrophysical} {Stochastic}
  background}} (\bibinfo {year} {2025}),\ \Eprint
  {https://arxiv.org/abs/2506.22542} {arXiv:2506.22542 [astro-ph.IM]}
  \BibitemShut {NoStop}%
\bibitem [{\citenamefont {Neal}(2011)}]{neal2011mcmc}%
  \BibitemOpen
  \bibfield  {author} {\bibinfo {author} {\bibfnamefont {R.~M.}\ \bibnamefont
  {Neal}},\ }in\ \href@noop {} {\emph {\bibinfo {booktitle} {Handbook of Markov
  Chain Monte Carlo}}},\ \bibinfo {editor} {edited by\ \bibinfo {editor}
  {\bibfnamefont {S.}~\bibnamefont {Brooks}}, \bibinfo {editor} {\bibfnamefont
  {A.}~\bibnamefont {Gelman}}, \bibinfo {editor} {\bibfnamefont {G.~L.}\
  \bibnamefont {Jones}},\ and\ \bibinfo {editor} {\bibfnamefont {X.-L.}\
  \bibnamefont {Meng}}}\ (\bibinfo  {publisher} {CRC Press},\ \bibinfo {year}
  {2011})\ pp.\ \bibinfo {pages} {113--162}\BibitemShut {NoStop}%
\bibitem [{\citenamefont {Buscicchio}\ \emph
  {et~al.}(2025{\natexlab{b}})\citenamefont {Buscicchio}, \citenamefont
  {Pozzoli}, \citenamefont {Chirico},\ and\ \citenamefont
  {Sesana}}]{buscicchio2025firstyear}%
  \BibitemOpen
  \bibfield  {author} {\bibinfo {author} {\bibfnamefont {R.}~\bibnamefont
  {Buscicchio}}, \bibinfo {author} {\bibfnamefont {F.}~\bibnamefont {Pozzoli}},
  \bibinfo {author} {\bibfnamefont {D.}~\bibnamefont {Chirico}},\ and\ \bibinfo
  {author} {\bibfnamefont {A.}~\bibnamefont {Sesana}},\ }\href
  {https://arxiv.org/abs/2511.03604} {\bibinfo {title} {The first year of lisa
  galactic foreground}} (\bibinfo {year} {2025}{\natexlab{b}}),\ \Eprint
  {https://arxiv.org/abs/2511.03604} {arXiv:2511.03604 [astro-ph.IM]}
  \BibitemShut {NoStop}%
\bibitem [{\citenamefont {Baghi}\ \emph {et~al.}(2026)\citenamefont {Baghi},
  \citenamefont {Karnesis},\ and\ \citenamefont
  {Bayle}}]{baghi2026statisticstimefrequencyaveragedspectra}%
  \BibitemOpen
  \bibfield  {author} {\bibinfo {author} {\bibfnamefont {Q.}~\bibnamefont
  {Baghi}}, \bibinfo {author} {\bibfnamefont {N.}~\bibnamefont {Karnesis}},\
  and\ \bibinfo {author} {\bibfnamefont {J.-B.}\ \bibnamefont {Bayle}},\ }\href
  {https://arxiv.org/abs/2602.12781} {\bibinfo {title} {Statistics of time and
  frequency-averaged spectra in gravitational-wave background searches}}
  (\bibinfo {year} {2026}),\ \Eprint {https://arxiv.org/abs/2602.12781}
  {arXiv:2602.12781 [gr-qc]} \BibitemShut {NoStop}%
\bibitem [{\citenamefont {Necula}\ \emph {et~al.}(2012)\citenamefont {Necula},
  \citenamefont {Klimenko},\ and\ \citenamefont {Mitselmakher}}]{Necula_2012}%
  \BibitemOpen
  \bibfield  {author} {\bibinfo {author} {\bibfnamefont {V.}~\bibnamefont
  {Necula}}, \bibinfo {author} {\bibfnamefont {S.}~\bibnamefont {Klimenko}},\
  and\ \bibinfo {author} {\bibfnamefont {G.}~\bibnamefont {Mitselmakher}},\
  }\href {https://doi.org/10.1088/1742-6596/363/1/012032} {\bibfield  {journal}
  {\bibinfo  {journal} {Journal of Physics: Conference Series}\ }\textbf
  {\bibinfo {volume} {363}},\ \bibinfo {pages} {012032} (\bibinfo {year}
  {2012})}\BibitemShut {NoStop}%
\bibitem [{\citenamefont {Cornish}(2020)}]{cornish2020Wavelets}%
  \BibitemOpen
  \bibfield  {author} {\bibinfo {author} {\bibfnamefont {N.~J.}\ \bibnamefont
  {Cornish}},\ }\href {https://doi.org/10.1103/PhysRevD.102.124038} {\bibfield
  {journal} {\bibinfo  {journal} {Physical Review D}\ }\textbf {\bibinfo
  {volume} {102}},\ \bibinfo {pages} {124038} (\bibinfo {year}
  {2020})}\BibitemShut {NoStop}%
\bibitem [{\citenamefont {Johnson}\ \emph {et~al.}(2026)\citenamefont
  {Johnson}, \citenamefont {Chatziioannou},\ and\ \citenamefont
  {Summers}}]{johnson2026wdmformalism}%
  \BibitemOpen
  \bibfield  {author} {\bibinfo {author} {\bibfnamefont {A.~D.}\ \bibnamefont
  {Johnson}}, \bibinfo {author} {\bibfnamefont {K.}~\bibnamefont
  {Chatziioannou}},\ and\ \bibinfo {author} {\bibfnamefont {J.}~\bibnamefont
  {Summers}},\ }\href {https://arxiv.org/abs/2606.21473} {\bibinfo {title} {The
  {WDM} time-frequency transform in gravitational-wave data analysis {I}:
  Formalism}} (\bibinfo {year} {2026}),\ \Eprint
  {https://arxiv.org/abs/2606.21473} {arXiv:2606.21473 [gr-qc]} \BibitemShut
  {NoStop}%
\bibitem [{\citenamefont {Vajpeyi}\ \emph
  {et~al.}(2026{\natexlab{b}})\citenamefont {Vajpeyi}, \citenamefont
  {Mentasti}, \citenamefont {Baghi}, \citenamefont {Burke},\ and\ \citenamefont
  {Speri}}]{vajpeyi2026wdm}%
  \BibitemOpen
  \bibfield  {author} {\bibinfo {author} {\bibfnamefont {A.}~\bibnamefont
  {Vajpeyi}}, \bibinfo {author} {\bibfnamefont {G.}~\bibnamefont {Mentasti}},
  \bibinfo {author} {\bibfnamefont {Q.}~\bibnamefont {Baghi}}, \bibinfo
  {author} {\bibfnamefont {O.}~\bibnamefont {Burke}},\ and\ \bibinfo {author}
  {\bibfnamefont {L.}~\bibnamefont {Speri}},\ }\href
  {https://arxiv.org/abs/2606.20269} {\bibinfo {title} {An explicit and
  differentiable {Wilson--Daubechies--Meyer} transform for gravitational-wave
  data analysis}} (\bibinfo {year} {2026}{\natexlab{b}}),\ \Eprint
  {https://arxiv.org/abs/2606.20269} {arXiv:2606.20269 [gr-qc]} \BibitemShut
  {NoStop}%
\bibitem [{\citenamefont {Häfner}\ and\ \citenamefont
  {Kirch}(2017)}]{HaefnerKirch}%
  \BibitemOpen
  \bibfield  {author} {\bibinfo {author} {\bibfnamefont {F.}~\bibnamefont
  {Häfner}}\ and\ \bibinfo {author} {\bibfnamefont {C.}~\bibnamefont
  {Kirch}},\ }\href {https://doi.org/10.1111/jtsa.12241} {\bibfield  {journal}
  {\bibinfo  {journal} {Journal of Time Series Analysis}\ }\textbf {\bibinfo
  {volume} {38}},\ \bibinfo {pages} {895} (\bibinfo {year} {2017})}\BibitemShut
  {NoStop}%
\bibitem [{\citenamefont {Tang}\ \emph {et~al.}(2026)\citenamefont {Tang},
  \citenamefont {Kirch}, \citenamefont {Lee},\ and\ \citenamefont
  {Meyer}}]{tang2026MovingPSD}%
  \BibitemOpen
  \bibfield  {author} {\bibinfo {author} {\bibfnamefont {Y.}~\bibnamefont
  {Tang}}, \bibinfo {author} {\bibfnamefont {C.}~\bibnamefont {Kirch}},
  \bibinfo {author} {\bibfnamefont {J.~E.}\ \bibnamefont {Lee}},\ and\ \bibinfo
  {author} {\bibfnamefont {R.}~\bibnamefont {Meyer}},\ }\href
  {https://doi.org/10.1080/01621459.2025.2594191} {\bibfield  {journal}
  {\bibinfo  {journal} {Journal of the American Statistical Association}\ ,\
  \bibinfo {pages} {1}} (\bibinfo {year} {2026})},\ \Eprint
  {https://arxiv.org/abs/2303.11561} {arXiv:2303.11561 [stat.ME]} \BibitemShut
  {NoStop}%
\bibitem [{\citenamefont {Thomson}(1982)}]{thomson1982spectrum}%
  \BibitemOpen
  \bibfield  {author} {\bibinfo {author} {\bibfnamefont {D.~J.}\ \bibnamefont
  {Thomson}},\ }\href {https://doi.org/10.1109/PROC.1982.12433} {\bibfield
  {journal} {\bibinfo  {journal} {Proceedings of the IEEE}\ }\textbf {\bibinfo
  {volume} {70}},\ \bibinfo {pages} {1055} (\bibinfo {year}
  {1982})}\BibitemShut {NoStop}%
\bibitem [{\citenamefont {Chatterji}\ \emph {et~al.}(2004)\citenamefont
  {Chatterji}, \citenamefont {Blackburn}, \citenamefont {Martin},\ and\
  \citenamefont {Katsavounidis}}]{chatterji2004qtransform}%
  \BibitemOpen
  \bibfield  {author} {\bibinfo {author} {\bibfnamefont {S.}~\bibnamefont
  {Chatterji}}, \bibinfo {author} {\bibfnamefont {L.}~\bibnamefont
  {Blackburn}}, \bibinfo {author} {\bibfnamefont {G.}~\bibnamefont {Martin}},\
  and\ \bibinfo {author} {\bibfnamefont {E.}~\bibnamefont {Katsavounidis}},\
  }\href {https://doi.org/10.1088/0264-9381/21/20/024} {\bibfield  {journal}
  {\bibinfo  {journal} {Classical and Quantum Gravity}\ }\textbf {\bibinfo
  {volume} {21}},\ \bibinfo {pages} {S1809} (\bibinfo {year}
  {2004})}\BibitemShut {NoStop}%
\bibitem [{\citenamefont
  {Cornish}(2026{\natexlab{a}})}]{cornishCovar2026wavelet}%
  \BibitemOpen
  \bibfield  {author} {\bibinfo {author} {\bibfnamefont {N.~J.}\ \bibnamefont
  {Cornish}},\ }\href {https://doi.org/10.1103/n5rl-fpqw} {\bibfield  {journal}
  {\bibinfo  {journal} {Physical Review D}\ }\textbf {\bibinfo {volume}
  {114}},\ \bibinfo {pages} {044035} (\bibinfo {year} {2026}{\natexlab{a}})},\
  \Eprint {https://arxiv.org/abs/2511.10632} {arXiv:2511.10632 [gr-qc]}
  \BibitemShut {NoStop}%
\bibitem [{\citenamefont
  {Cornish}(2026{\natexlab{b}})}]{cornish2026modelingnonstationarynoiseapplications}%
  \BibitemOpen
  \bibfield  {author} {\bibinfo {author} {\bibfnamefont {N.~J.}\ \bibnamefont
  {Cornish}},\ }\href {https://doi.org/10.1103/vfvw-rhvf} {\bibfield  {journal}
  {\bibinfo  {journal} {Physical Review D}\ }\textbf {\bibinfo {volume}
  {114}},\ \bibinfo {pages} {064054} (\bibinfo {year} {2026}{\natexlab{b}})},\
  \Eprint {https://arxiv.org/abs/2607.13168} {arXiv:2607.13168 [gr-qc]}
  \BibitemShut {NoStop}%
\bibitem [{\citenamefont {Moore}\ and\ \citenamefont
  {Kinowski}(2026)}]{Moore_WDM_GW_wavelets}%
  \BibitemOpen
  \bibfield  {author} {\bibinfo {author} {\bibfnamefont {C.~J.}\ \bibnamefont
  {Moore}}\ and\ \bibinfo {author} {\bibfnamefont {T.}~\bibnamefont
  {Kinowski}},\ }\href {https://doi.org/10.5281/zenodo.21916336} {\bibinfo
  {title} {{WDM\_GW\_wavelets}}} (\bibinfo {year} {2026})\BibitemShut {NoStop}%
\bibitem [{\citenamefont {Digman}\ and\ \citenamefont
  {Breivik}(2022)}]{matt_code_WDMWaveletTransforms}%
  \BibitemOpen
  \bibfield  {author} {\bibinfo {author} {\bibfnamefont {M.}~\bibnamefont
  {Digman}}\ and\ \bibinfo {author} {\bibfnamefont {K.}~\bibnamefont
  {Breivik}},\ }\href {https://github.com/XGI-MSU/WDMWaveletTransforms}
  {\bibinfo {title} {{WDMWaveletTransforms}}} (\bibinfo {year}
  {2022})\BibitemShut {NoStop}%
\bibitem [{\citenamefont {Cornish}()}]{cornish_code_WDM_Transform}%
  \BibitemOpen
  \bibfield  {author} {\bibinfo {author} {\bibfnamefont {N.}~\bibnamefont
  {Cornish}},\ }\href
  {https://github.com/eXtremeGravityInstitute/WDM_Transform} {\bibinfo {title}
  {{WDM\_Transform}}}\BibitemShut {NoStop}%
\bibitem [{\citenamefont {Farr}(2025)}]{farr_code_WDMWavelets_jl}%
  \BibitemOpen
  \bibfield  {author} {\bibinfo {author} {\bibfnamefont {W.~M.}\ \bibnamefont
  {Farr}},\ }\href {https://github.com/farr/WDMWavelets.jl} {\bibinfo {title}
  {{WDMWavelets.jl}}} (\bibinfo {year} {2025})\BibitemShut {NoStop}%
\bibitem [{\citenamefont {{Johnson}}\ \emph {et~al.}(2026)\citenamefont
  {{Johnson}}, \citenamefont {{Chatziioannou}},\ and\ \citenamefont
  {{Rosati}}}]{JohnsonWDMII2026}%
  \BibitemOpen
  \bibfield  {author} {\bibinfo {author} {\bibfnamefont {A.~D.}\ \bibnamefont
  {{Johnson}}}, \bibinfo {author} {\bibfnamefont {K.}~\bibnamefont
  {{Chatziioannou}}},\ and\ \bibinfo {author} {\bibfnamefont {R.}~\bibnamefont
  {{Rosati}}},\ }\href {https://doi.org/10.48550/arXiv.2610.19002} {\bibfield
  {journal} {\bibinfo  {journal} {arXiv e-prints}\ ,\ \bibinfo {eid}
  {arXiv:2610.19002}} (\bibinfo {year} {2026})},\ \Eprint
  {https://arxiv.org/abs/2610.19002} {arXiv:2610.19002 [gr-qc]} \BibitemShut
  {NoStop}%
\bibitem [{\citenamefont {Eilers}\ and\ \citenamefont
  {Marx}(2003)}]{eilers2003multivariate}%
  \BibitemOpen
  \bibfield  {author} {\bibinfo {author} {\bibfnamefont {P.~H.}\ \bibnamefont
  {Eilers}}\ and\ \bibinfo {author} {\bibfnamefont {B.~D.}\ \bibnamefont
  {Marx}},\ }\href@noop {} {\bibfield  {journal} {\bibinfo  {journal}
  {Chemometrics and Intelligent Laboratory Systems}\ }\textbf {\bibinfo
  {volume} {66}},\ \bibinfo {pages} {159} (\bibinfo {year} {2003})}\BibitemShut
  {NoStop}%
\bibitem [{\citenamefont {Wood}(2017{\natexlab{a}})}]{wood2017generalized}%
  \BibitemOpen
  \bibfield  {author} {\bibinfo {author} {\bibfnamefont {S.~N.}\ \bibnamefont
  {Wood}},\ }\href@noop {} {\emph {\bibinfo {title} {Generalized Additive
  Models: An Introduction with R}}},\ \bibinfo {edition} {2nd}\ ed.\ (\bibinfo
  {publisher} {CRC Press},\ \bibinfo {year} {2017})\BibitemShut {NoStop}%
\bibitem [{\citenamefont {Wood}(2017{\natexlab{b}})}]{wood2017derivative}%
  \BibitemOpen
  \bibfield  {author} {\bibinfo {author} {\bibfnamefont {S.~N.}\ \bibnamefont
  {Wood}},\ }\href {https://doi.org/10.1007/s11222-016-9666-x} {\bibfield
  {journal} {\bibinfo  {journal} {Statistics and Computing}\ }\textbf {\bibinfo
  {volume} {27}},\ \bibinfo {pages} {985} (\bibinfo {year}
  {2017}{\natexlab{b}})}\BibitemShut {NoStop}%
\bibitem [{\citenamefont {Hoffman}\ \emph {et~al.}(2014)\citenamefont
  {Hoffman}, \citenamefont {Gelman} \emph {et~al.}}]{hoffman2014_NUTS}%
  \BibitemOpen
  \bibfield  {author} {\bibinfo {author} {\bibfnamefont {M.~D.}\ \bibnamefont
  {Hoffman}}, \bibinfo {author} {\bibfnamefont {A.}~\bibnamefont {Gelman}},
  \emph {et~al.},\ }\href@noop {} {\bibfield  {journal} {\bibinfo  {journal}
  {J. Mach. Learn. Res.}\ }\textbf {\bibinfo {volume} {15}},\ \bibinfo {pages}
  {1593} (\bibinfo {year} {2014})}\BibitemShut {NoStop}%
\bibitem [{\citenamefont {Phan}\ \emph {et~al.}(2019)\citenamefont {Phan},
  \citenamefont {Pradhan},\ and\ \citenamefont
  {Jankowiak}}]{phan2019composable}%
  \BibitemOpen
  \bibfield  {author} {\bibinfo {author} {\bibfnamefont {D.}~\bibnamefont
  {Phan}}, \bibinfo {author} {\bibfnamefont {N.}~\bibnamefont {Pradhan}},\ and\
  \bibinfo {author} {\bibfnamefont {M.}~\bibnamefont {Jankowiak}},\ }\href@noop
  {} {\bibinfo {title} {Composable effects for flexible and
  acceleration-efficient probabilistic programming}} (\bibinfo {year} {2019}),\
  \Eprint {https://arxiv.org/abs/1912.11554} {arXiv:1912.11554 [cs.PL]}
  \BibitemShut {NoStop}%
\bibitem [{\citenamefont {Bradbury}\ \emph {et~al.}(2018)\citenamefont
  {Bradbury}, \citenamefont {Frostig}, \citenamefont {Hawkins}, \citenamefont
  {Johnson}, \citenamefont {Leary} \emph {et~al.}}]{jax2018github}%
  \BibitemOpen
  \bibfield  {author} {\bibinfo {author} {\bibfnamefont {J.}~\bibnamefont
  {Bradbury}}, \bibinfo {author} {\bibfnamefont {R.}~\bibnamefont {Frostig}},
  \bibinfo {author} {\bibfnamefont {P.}~\bibnamefont {Hawkins}}, \bibinfo
  {author} {\bibfnamefont {M.~J.}\ \bibnamefont {Johnson}}, \bibinfo {author}
  {\bibfnamefont {C.}~\bibnamefont {Leary}}, \emph {et~al.},\ }\href
  {http://github.com/google/jax} {\bibinfo {title} {{JAX}: composable
  transformations of {P}ython+{N}um{P}y programs}} (\bibinfo {year}
  {2018})\BibitemShut {NoStop}%
\bibitem [{\citenamefont {Neal}(2003)}]{neal2003slice}%
  \BibitemOpen
  \bibfield  {author} {\bibinfo {author} {\bibfnamefont {R.~M.}\ \bibnamefont
  {Neal}},\ }\href {https://doi.org/10.1214/aos/1056562461} {\bibfield
  {journal} {\bibinfo  {journal} {The Annals of Statistics}\ }\textbf {\bibinfo
  {volume} {31}},\ \bibinfo {pages} {705} (\bibinfo {year} {2003})}\BibitemShut
  {NoStop}%
\bibitem [{\citenamefont {Papaspiliopoulos}\ \emph {et~al.}(2007)\citenamefont
  {Papaspiliopoulos}, \citenamefont {Roberts},\ and\ \citenamefont
  {Sk{\"o}ld}}]{papaspiliopoulos2007general}%
  \BibitemOpen
  \bibfield  {author} {\bibinfo {author} {\bibfnamefont {O.}~\bibnamefont
  {Papaspiliopoulos}}, \bibinfo {author} {\bibfnamefont {G.~O.}\ \bibnamefont
  {Roberts}},\ and\ \bibinfo {author} {\bibfnamefont {M.}~\bibnamefont
  {Sk{\"o}ld}},\ }\href {https://doi.org/10.1214/088342307000000014} {\bibfield
   {journal} {\bibinfo  {journal} {Statistical Science}\ }\textbf {\bibinfo
  {volume} {22}},\ \bibinfo {pages} {59} (\bibinfo {year} {2007})}\BibitemShut
  {NoStop}%
\bibitem [{\citenamefont {Betancourt}\ and\ \citenamefont
  {Girolami}(2015)}]{betancourt2015hamiltonian}%
  \BibitemOpen
  \bibfield  {author} {\bibinfo {author} {\bibfnamefont {M.}~\bibnamefont
  {Betancourt}}\ and\ \bibinfo {author} {\bibfnamefont {M.}~\bibnamefont
  {Girolami}},\ }in\ \href {https://doi.org/10.1201/b18502-5} {\emph {\bibinfo
  {booktitle} {Current Trends in Bayesian Methodology with Applications}}},\
  \bibinfo {editor} {edited by\ \bibinfo {editor} {\bibfnamefont {S.~K.}\
  \bibnamefont {Upadhyay}}, \bibinfo {editor} {\bibfnamefont {U.}~\bibnamefont
  {Singh}}, \bibinfo {editor} {\bibfnamefont {D.~K.}\ \bibnamefont {Dey}},\
  and\ \bibinfo {editor} {\bibfnamefont {A.}~\bibnamefont {Loganathan}}}\
  (\bibinfo  {publisher} {Chapman and Hall/CRC},\ \bibinfo {year} {2015})\ pp.\
  \bibinfo {pages} {79--101}\BibitemShut {NoStop}%
\bibitem [{\citenamefont {{Gelman}}\ and\ \citenamefont
  {{Rubin}}(1992)}]{gelman1992inference}%
  \BibitemOpen
  \bibfield  {author} {\bibinfo {author} {\bibfnamefont {A.}~\bibnamefont
  {{Gelman}}}\ and\ \bibinfo {author} {\bibfnamefont {D.~B.}\ \bibnamefont
  {{Rubin}}},\ }\href {https://doi.org/10.1214/ss/1177011136} {\bibfield
  {journal} {\bibinfo  {journal} {Statistical Science}\ }\textbf {\bibinfo
  {volume} {7}},\ \bibinfo {pages} {457} (\bibinfo {year} {1992})}\BibitemShut
  {NoStop}%
\bibitem [{\citenamefont {Vehtari}\ \emph {et~al.}(2021)\citenamefont
  {Vehtari}, \citenamefont {Gelman}, \citenamefont {Simpson}, \citenamefont
  {Carpenter},\ and\ \citenamefont {B{\"u}rkner}}]{vehtari2021rank}%
  \BibitemOpen
  \bibfield  {author} {\bibinfo {author} {\bibfnamefont {A.}~\bibnamefont
  {Vehtari}}, \bibinfo {author} {\bibfnamefont {A.}~\bibnamefont {Gelman}},
  \bibinfo {author} {\bibfnamefont {D.}~\bibnamefont {Simpson}}, \bibinfo
  {author} {\bibfnamefont {B.}~\bibnamefont {Carpenter}},\ and\ \bibinfo
  {author} {\bibfnamefont {P.-C.}\ \bibnamefont {B{\"u}rkner}},\ }\href
  {https://doi.org/10.1214/20-BA1221} {\bibfield  {journal} {\bibinfo
  {journal} {Bayesian Analysis}\ }\textbf {\bibinfo {volume} {16}},\ \bibinfo
  {pages} {667} (\bibinfo {year} {2021})}\BibitemShut {NoStop}%
\bibitem [{\citenamefont {{Betancourt}}(2016)}]{Betancourt:2016:arXiv}%
  \BibitemOpen
  \bibfield  {author} {\bibinfo {author} {\bibfnamefont {M.}~\bibnamefont
  {{Betancourt}}},\ }\href {https://doi.org/10.48550/arXiv.1604.00695}
  {\bibfield  {journal} {\bibinfo  {journal} {arXiv e-prints}\ ,\ \bibinfo
  {eid} {arXiv:1604.00695}} (\bibinfo {year} {2016})},\ \Eprint
  {https://arxiv.org/abs/1604.00695} {arXiv:1604.00695 [stat.ME]} \BibitemShut
  {NoStop}%
\bibitem [{\citenamefont {Bayle}\ and\ \citenamefont
  {Hartwig}(2023)}]{bayle2023lisainstrument}%
  \BibitemOpen
  \bibfield  {author} {\bibinfo {author} {\bibfnamefont {J.-B.}\ \bibnamefont
  {Bayle}}\ and\ \bibinfo {author} {\bibfnamefont {O.}~\bibnamefont
  {Hartwig}},\ }\href {https://doi.org/10.1103/PhysRevD.107.083019} {\bibfield
  {journal} {\bibinfo  {journal} {Physical Review D}\ }\textbf {\bibinfo
  {volume} {107}},\ \bibinfo {pages} {083019} (\bibinfo {year}
  {2023})}\BibitemShut {NoStop}%
\bibitem [{\citenamefont {Staab}\ \emph {et~al.}(2023)\citenamefont {Staab},
  \citenamefont {Bayle},\ and\ \citenamefont {Hartwig}}]{staab2023pytdi}%
  \BibitemOpen
  \bibfield  {author} {\bibinfo {author} {\bibfnamefont {M.}~\bibnamefont
  {Staab}}, \bibinfo {author} {\bibfnamefont {J.-B.}\ \bibnamefont {Bayle}},\
  and\ \bibinfo {author} {\bibfnamefont {O.}~\bibnamefont {Hartwig}},\ }\href
  {https://doi.org/10.5281/zenodo.6351736} {\bibinfo {title} {{PyTDI}}}
  (\bibinfo {year} {2023})\BibitemShut {NoStop}%
\bibitem [{\citenamefont {{G{\'o}rski}}\ \emph {et~al.}(2005)\citenamefont
  {{G{\'o}rski}}, \citenamefont {{Hivon}}, \citenamefont {{Banday}},
  \citenamefont {{Wandelt}}, \citenamefont {{Hansen}}, \citenamefont
  {{Reinecke}},\ and\ \citenamefont {{Bartelmann}}}]{Gorski2005}%
  \BibitemOpen
  \bibfield  {author} {\bibinfo {author} {\bibfnamefont {K.~M.}\ \bibnamefont
  {{G{\'o}rski}}}, \bibinfo {author} {\bibfnamefont {E.}~\bibnamefont
  {{Hivon}}}, \bibinfo {author} {\bibfnamefont {A.~J.}\ \bibnamefont
  {{Banday}}}, \bibinfo {author} {\bibfnamefont {B.~D.}\ \bibnamefont
  {{Wandelt}}}, \bibinfo {author} {\bibfnamefont {F.~K.}\ \bibnamefont
  {{Hansen}}}, \bibinfo {author} {\bibfnamefont {M.}~\bibnamefont
  {{Reinecke}}},\ and\ \bibinfo {author} {\bibfnamefont {M.}~\bibnamefont
  {{Bartelmann}}},\ }\href {https://doi.org/10.1086/427976} {\bibfield
  {journal} {\bibinfo  {journal} {\apj}\ }\textbf {\bibinfo {volume} {622}},\
  \bibinfo {pages} {759} (\bibinfo {year} {2005})},\ \Eprint
  {https://arxiv.org/abs/astro-ph/0409513} {arXiv:astro-ph/0409513 [astro-ph]}
  \BibitemShut {NoStop}%
\bibitem [{\citenamefont {Prince}\ \emph {et~al.}(2002)\citenamefont {Prince},
  \citenamefont {Tinto}, \citenamefont {Larson},\ and\ \citenamefont
  {Armstrong}}]{prince2002lisa}%
  \BibitemOpen
  \bibfield  {author} {\bibinfo {author} {\bibfnamefont {T.~A.}\ \bibnamefont
  {Prince}}, \bibinfo {author} {\bibfnamefont {M.}~\bibnamefont {Tinto}},
  \bibinfo {author} {\bibfnamefont {S.~L.}\ \bibnamefont {Larson}},\ and\
  \bibinfo {author} {\bibfnamefont {J.~W.}\ \bibnamefont {Armstrong}},\ }\href
  {https://doi.org/10.1103/PhysRevD.66.122002} {\bibfield  {journal} {\bibinfo
  {journal} {Phys. Rev. D}\ }\textbf {\bibinfo {volume} {66}},\ \bibinfo
  {pages} {122002} (\bibinfo {year} {2002})}\BibitemShut {NoStop}%
\bibitem [{\citenamefont {Colpi}\ \emph {et~al.}(2024)\citenamefont {Colpi}
  \emph {et~al.}}]{LISA:2024hlh}%
  \BibitemOpen
  \bibfield  {author} {\bibinfo {author} {\bibfnamefont {M.}~\bibnamefont
  {Colpi}} \emph {et~al.} (\bibinfo {collaboration} {LISA}),\ }\href@noop {}
  {\bibinfo {title} {{LISA Definition Study Report}}} (\bibinfo {year}
  {2024}),\ \Eprint {https://arxiv.org/abs/2402.07571} {arXiv:2402.07571
  [astro-ph.CO]} \BibitemShut {NoStop}%
\bibitem [{\citenamefont {Burke}\ \emph
  {et~al.}(2026{\natexlab{b}})\citenamefont {Burke}, \citenamefont {Pozzoli},
  \citenamefont {Muratore},\ and\ \citenamefont
  {Gair}}]{burke2026mindsecondgapimpact}%
  \BibitemOpen
  \bibfield  {author} {\bibinfo {author} {\bibfnamefont {O.}~\bibnamefont
  {Burke}}, \bibinfo {author} {\bibfnamefont {F.}~\bibnamefont {Pozzoli}},
  \bibinfo {author} {\bibfnamefont {M.}~\bibnamefont {Muratore}},\ and\
  \bibinfo {author} {\bibfnamefont {J.~R.}\ \bibnamefont {Gair}},\ }\href
  {https://arxiv.org/abs/2609.31550} {\bibinfo {title} {Mind the second gap:
  The impact of gaps on noise and signal parameter inference}} (\bibinfo {year}
  {2026}{\natexlab{b}}),\ \Eprint {https://arxiv.org/abs/2609.31550}
  {arXiv:2609.31550 [gr-qc]} \BibitemShut {NoStop}%
\bibitem [{\citenamefont {Burke}\ \emph
  {et~al.}(2026{\natexlab{c}})\citenamefont {Burke}, \citenamefont {Muratore},\
  and\ \citenamefont {Woan}}]{burke_tdi_gaps}%
  \BibitemOpen
  \bibfield  {author} {\bibinfo {author} {\bibfnamefont {O.}~\bibnamefont
  {Burke}}, \bibinfo {author} {\bibfnamefont {M.}~\bibnamefont {Muratore}},\
  and\ \bibinfo {author} {\bibfnamefont {G.}~\bibnamefont {Woan}},\ }\href
  {https://doi.org/10.1103/xpvf-syrw} {\bibfield  {journal} {\bibinfo
  {journal} {Phys. Rev. Appl.}\ }\textbf {\bibinfo {volume} {25}},\ \bibinfo
  {pages} {034041} (\bibinfo {year} {2026}{\natexlab{c}})}\BibitemShut
  {NoStop}%
\bibitem [{\citenamefont {Lee}\ and\ \citenamefont
  {Durb{\'a}n}(2011)}]{lee2011anova}%
  \BibitemOpen
  \bibfield  {author} {\bibinfo {author} {\bibfnamefont {D.-J.}\ \bibnamefont
  {Lee}}\ and\ \bibinfo {author} {\bibfnamefont {M.}~\bibnamefont
  {Durb{\'a}n}},\ }\href {https://doi.org/10.1177/1471082X1001100104}
  {\bibfield  {journal} {\bibinfo  {journal} {Statistical Modelling}\ }\textbf
  {\bibinfo {volume} {11}},\ \bibinfo {pages} {49} (\bibinfo {year}
  {2011})}\BibitemShut {NoStop}%
\bibitem [{\citenamefont {Lee}\ \emph {et~al.}(2013)\citenamefont {Lee},
  \citenamefont {Durb{\'a}n},\ and\ \citenamefont {Eilers}}]{lee2013nested}%
  \BibitemOpen
  \bibfield  {author} {\bibinfo {author} {\bibfnamefont {D.-J.}\ \bibnamefont
  {Lee}}, \bibinfo {author} {\bibfnamefont {M.}~\bibnamefont {Durb{\'a}n}},\
  and\ \bibinfo {author} {\bibfnamefont {P.~H.~C.}\ \bibnamefont {Eilers}},\
  }\href {https://doi.org/10.1016/j.csda.2012.11.013} {\bibfield  {journal}
  {\bibinfo  {journal} {Computational Statistics \& Data Analysis}\ }\textbf
  {\bibinfo {volume} {61}},\ \bibinfo {pages} {22} (\bibinfo {year}
  {2013})}\BibitemShut {NoStop}%
\bibitem [{\citenamefont {Gu}(2014)}]{gu2014ssanova}%
  \BibitemOpen
  \bibfield  {author} {\bibinfo {author} {\bibfnamefont {C.}~\bibnamefont
  {Gu}},\ }\href {https://doi.org/10.18637/jss.v058.i05} {\bibfield  {journal}
  {\bibinfo  {journal} {Journal of Statistical Software}\ }\textbf {\bibinfo
  {volume} {58}},\ \bibinfo {pages} {1} (\bibinfo {year} {2014})}\BibitemShut
  {NoStop}%
\bibitem [{\citenamefont {Baghi}\ \emph {et~al.}(2019)\citenamefont {Baghi},
  \citenamefont {Thorpe}, \citenamefont {Slutsky}, \citenamefont {Baker},
  \citenamefont {Canton}, \citenamefont {Korsakova},\ and\ \citenamefont
  {Karnesis}}]{baghi2019gaps}%
  \BibitemOpen
  \bibfield  {author} {\bibinfo {author} {\bibfnamefont {Q.}~\bibnamefont
  {Baghi}}, \bibinfo {author} {\bibfnamefont {J.~I.}\ \bibnamefont {Thorpe}},
  \bibinfo {author} {\bibfnamefont {J.}~\bibnamefont {Slutsky}}, \bibinfo
  {author} {\bibfnamefont {J.}~\bibnamefont {Baker}}, \bibinfo {author}
  {\bibfnamefont {T.~D.}\ \bibnamefont {Canton}}, \bibinfo {author}
  {\bibfnamefont {N.}~\bibnamefont {Korsakova}},\ and\ \bibinfo {author}
  {\bibfnamefont {N.}~\bibnamefont {Karnesis}},\ }\href
  {https://doi.org/10.1103/PhysRevD.100.022003} {\bibfield  {journal} {\bibinfo
   {journal} {Physical Review D}\ }\textbf {\bibinfo {volume} {100}},\ \bibinfo
  {pages} {022003} (\bibinfo {year} {2019})}\BibitemShut {NoStop}%
\bibitem [{\citenamefont {Pearson}\ and\ \citenamefont
  {Cornish}(2026)}]{gapsnextgen2026}%
  \BibitemOpen
  \bibfield  {author} {\bibinfo {author} {\bibfnamefont {N.}~\bibnamefont
  {Pearson}}\ and\ \bibinfo {author} {\bibfnamefont {N.~J.}\ \bibnamefont
  {Cornish}},\ }\href {https://doi.org/10.1103/htg6-w4yy} {\bibfield  {journal}
  {\bibinfo  {journal} {Physical Review D}\ }\textbf {\bibinfo {volume}
  {113}},\ \bibinfo {pages} {064033} (\bibinfo {year} {2026})},\ \Eprint
  {https://arxiv.org/abs/2509.05479} {arXiv:2509.05479 [gr-qc]} \BibitemShut
  {NoStop}%
\bibitem [{\citenamefont {Pearson}\ \emph {et~al.}(2026)\citenamefont
  {Pearson}, \citenamefont {{\c C}al{\i}{\c s}kan}, \citenamefont {Bini},\ and\
  \citenamefont {Cornish}}]{pearson2026compact}%
  \BibitemOpen
  \bibfield  {author} {\bibinfo {author} {\bibfnamefont {N.}~\bibnamefont
  {Pearson}}, \bibinfo {author} {\bibfnamefont {M.}~\bibnamefont {{\c
  C}al{\i}{\c s}kan}}, \bibinfo {author} {\bibfnamefont {S.}~\bibnamefont
  {Bini}},\ and\ \bibinfo {author} {\bibfnamefont {N.~J.}\ \bibnamefont
  {Cornish}},\ }\href {https://arxiv.org/abs/2609.05601} {\bibinfo {title} {A
  compact time-frequency representation for gravitational-wave data analysis}}
  (\bibinfo {year} {2026}),\ \Eprint {https://arxiv.org/abs/2609.05601}
  {arXiv:2609.05601 [gr-qc]} \BibitemShut {NoStop}%
\bibitem [{\citenamefont {Ribatet}\ \emph {et~al.}(2012)\citenamefont
  {Ribatet}, \citenamefont {Cooley},\ and\ \citenamefont
  {Davison}}]{ribatet2012bayesian}%
  \BibitemOpen
  \bibfield  {author} {\bibinfo {author} {\bibfnamefont {M.}~\bibnamefont
  {Ribatet}}, \bibinfo {author} {\bibfnamefont {D.}~\bibnamefont {Cooley}},\
  and\ \bibinfo {author} {\bibfnamefont {A.~C.}\ \bibnamefont {Davison}},\
  }\href {https://doi.org/10.5705/ss.2009.248} {\bibfield  {journal} {\bibinfo
  {journal} {Statistica Sinica}\ }\textbf {\bibinfo {volume} {22}},\ \bibinfo
  {pages} {813} (\bibinfo {year} {2012})}\BibitemShut {NoStop}%
\bibitem [{\citenamefont {Vajpeyi}\ \emph
  {et~al.}(2026{\natexlab{c}})\citenamefont {Vajpeyi}, \citenamefont {Meyer},
  \citenamefont {Maturana-Russel},\ and\ \citenamefont
  {Liu}}]{multivariatepspline2026}%
  \BibitemOpen
  \bibfield  {author} {\bibinfo {author} {\bibfnamefont {A.}~\bibnamefont
  {Vajpeyi}}, \bibinfo {author} {\bibfnamefont {R.}~\bibnamefont {Meyer}},
  \bibinfo {author} {\bibfnamefont {P.}~\bibnamefont {Maturana-Russel}},\ and\
  \bibinfo {author} {\bibfnamefont {J.}~\bibnamefont {Liu}},\ }\href
  {https://arxiv.org/abs/2607.04833} {\bibinfo {title} {Multivariate {B}ayesian
  {P}-spline estimation of spectral density matrices, with application to
  {LISA} {TDI} noise}} (\bibinfo {year} {2026}{\natexlab{c}}),\ \Eprint
  {https://arxiv.org/abs/2607.04833} {arXiv:2607.04833 [stat.ME]} \BibitemShut
  {NoStop}%
\bibitem [{\citenamefont {Gupta}\ and\ \citenamefont
  {Cornish}(2024)}]{gupta2024bayesian}%
  \BibitemOpen
  \bibfield  {author} {\bibinfo {author} {\bibfnamefont {T.}~\bibnamefont
  {Gupta}}\ and\ \bibinfo {author} {\bibfnamefont {N.~J.}\ \bibnamefont
  {Cornish}},\ }\href {https://doi.org/10.1103/PhysRevD.109.064040} {\bibfield
  {journal} {\bibinfo  {journal} {Phys. Rev. D}\ }\textbf {\bibinfo {volume}
  {109}},\ \bibinfo {pages} {064040} (\bibinfo {year} {2024})},\ \Eprint
  {https://arxiv.org/abs/2312.11808} {arXiv:2312.11808 [gr-qc]} \BibitemShut
  {NoStop}%
\bibitem [{\citenamefont {Lang}\ and\ \citenamefont
  {Brezger}(2004)}]{lang2004bayesian}%
  \BibitemOpen
  \bibfield  {author} {\bibinfo {author} {\bibfnamefont {S.}~\bibnamefont
  {Lang}}\ and\ \bibinfo {author} {\bibfnamefont {A.}~\bibnamefont {Brezger}},\
  }\href@noop {} {\bibfield  {journal} {\bibinfo  {journal} {Journal of
  Computational and Graphical Statistics}\ }\textbf {\bibinfo {volume} {13}},\
  \bibinfo {pages} {183} (\bibinfo {year} {2004})}\BibitemShut {NoStop}%
\bibitem [{\citenamefont {Boer}(2023)}]{boer2023tensor}%
  \BibitemOpen
  \bibfield  {author} {\bibinfo {author} {\bibfnamefont {M.~P.}\ \bibnamefont
  {Boer}},\ }\href@noop {} {\bibfield  {journal} {\bibinfo  {journal}
  {Statistical Modelling}\ }\textbf {\bibinfo {volume} {23}},\ \bibinfo {pages}
  {465} (\bibinfo {year} {2023})}\BibitemShut {NoStop}%
\end{thebibliography}
\end{document}